%% file: main.tex
\documentclass[10pt,letterpaper]{article}
\usepackage[T1]{fontenc}
\usepackage{iclr2027_conference,times}
\usepackage{amsmath,amssymb,booktabs,longtable,tabularx,multirow}
\usepackage{graphicx,xcolor,tikz,listings}
\usepackage{placeins,float}
\usepackage{hyperref,url}
\usetikzlibrary{arrows.meta,positioning,fit,calc}
\input{paper_commands}

\title{CEO Arena: Evaluating Long-Horizon\\Multi-Agent Decision-Making in Competitive Markets}
\input{arxiv_setup}
\begin{document}
\arxivmaketitle
\begin{abstract}
Long-horizon competition tests agents’ ability to coordinate business decisions under uncertainty and adapt to changing rival strategies. We introduce \textbf{CEO Arena}, a benchmark that uses matched replacement evaluation to assess operating returns alongside an agent’s effects on rivals and the market. Each CEO agent is compared with a reference policy in the same company under the same economic seed, holding other agents’ identities and assignments fixed while all agents adapt. In a shared eight-company market spanning 500 simulated days, CEOs make sequential decisions on pricing, procurement, marketing, research and development, and service using private company information and noisy market signals, under resource constraints and delayed feedback. We evaluate eight LLM-based CEO agents in 27 main runs and 26 robustness runs. In the main evaluation, most agents have negative mean returns, and private gains can accompany market losses. Robustness analyses suggest that aggregate patterns extend beyond the original rule-based baseline; four of the 56 directed pairs show relatively stable effects. Memory, action, and accounting traces suggest demand capture and rivals’ pricing and spending responses as possible explanations. \textbf{CEO Arena provides a controlled testbed for studying long-horizon agent competition, strategic interaction, and market externalities.}

\end{abstract}
\input{figures/figure1}
\input{sections/introduction}
\input{sections/related_work}
\input{sections/environment}

\input{sections/evaluation}
\input{sections/experiments}
\input{sections/discussion}
\label{main-text-end}
\input{sections/statements}
\bibliography{research/verified_ml,research/verified_economics}
\bibliographystyle{iclr2027_conference}
\appendix
\input{appendix/metrics}
\input{appendix/protocol}
\input{appendix/rule_search}

\input{appendix/economy}
\input{appendix/parameter_checks}
\input{appendix/trajectory_evidence}
\input{appendix/reproducibility}
\end{document}

%% file: paper_commands.tex
\newif\ifdraftpaper
\draftpaperfalse
\definecolor{ArenaGold}{HTML}{D4A600}
\definecolor{ArenaInk}{HTML}{18232F}
\definecolor{ArenaTeal}{HTML}{187F85}
\definecolor{ArenaGray}{HTML}{697586}
\definecolor{ArenaLight}{HTML}{F4F6F8}
\definecolor{ArenaBlue}{HTML}{2166AC}
\definecolor{ArenaRed}{HTML}{B2182B}
\definecolor{DraftInk}{HTML}{8A6418}
\newcommand{\arena}{CEO Arena}

\newcommand{\codename}[1]{\texttt{#1}}

\newcolumntype{Y}{>{\raggedright\arraybackslash}X}
\hypersetup{colorlinks=true,linkcolor=ArenaTeal,citecolor=ArenaTeal,urlcolor=ArenaTeal,pdftitle={CEO Arena: Evaluating Long-Horizon Multi-Agent Decision-Making in Competitive Markets}}

\usepackage{tcolorbox}
\tcbuselibrary{breakable,skins,listings}
\tcbset{arenabox/.style={enhanced,breakable,colback=ArenaLight,colframe=ArenaInk!88,coltitle=white,
  fonttitle=\bfseries\small,arc=1.8mm,boxrule=0.6pt,left=3mm,right=3mm,top=2mm,bottom=2mm,
  toptitle=0.6mm,bottomtitle=0.6mm}}
\newtcolorbox{promptbox}[1]{arenabox,title={#1},after={\par\suppressfloats[t]},
  before upper={\footnotesize\ttfamily\raggedright\setlength{\parskip}{1.2pt}\obeylines}}
\newtcblisting{codebox}[1]{arenabox,title={#1},listing only,
  listing options={language=Python,basicstyle=\ttfamily\footnotesize,frame=none,columns=fullflexible,keepspaces=true,
    keywordstyle=\bfseries,showstringspaces=false,aboveskip=0pt,belowskip=0pt}}

%% file: arxiv_setup.tex
\iclrfinalcopy
\newcommand{\authoraffil}[1]{\raisebox{0.5ex}{\fontsize{7}{8}\selectfont #1}}
\author{%
\begin{minipage}[t]{\dimexpr\textwidth-2\tabcolsep\relax}
\raggedright
\normalsize\bfseries
An Yan\authoraffil{1,2}\quad
Yu Huo\authoraffil{1,3}\quad
Zhiwei Shang\authoraffil{1,4}\quad
Yiran Peng\authoraffil{5}\quad
Chenglin Wu\authoraffil{1,*}\\[0.4em]
\small\normalfont
\authoraffil{1}DeepWisdom;
\authoraffil{2}Fudan University;
\authoraffil{3}The Chinese University of Hong Kong;
\authoraffil{4}The Chinese University of Hong Kong, Shenzhen;
\authoraffil{5}The Hong Kong University of Science and Technology (Guangzhou)
\end{minipage}%
}
\hypersetup{
  pdfauthor={An Yan, Yu Huo, Zhiwei Shang, Yiran Peng, Chenglin Wu},
  pdfsubject={Long-horizon multi-agent competition and evaluation},
  pdfkeywords={}
}
\newcommand{\arxivmaketitle}{%
  \maketitle\lhead{Preprint}%
  \begingroup
  \renewcommand{\thefootnote}{\fnsymbol{footnote}}%
  \footnotetext[1]{Corresponding author: Chenglin Wu
    (\href{mailto:alexanderwu@deepwisdom.ai}{\nolinkurl{alexanderwu@deepwisdom.ai}}).}%
  \endgroup
}

%% file: figures/figure1.tex
\begingroup
\setlength{\intextsep}{0pt}
\begin{figure}[H]
\centering
\includegraphics[width=\linewidth]{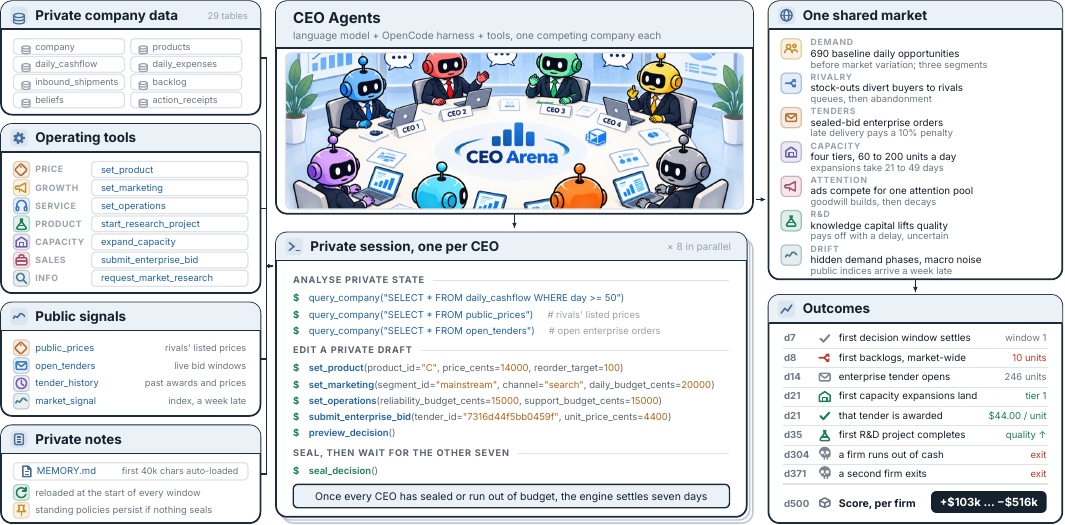}
\caption{\textbf{CEO Arena system overview.} CEO agents use private company data and tools to make operating decisions in a shared market. Each decision window collects their committed policies before advancing the market and returning feedback. The right-hand timeline shows the seed-11 reference run.}
\label{fig:overview}
\end{figure}
\endgroup

%% file: sections/introduction.tex
\section{Introduction}
\label{sec:introduction}
Agent benchmarks evaluate individual agents' tool use and task completion in interactive environments \citep{liu2024agentbench,zhou2024webarena,mialon2023gaia,xie2024osworld}. Evaluation also covers sustained workflows in software engineering, machine-learning experimentation, research replication, and workplace operations \citep{jimenez2023swebench,chan2024mlebench,starace2025paperbench,xu2024agentcompany}. Long-horizon multi-agent evaluations extend this focus to interacting agents' task outcomes and individual contributions \citep{zhu-etal-2025-multiagentbench}.

In such systems, an agent's decisions can generate externalities for other agents and alter system-level outcomes. Their direction, mechanisms, and stability across runs require separate evaluation. Long-horizon business operation provides a concrete setting for studying these effects. Existing benchmarks examine sustained management and resource allocation \citep{backlund2025vending,han2026enterprisearena,chen2026ceobench}, while competitive settings place autonomous firms in shared markets \citep{andonVendingArena,sugiura2026coffeebench,zheng2026marketbench}.

To jointly evaluate private performance and externalities in long-horizon multi-agent systems, we introduce \arena{} and its matched replacement evaluation. We compare each focal agent with a reference policy at the same seat and economic seed, holding the other agents' identities and seats fixed while all agents adapt. We report private gains, each rival's change, and total market change. The protocol runs in a 500-day, eight-company market with private observations, resource constraints, and delayed feedback; LLM-based CEOs share an OpenCode-based harness.

Across nine lineups and three economic seeds, GPT-6 Astra and GPT-5.6 Sol rank first and second in Mean Score. Replacing Rule with Astra improves both private and total market returns in all three seeds, whereas Sol improves private returns in all three but reduces total market returns on average. We also identify four comparatively stable directed effects between agents. Repeated runs and No Action comparisons assess stability; memory and action traces inform mechanism hypotheses.

Our contributions are:
\begin{itemize}
\item \textbf{Matched replacement evaluation.} A protocol separating private gains, directed and aggregate externalities, and total market changes through same-seat reference-policy comparisons.
\item \textbf{Shared-market testbed.} A long-horizon market with private observations, resource constraints, delayed feedback, and adaptive rivals. It tests long-term planning under uncertainty, information gathering from noisy signals, adaptation to changing markets, and coordination of business decisions toward a firm’s goal.
\item \textbf{Empirical evidence.} Private gains can accompany contrasting externalities, with repeated-run and reference-policy checks and hypotheses grounded in memory and actions.
\end{itemize}

%% file: sections/related_work.tex
\section{Related Work}
\label{sec:related-work}
\paragraph{Interactive and long-horizon agents.}
Benchmarks assess sequential action and tool use \citep{shridhar2020alfworld,wang2022scienceworld,qin2023toolllm,wang2023mint}, broad task performance \citep{liu2024agentbench,ma2024agentboard,mialon2023gaia,xi2024agentgym}, and interaction through web, computer, and service interfaces \citep{zhou2024webarena,drouin2024workarena,xie2024osworld,yao2025taubench}. Software engineering, machine-learning experimentation, research replication, and workplace operations extend evaluation to sustained workflows \citep{jimenez2023swebench,huang2023mlagentbench,chan2024mlebench,starace2025paperbench,xu2024agentcompany}. Reasoning, memory, and workflow methods aim to improve these capabilities \citep{yao2022react,shinn2023reflexion,wang2023voyager,zhang2025aflow}.

\paragraph{Multi-agent systems.}
Multi-agent frameworks organize roles and communication for collaborative work \citep{wu2023autogen,li2023camel,hong2023metagpt}; social simulations study interactions among agents \citep{park2023generative,zhou2023sotopia,piao2026agentsociety}. Economic research models agents' decisions and institutions \citep{zheng2022aieconomist,tesfatsion2002ace,horton2023homosilicus}. Game environments examine strategic interaction and resource competition \citep{lanctot2019openspiel,terry2021pettingzoo,suarez2019neuralmmo,baker2019autocurricula,vinyals2019alphastar,bakhtin2022nopress,fair2022cicero}. LLMArena evaluates game performance \citep{chen-etal-2024-llmarena}; MultiAgentBench measures task outcomes, individual contributions, and coordination \citep{zhu-etal-2025-multiagentbench}. Melting Pot supplements focal-population performance with background-population return and inequality metrics to study externalities \citep{leibo2021meltingpot}. Empirical game theory compares strategy profiles \citep{wellman2025egta}, while interference analysis formalizes outcome dependence across units \citep{hudgens2008interference,aronow2017interference}.

\paragraph{Competitive business-agent evaluation.}
Vending-Bench, EnterpriseArena, and CEO-Bench study sustained business operation and resource allocation \citep{backlund2025vending,han2026enterprisearena,chen2026ceobench}. Vending-Bench Arena scores agents competing in a shared market \citep{andonVendingArena}; CoffeeBench evaluates a focal roaster against fixed background agent types \citep{sugiura2026coffeebench}; Market-Bench reports individual metrics alongside market inequality and concentration \citep{zheng2026marketbench}. CEO Arena compares an agent with a reference policy at the same seat and economic seed, holding the other agents' identities and seats fixed. These comparisons separate private gains, directed and aggregate externalities, and total market changes while allowing all firms to adapt.

%% file: sections/environment.tex
\section{The Design Details of CEO Arena}
\label{sec:environment}
\label{sec:benchmark}
\subsection{Design Objectives and Task Overview}
\arena{} places eight firms with identical endowments in a shared product market for 500 simulated days. The engine controls customers, suppliers, and settlement; CEO agents set prices, procurement targets, budgets, projects, and bids. The market has three product tiers and three customer segments. Figure~\ref{fig:overview} shows the interaction loop.

\subsection{Shared Market and Economic Dynamics}
\begin{figure}[t]
\centering
\includegraphics[width=\linewidth]{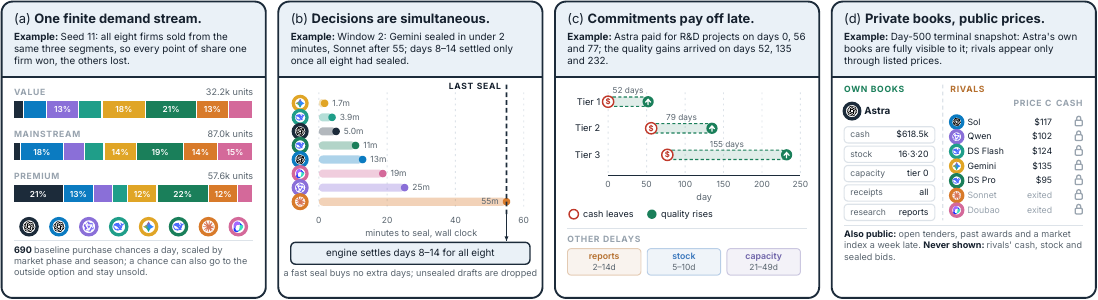}
\caption{\textbf{Shared-market mechanics.} In the seed-11 reference run, the panels show realized sales by segment, one same-version decision window, Astra's delayed projects, and a terminal-state comparison of private and public information. The 690 daily opportunities are a baseline before market variation; Appendix~\ref{app:economy} specifies delay rules and parameter ranges.}
\label{fig:world-mechanics}
\end{figure}

Figure~\ref{fig:world-mechanics} illustrates these mechanisms with examples from the seed-11 reference run.

Finite daily purchase opportunities for segment $h$ are allocated among eligible offers $\mathcal A_{ht}$ and an outside option:
\begin{equation}
P_{ijht}=\frac{\exp(\widetilde v_{ijht})}
{\exp(\widetilde v_h^{\mathrm{out}})+\sum_{(k,r)\in\mathcal A_{ht}}\exp(\widetilde v_{krht})}.
\label{eq:market-choice}
\end{equation}
$P_{ijht}$ is the probability of selecting firm $i$'s product $j$ on day $t$, with stabilized utility $\widetilde v$. The shared logit denominator couples firms' demand; the outside option lets total sales vary \citep{train2009discrete}. Appendix~\ref{app:demand} specifies utilities and allocation.

Inventory, capacity, and cash constrain operations. Agents revise policies every seven days; commitments can span multiple windows (Table~\ref{tab:actions}; Appendices~\ref{app:parameters}--\ref{app:parameter-checks}).

\FloatBarrier
\subsection{CEO Agents and Interaction Protocol}
Each LLM-based CEO uses the shared OpenCode harness, typed tools, and persistent private memory. Within a window, it queries records, stages a private draft, and seals it with \codename{seal\_decision}. New windows reset the conversation, loading \texttt{MEMORY.md} and a factual briefing.

Agents observe their own records and public market information, but not rivals' private states or future shocks. Paid reports provide delayed, noisy estimates of market conditions.

Active CEOs decide from the same world version; settlement waits for their windows to close. Unsealed drafts are discarded; standing policies continue. Appendix~\ref{app:protocol} specifies tools, persistence, and decision budgets.
\input{tables/table1}

%% file: tables/table1.tex
\begin{table}[t]
\centering\small
\caption{Company capabilities and their consequences. ``Next window'' means the next committed policy period; daily execution and longer project delays occur inside the engine. Appendix~\ref{app:economy} gives formulas and sources; Appendix~\ref{app:protocol} specifies tools.}
\label{tab:actions}
\begin{tabularx}{\linewidth}{@{}>{\raggedright\arraybackslash}p{1.9cm}Y Y >{\raggedright\arraybackslash}p{2.35cm}@{}}
\toprule
Capability & Evidence and action & Direct / competitive consequence & Delay \\
\midrule
Analysis & Query private SQL and public quotes & Diagnose policy outcomes; no direct state change & Immediate \\
Pricing & Set product price and promotion & Margin versus jointly allocated demand & Next window \\
Procurement & Inspect inventory; set reorder target & Cash tied in stock; shortages redirect demand & 5--10 days \\
Marketing & Set segment--channel budgets & Attention share and persistent awareness & Daily carryover \\
Innovation & Fund daily development or a project & Product quality changes relative attractiveness & 7 days / 45--150 day medians \\
Operations & Fund reliability and customer support & Failures, queues, satisfaction, future demand & Daily carryover \\
Capacity & Expand, downsize, or cancel construction & Throughput, overhead and sunk-cost trade-offs & 21--49 / 7 days \\
Enterprise sales & Inspect tenders; submit sealed bids & Shared orders with delivery duties and penalties & 7-day bid / delivery \\
Market research & Buy a segment report & Better private estimates, not product quality & 2--14 days \\
\bottomrule
\end{tabularx}
\end{table}

%% file: sections/evaluation.tex
\subsection{Evaluation Design and Metrics}
\label{sec:tracks}
Let $\mathcal M$ contain the eight LLM-based CEOs and $R$ denote Rule CEO. The reference lineup is $L_{-R}=\mathcal M$; in $L_{-B}$, Rule takes agent $B$'s seat while the other seven identities and seats remain fixed. Each world runs independently, allowing rivals to adapt. Comparisons remain conditional on the baseline, lineup, and seed.

A company's terminal Score is
\begin{equation}
S_i=\underbrace{C_{i,T}}_{\text{cash}}
+\underbrace{\sum_j v_j(H_{ij,T}+I^{\mathrm{in}}_{ij,T})}_{\text{inventory salvage}}
-\underbrace{V_{i,0}}_{\text{initial value}}.
\label{eq:score}
\end{equation}
$H$ and $I^{\mathrm{in}}$ denote on-hand and paid inbound stock, $v_j$ the product's liquidation value, and $V_{i,0}$ initial enterprise value. Appendix~\ref{app:metrics} details valuation, including bankrupt firms. Mean Score equally averages a candidate's outcomes across its included lineups and seeds and determines the ranking.

\paragraph{Private returns and externalities.}
For seed $s$, let $S_{A,-B,s}$ be participant $A$'s Score in $L_{-B}$ and $Y_{-B,s}$ the sum of company Scores there. The private difference is $\Delta^{\mathrm{private}}_{B,s}=S_{B,-R,s}-S_{R,-B,s}$; the directed externality on rival $A$ is $\theta_{B\rightarrow A,s}=S_{A,-R,s}-S_{A,-B,s}$. Let $\Delta^{\mathrm{others}}_{B,s}$ sum the seven directed externalities. The market difference is
\begin{equation}
\Delta^{\mathrm{system}}_{B,s}
=Y_{-R,s}-Y_{-B,s}
=\Delta^{\mathrm{private}}_{B,s}
+\Delta^{\mathrm{others}}_{B,s}.
\label{eq:market-decomposition}
\end{equation}

%% file: sections/experiments.tex
\section{Experiments and Results}
\label{sec:experiments}
\subsection{Experimental Design}
\label{sec:experimental-setup}
We evaluate CEO agents instantiated with eight language models under a shared OpenCode-based harness. The main evaluation compares these agents and Rule CEO in nine lineups: one reference lineup and eight same-seat Rule replacements. Each runs for 500 days under seeds 11, 29, and 47, giving 27 markets and 216 company outcomes (24 per agent). We identify LLM-based agents by their underlying model names. Agent performance depends not only on operating decisions but also on interaction reliability. Model routes, seats, and inference settings are recorded in Appendix~\ref{app:models}.

Robustness analyses use two further evaluations of all nine lineups at seed 11 (18 runs) and eight No Action replacements. The repeated lineups test the stability of operating returns and matched Score differences; the No Action lineups test sensitivity to baseline policy using the same reference controls. These runs retain the economic environment, scoring rules, and model inference parameters. Main-evaluation estimates weight the three seeds equally; Appendix~\ref{app:metrics} defines the repeated and baseline comparisons.

The LLM-based agents share the system prompt, tools, and private-memory policy. Each window allows 32 accounting-valid replies, 131,072 generated tokens including reasoning, and 60 active minutes (Appendix~\ref{app:protocol}). Rule is a fixed operating policy selected in a separate 75-world, model-free search (Appendix~\ref{app:rule-search}). No Action seals an empty decision, preserving initial standing policies while normal settlement continues (Appendix~\ref{app:noop}). Both baselines bypass LLM inference.

All 53 runs reached day 500 and passed model-free replay (Appendix~\ref{app:repro}). Runs with unsealed decisions remain in the analysis. Across 424 company trajectories, successfully sealed decisions with complete reported usage contain 325,749 model replies, 799,569 tool calls, and 8.25 billion tokens (Appendix~\ref{app:workload}).

\paragraph{Illustrative rollouts.}
Figure~\ref{fig:cash-seed11} follows the eight companies in the reference market at seed 11. Figure~\ref{fig:cash-overview} groups cash paths by agent; Appendix~\ref{app:trajectory-evidence} shows each market.
\input{figures/cash_seed11}
\input{figures/cash_overview}

The trajectories illustrate liquidity during the simulations. Quantitative evaluation uses terminal Score (Equation~\ref{eq:score}), which adds inventory liquidation value to final cash and subtracts initial enterprise value. We rank CEO agents by Mean Score across eight lineups and three seeds (24 main-evaluation runs per agent).

\subsection{Operating Performance}
\input{tables/table2}
Only Astra, Sol, and Qwen earn positive Mean Scores (Table~\ref{tab:performance}): \$116.94k, \$98.44k, and \$8.04k. Rule averages $-\$47.30$k, ahead of the other five LLM-based agents. The  ordering is unchanged when terminal salvage is removed from all 216 outcomes (Table~\ref{tab:terminal-valuation}). Sixteen of 216 firms go bankrupt, while 145 end with negative Score; liquidity and profitability are distinct. Figure~\ref{fig:performance} shows lineup-level variation. Appendix~\ref{app:trajectory-evidence} reports behavioral and trajectory evidence.
\input{figures/figure2}

Averaging each agent's eight-lineup Scores across the three seed-11 runs reproduces the full nine-agent ranking in Table~\ref{tab:performance}, although individual runs contain rank swaps. Astra and Sol retain positive eight-lineup means in every run; Rule stays ahead of the bottom five. Qwen's eight-lineup mean changes sign, which means its small positive main-evaluation mean is less stable. Appendix~\ref{app:operating-repeat} reports per-run scores and ranks.

\subsection{Externalities}
\label{sec:externalities}
Matched contrasts compare an entrant with Rule CEO at the same seat and seed, holding the other seven agents' identities and seats fixed. Private and market Score differences subtract the corresponding Rule outcomes.

\input{figures/figure3}

Private gains need not raise market returns (Figure~\ref{fig:value}A; Table~\ref{tab:value}). Astra's private and market differences average \$116.36k and \$466.68k, both positive in all three seeds. Sol's private difference is positive in all three (mean \$131.55k), but its market difference averages $-\$89.66$k and is negative in two. All three reference markets have negative total Score, so Astra's positive market difference means a smaller aggregate loss than under the matched Rule replacement.

\input{tables/table3}

Figure~\ref{fig:value}B reports the three-seed mean directed Score difference for each of the 56 entrant--rival pairs. The largest negative and positive differences are Sol $\rightarrow$ Sonnet ($-\$195.66$k) and Qwen $\rightarrow$ DeepSeek Pro ($+\$346.29$k), respectively. These relationships are directional and need not be reciprocal.

Removing terminal salvage preserves the sign of every entrant's three-seed mean market Score difference and every entrant--rival pair's three-seed mean directed Score difference (Table~\ref{tab:cash-only-contrasts}). At seed 11, we repeat entrant comparisons against No Action, which preserves initial standing policies rather than Rule's active pricing, replenishment, and spending policy. Both baselines use the same three-run reference-control mean. Seven of eight entrants retain the sign of their market Score difference (Figure~\ref{fig:value}C), supporting generalization of the aggregate pattern beyond the selected Rule policy. Appendix~\ref{app:baseline-results} reports all comparisons. Across the three main seeds and two seed-11 repeats, four of 56 directed pairs have $|\mathrm{mean}|>\mathrm{SD}$ (Figure~\ref{fig:value}D). Three retain their sign in all five observations; the fourth does so in four. Appendix~\ref{app:repeat-results} reports all 56 pairs.

\subsection{Mechanism Hypotheses}
\label{sec:mechanism-hypotheses}
We use main-evaluation memory, sealed actions, and accounts to propose mechanisms (Appendix~\ref{app:mechanisms}). Astra repeatedly trims inventory and unused capacity; Sol sustains greater sales with larger marketing and capacity outlays, then downsizes as demand weakens. Qwen uses explicit reversal rules for prices and budgets; its 11 profitable outcomes average higher sales at comparable marketing expenditure. These patterns suggest that aligning persistent spending with realized demand helps explain positive mean returns.

Aggregate externalities reflect both demand capture and rivals' spending. Relative to Rule, Sol adds more own consumer orders than Astra in every seed. On average, rivals' spending reductions exceed revenue losses with Astra, while revenue losses exceed savings with Sol. This difference is consistent with stronger competitive pressure from Sol and more favorable spending adjustments alongside Astra.

The four directed pairs suggest different channels. Doubao follows DeepSeek Pro's prices, sometimes adding further discounts; Sonnet spends more on development and innovation with Sol in all three seeds. With Astra, Flash spends less on innovation in every seed, while Pro's gains combine lower spending at seed 11 with greater sales at seeds 29 and 47. These traces support price-following and investment responses as candidate channels.

%% file: figures/cash_seed11.tex
\begin{figure}[t]
\centering
\includegraphics[width=\linewidth]{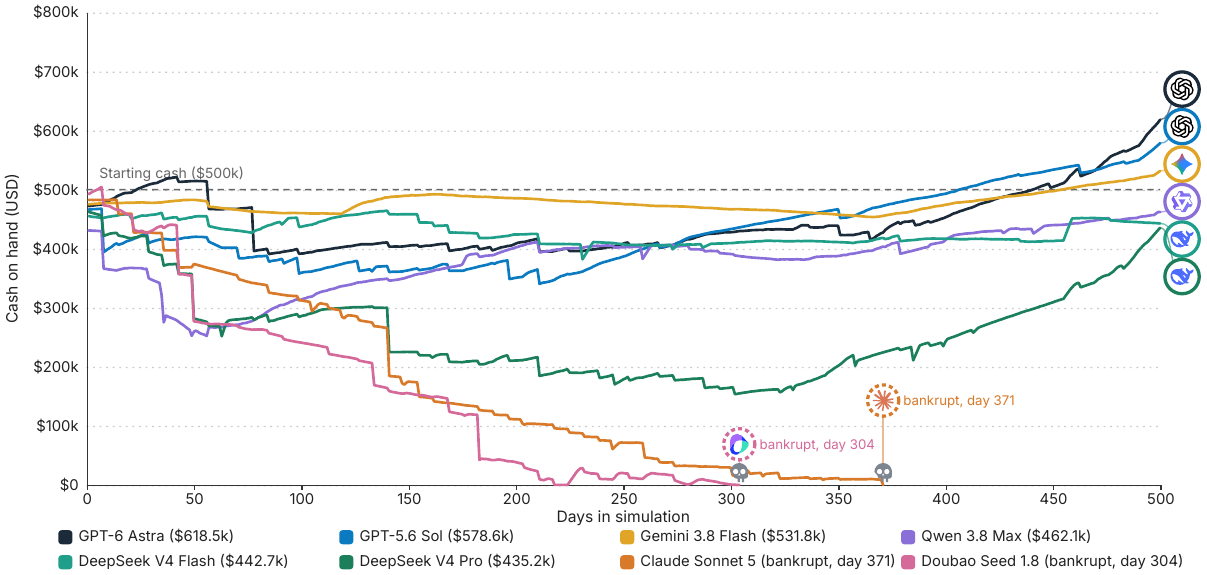}
\caption{\textbf{An illustrative shared-market run.} Daily cash balances of the eight agent-controlled companies in the main-evaluation reference lineup at seed 11. The dashed line marks the initial \$500,000 cash balance; markers indicate bankruptcy. Legend parentheses give terminal cash on hand or, for bankrupt companies, the bankruptcy day.}
\label{fig:cash-seed11}
\end{figure}

%% file: figures/cash_overview.tex
\begin{figure}[t]
\centering
\includegraphics[width=.90\linewidth]{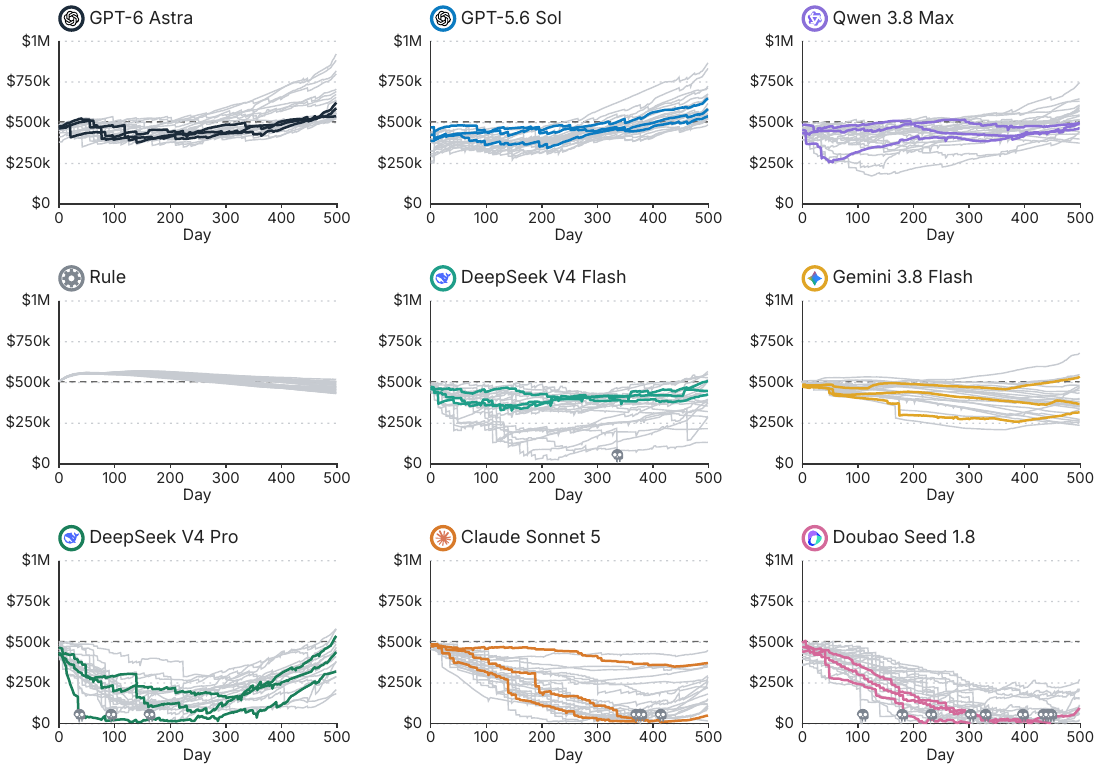}
\caption{\textbf{Cash trajectories across the main evaluation.} Each LLM-based agent panel shows 24 company trajectories: three colored curves from the reference markets at seeds 11, 29, and 47, and 21 gray curves from seven lineups in which Rule replaces a rival. The Rule panel contains its 24 replacement-market trajectories. Dashed lines mark initial cash; markers indicate bankruptcy.}
\label{fig:cash-overview}
\end{figure}

%% file: tables/table2.tex
\begin{table}[t]
\centering\small
\setlength{\tabcolsep}{4pt}
\caption{\textbf{Operating performance.} Amounts are USD thousands; each CEO agent has 24 outcomes. Rank uses Mean Score. Bankruptcy rate counts cash-negative exits.}
\label{tab:performance}
\begin{tabular}{@{}rlrr@{}}
\toprule
Rank & CEO & Mean Score & \shortstack{Bankruptcy\\rate} \\
\midrule
1 & GPT-6 Astra & 116.94 & 0.0\% \\
2 & GPT-5.6 Sol & 98.44 & 0.0\% \\
3 & Qwen 3.8 Max & 8.04 & 0.0\% \\
4 & Rule CEO & -47.30 & 0.0\% \\
5 & DeepSeek V4 Flash & -114.50 & 4.2\% \\
6 & Gemini 3.8 Flash & -119.79 & 0.0\% \\
7 & DeepSeek V4 Pro & -159.33 & 12.5\% \\
8 & Claude Sonnet 5 & -314.36 & 12.5\% \\
9 & Doubao Seed 1.8 & -434.88 & 37.5\% \\
\bottomrule
\end{tabular}
\end{table}

%% file: figures/figure2.tex
\begin{figure}[t]
\centering
\includegraphics[width=\linewidth]{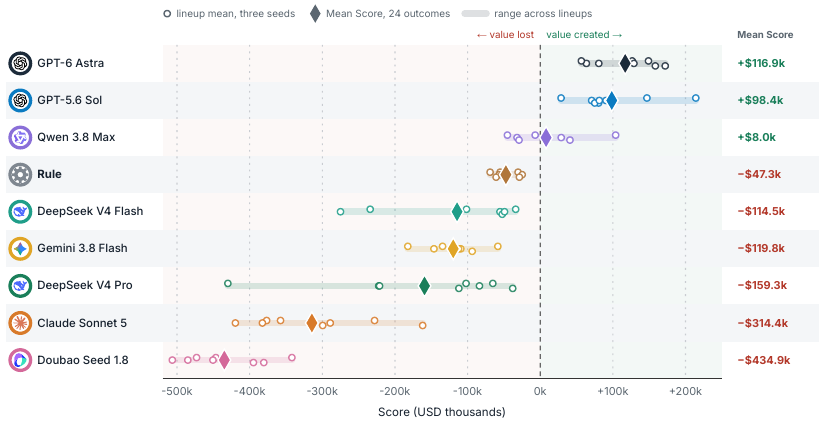}
\caption{\textbf{Operating performance in the main evaluation.} Open circles show each agent's eight three-seed lineup means; pale bars span their minimum and maximum. Diamonds and right-hand values give Mean Score across 24 company outcomes. Rows follow Mean Score rank.}
\label{fig:performance}
\end{figure}

%% file: figures/figure3.tex
\begin{figure}[t]
\centering
\includegraphics[width=\linewidth]{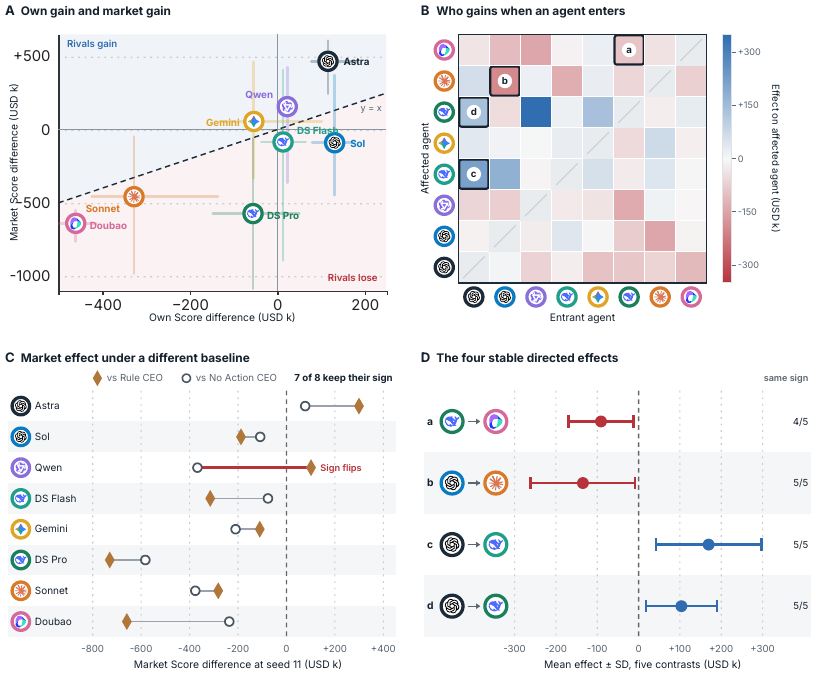}
\caption{\textbf{Externalities across agents.} A, B, and D use matched Rule comparisons. (A) Logos mark three-seed mean private and market Score differences. Pale crosses span the minimum--maximum across seeds on each axis; shading above/below $y=x$ denotes aggregate gains/losses for rivals. (B) Entrant agents are columns and affected agents rows; blue/red denotes positive/negative directed three-seed mean differences. Diagonal cells are undefined; boxes link to D. (C) Seed-11 market differences against Rule (diamonds) and No Action (open circles); seven of eight signs agree. (D) Four of 56 directed pairs satisfy $|\mathrm{mean}|>\mathrm{SD}$ across three main seeds and two seed-11 repeats. Bars show sample SD; counts show sign agreement.}
\label{fig:value}
\end{figure}

%% file: tables/table3.tex
\begin{table}[t]
\centering\small
\setlength{\tabcolsep}{3pt}
\caption{\textbf{Private returns and externalities.} Main-evaluation three-seed means (USD thousands), subtracting matched Rule outcomes from agent outcomes. Mean externality equals the total divided by seven. Private plus total externality equals the market difference before rounding. The final column counts positive market differences, not profitable markets.}
\label{tab:value}
\begin{tabular}{@{}lrrrrrc@{}}
\toprule
Entrant Agent & \shortstack{Private Score\\difference} & \shortstack{Mean externality\\per rival} & \shortstack{Total\\externality} & \shortstack{Market Score\\difference} & \multicolumn{2}{c}{\shortstack{Positive market-\\difference seeds}} \\
\midrule
GPT-6 Astra & 116.36 & 50.05 & 350.33 & 466.68 & \multicolumn{2}{c}{3/3} \\
GPT-5.6 Sol & 131.55 & -31.60 & -221.22 & -89.66 & \multicolumn{2}{c}{1/3} \\
Qwen 3.8 Max & 23.28 & 19.09 & 133.64 & 156.92 & \multicolumn{2}{c}{2/3} \\
DeepSeek V4 Flash & 13.62 & -13.97 & -97.82 & -84.20 & \multicolumn{2}{c}{2/3} \\
Gemini 3.8 Flash & -53.09 & 15.55 & 108.87 & 55.78 & \multicolumn{2}{c}{2/3} \\
DeepSeek V4 Pro & -54.96 & -74.04 & -518.29 & -573.25 & \multicolumn{2}{c}{0/3} \\
Claude Sonnet 5 & -327.00 & -18.72 & -131.05 & -458.05 & \multicolumn{2}{c}{0/3} \\
Doubao Seed 1.8 & -460.22 & -25.67 & -179.67 & -639.88 & \multicolumn{2}{c}{0/3} \\
\bottomrule
\end{tabular}
\end{table}

%% file: sections/discussion.tex
\section{Discussion and Limitations}
\label{sec:limitations}
Rule was selected with nine firms but evaluated with eight under the same 690 baseline opportunities, potentially easing competition. Although its upper-half standing in both settings and the seven-of-eight market-sign agreement under No Action suggest limited impact on our qualitative findings, rule’s eight-firm optimality and effect magnitudes remain untested.

Our externality analysis compares rivals’ and market Scores under matched CEO replacements and proposes mechanism hypotheses for the observed differences, but does not provide a comprehensive causal analysis. Identifying the causal mechanisms will require further targeted interventions and causal-inference analysis.

We characterize robustness using means, sample standard deviations, and sign consistency. Informative confidence intervals would require more extensive replication.

\arena{} models resource constraints, delayed investments, noisy observations, and adaptive rivals, but remains stylized. Firms share a product catalog and initial endowments; demand and costs follow programmed rules, and decisions use fixed tools. We have not tested whether rankings and externality patterns hold under different demand, liquidity, delay, or observation noise settings. The simulator omits financing, labor markets, regulation, direct inter-agent communication, and shared supplier scarcity. Transfer to real firms remains untested.

\section{Conclusion}
\label{sec:conclusion}
\arena{} uses matched replacement evaluation to separate private performance from externalities in long-horizon multi-agent competition. Our results show that private gains can accompany market losses and there are some relatively stable directed externalities between agents. Memory and action traces suggest demand capture and rivals' spending responses as candidate mechanisms for targeted testing. Evaluating agents in multi-agent systems therefore requires assessing both their own performance and the direction, distribution, and stability of their externalities.

%% file: sections/statements.tex
\section*{Reproducibility Statement}
Appendices~\ref{app:metrics}--\ref{app:repro} specify the metrics, tools, baselines, economic equations, parameters, and execution rules. The artifact contains the 27-run main evaluation, its 216 company outcomes, the 26 additional-run score analyses, and the complete repeated and baseline comparisons. Source and configuration hashes bind the evidence to the evaluated implementation.

\section*{Ethics Statement}
The experiment uses synthetic firms and programmatic customers. It involves no human participants. Company return is a task objective; it excludes consumer welfare and broader social costs. Results therefore do not establish that these policies are suitable for autonomous operation of real organizations. Shared artifacts exclude credentials and private provider reasoning.

\section*{AI Use Statement}
Generative AI assisted implementation, analysis code, parameter discussions, literature search, writing, and translation. The evaluated CEOs also use language models under the reported protocol. Numerical results come from retained experiment records. Source comparisons, numerical reconciliation, and replay support the verification of AI-assisted work. The authors are responsible for checking the code, references, interpretations, and final manuscript.

%% file: appendix/metrics.tex
\section{Exact Evaluation Definitions}
\label{app:metrics}
\paragraph{Lineups and notation.}
Let $\mathcal M=\{M_1,\ldots,M_8\}$ be the LLM-based agent pool, $\mathcal C=\mathcal M\cup\{R\}$, and $\mathcal S=\{11,29,47\}$. We use the complete leave-one-out design
\begin{equation}
L_{-j}=\mathcal C\setminus\{j\},\quad j\in\mathcal C;\qquad
|L_{-j}|=8,\quad |\mathcal L|=9,\quad |\mathcal L\times\mathcal S|=27.
\end{equation}
Agent $M_i$ keeps seat $i$; Rule inherits that seat in $L_{-M_i}$. Each candidate appears in eight lineups and 24 outcomes. Company-specific random streams remain keyed by fixed seats, while market-level streams are shared. No Action is outside this formal candidate pool.

\paragraph{Terminal Score.}
Equation~\ref{eq:score} uses the same valuation for every firm at $T=500$. With on-hand stock $H$, paid inbound stock $I^{\mathrm{in}}$, and liquidation price $v_j$,
\begin{align}
V_{i,0}&=C_{i,0}+\sum_j v_jH_{ij,0},\\
S_{i,-b,s}&=C_{i,T}+\sum_jv_j(H_{ij,T}+I^{\mathrm{in}}_{ij,T})-V_{i,0}.
\label{eq:score-detail}
\end{align}
The initial basis is $500{,}000+600(8)+400(15)+200(26)=516{,}000$ dollars. Initial inventory is valued at liquidation price. Capacity, knowledge, goodwill, and unfinished projects have no separate terminal value. Bankruptcy stops operating decisions and trading; terminal valuation still includes retained on-hand and paid inbound stock.

\paragraph{Operating returns.}
For candidate $i$, the lineup mean and overall mean are
\begin{equation}
\bar S_{i,-b}=\frac{1}{3}\sum_{s\in\mathcal S}S_{i,-b,s},\qquad
\bar S_i=\frac{1}{8}\sum_{b\ne i}\bar S_{i,-b}
=\frac{1}{24}\sum_{b\ne i}\sum_s S_{i,-b,s}.
\label{eq:a}
\end{equation}
Only $\bar S_i$ determines rank; exact ties use the model identifier.

\paragraph{Private returns and externalities.}
For entrant $B\in\mathcal M$, matched differences use the same seed and seat:
\begin{align}
Y_{-b,s}&=\sum_{i\in L_{-b}}S_{i,-b,s},\\
\Delta^{\mathrm{private}}_{B,s}&=S_{B,-R,s}-S_{R,-B,s},\\
\theta_{B\rightarrow A,s}&=S_{A,-R,s}-S_{A,-B,s},\qquad A\in\mathcal M\setminus\{B\},\\
\Delta^{\mathrm{others}}_{B,s}&=\sum_{A\ne B}\theta_{B\rightarrow A,s},\\
\Delta^{\mathrm{system}}_{B,s}&=Y_{-R,s}-Y_{-B,s}
=\Delta^{\mathrm{private}}_{B,s}+\Delta^{\mathrm{others}}_{B,s}.
\label{eq:c}
\end{align}
All main-evaluation means in Tables~\ref{tab:performance}--\ref{tab:value} and Figures~\ref{fig:performance} and~\ref{fig:value}A--B give equal weight to the three seeds. Figure~\ref{fig:value}B places $\bar\theta_{B\rightarrow A}$ at row $A$, column $B$; diagonal cells are undefined. Table~\ref{tab:value}'s mean externality per rival is $\bar\Delta^{\mathrm{others}}_B/7$. Its positive-seed count is $\sum_s\mathbf1[\Delta^{\mathrm{system}}_{B,s}>0]$. The eight entrant comparisons share the same reference market within each seed. Each directed difference includes the subsequent responses of all seven rivals, so the 56 cells are neither independent samples nor isolated bilateral interventions.

\subsection{Stability of Operating Returns}
\label{app:operating-repeat}
Table~\ref{tab:operating-repeat} reports each agent's mean Score over its eight lineups in the main seed-11 run and two repeats. Parentheses give ranks within each run. This fixed-seed comparison does not replace the three-seed main-evaluation ranking in Table~\ref{tab:performance}.

\begin{table}[t]
\centering\small
\setlength{\tabcolsep}{5pt}
\caption{Operating performance across three seed-11 runs (USD thousands). Each value averages eight lineups; parentheses show ranks within each run.}
\label{tab:operating-repeat}
\begin{tabular}{@{}lrrr@{}}
\toprule CEO agent & F11 & R1 & R2\\\midrule
GPT-6 Astra & $91.14\,(1)$ & $54.97\,(1)$ & $105.44\,(1)$\\
GPT-5.6 Sol & $87.21\,(2)$ & $17.00\,(3)$ & $66.98\,(2)$\\
Qwen 3.8 Max & $-20.41\,(3)$ & $26.05\,(2)$ & $-51.57\,(4)$\\
Rule CEO & $-61.69\,(4)$ & $-54.80\,(4)$ & $-50.10\,(3)$\\
DeepSeek V4 Flash & $-106.45\,(5)$ & $-79.47\,(5)$ & $-148.50\,(7)$\\
Gemini 3.8 Flash & $-152.67\,(7)$ & $-97.07\,(6)$ & $-136.08\,(5)$\\
DeepSeek V4 Pro & $-127.94\,(6)$ & $-149.44\,(7)$ & $-139.69\,(6)$\\
Claude Sonnet 5 & $-318.43\,(8)$ & $-176.32\,(8)$ & $-195.31\,(8)$\\
Doubao Seed 1.8 & $-431.24\,(9)$ & $-434.77\,(9)$ & $-459.76\,(9)$\\
\bottomrule
\end{tabular}
\end{table}

\subsection{Stability of Directed Externalities}
\label{app:repeat-results}
For each directed pair, the five contrasts use main-evaluation seeds 11, 29, and 47, followed by the two repeated seed-11 runs. Each repeated reference control is compared with the replacement run having the same repetition index. Let $d_r$ denote one contrast. We report
\begin{equation}
\bar d=\frac{1}{5}\sum_{r=1}^{5}d_r,\qquad
s_d=\sqrt{\frac{1}{4}\sum_{r=1}^{5}(d_r-\bar d)^2}.
\end{equation}
Figure~\ref{fig:value}D displays all pairs satisfying $|\bar d|>s_d$: four of 56. This is a descriptive display criterion, not a significance test. The five observations give seed 11 weight $3/5$ and each other seed weight $1/5$; they are not five independent economic seeds and do not replace the main-evaluation means. Error bars are sample SDs, not confidence intervals. Shared controls and markets make pairwise contrasts dependent.

Table~\ref{tab:repeat-full} reports every pair. Agent indices match Figure~\ref{fig:value}: 1 Astra, 2 Sol, 3 Qwen, 4 DeepSeek Flash, 5 Gemini, 6 DeepSeek Pro, 7 Sonnet, 8 Doubao. F11/F29/F47 identify main-evaluation seeds; R1/R2 are the repeated seed-11 runs. Sign counts compare each observation with the five-observation mean's sign. Bold pairs meet the display criterion.
\begingroup
\small\setlength{\tabcolsep}{3pt}
\begin{longtable}{@{}lrrrrrrrr@{}}
\caption{All 56 directed externalities across five observations (USD thousands).}\label{tab:repeat-full}\\
\toprule Pair & F11 & F29 & F47 & R1 & R2 & Mean & SD & Sign\\\midrule
\endfirsthead
\multicolumn{9}{l}{\small Table~\ref{tab:repeat-full} continued.}\\
\toprule Pair & F11 & F29 & F47 & R1 & R2 & Mean & SD & Sign\\\midrule
\endhead
\midrule\multicolumn{9}{r}{\small Continued on next page.}\\\endfoot
\bottomrule\endlastfoot
1 $\to$ 2 & $58.85$ & $-82.49$ & $-7.86$ & $2.99$ & $180.06$ & $30.31$ & $97.68$ & 3/5 \\
1 $\to$ 3 & $-50.24$ & $-90.48$ & $-41.60$ & $47.59$ & $-67.60$ & $-40.47$ & $52.65$ & 4/5 \\
\textbf{1 $\to$ 4} & $75.01$ & $293.35$ & $290.29$ & $6.43$ & $182.75$ & $169.57$ & $128.09$ & 5/5 \\
1 $\to$ 5 & $131.04$ & $19.32$ & $-39.45$ & $-13.86$ & $-8.97$ & $17.62$ & $66.75$ & 2/5 \\
\textbf{1 $\to$ 6} & $233.33$ & $149.53$ & $29.61$ & $59.90$ & $45.48$ & $103.57$ & $86.17$ & 5/5 \\
1 $\to$ 7 & $-81.83$ & $224.64$ & $43.64$ & $-28.45$ & $-47.95$ & $22.01$ & $122.22$ & 2/5 \\
1 $\to$ 8 & $-0.29$ & $4.69$ & $-108.07$ & $-242.17$ & $28.95$ & $-63.38$ & $113.03$ & 3/5 \\
2 $\to$ 1 & $62.46$ & $58.11$ & $-99.80$ & $75.38$ & $-65.61$ & $6.11$ & $82.22$ & 3/5 \\
2 $\to$ 3 & $34.90$ & $-103.30$ & $-150.14$ & $-22.82$ & $152.47$ & $-17.78$ & $118.97$ & 3/5 \\
2 $\to$ 4 & $111.97$ & $424.59$ & $-0.40$ & $-125.01$ & $-207.29$ & $40.77$ & $246.53$ & 2/5 \\
2 $\to$ 5 & $89.25$ & $-100.40$ & $47.20$ & $-163.89$ & $-202.58$ & $-66.09$ & $128.78$ & 3/5 \\
2 $\to$ 6 & $-11.85$ & $-43.60$ & $-81.95$ & $-2.56$ & $22.55$ & $-23.48$ & $40.36$ & 4/5 \\
\textbf{2 $\to$ 7} & $-340.16$ & $-73.02$ & $-173.80$ & $-38.60$ & $-47.90$ & $-134.70$ & $126.80$ & 5/5 \\
2 $\to$ 8 & $-279.14$ & $34.37$ & $-68.94$ & $-69.48$ & $28.65$ & $-70.91$ & $126.85$ & 3/5 \\
3 $\to$ 1 & $40.14$ & $-67.34$ & $-169.14$ & $4.57$ & $17.26$ & $-34.90$ & $85.12$ & 2/5 \\
3 $\to$ 2 & $11.41$ & $64.92$ & $-135.57$ & $68.05$ & $-51.39$ & $-8.52$ & $86.07$ & 2/5 \\
3 $\to$ 4 & $151.07$ & $-60.31$ & $-109.51$ & $-107.24$ & $-80.04$ & $-41.20$ & $109.39$ & 4/5 \\
3 $\to$ 5 & $258.40$ & $-127.35$ & $-122.23$ & $15.03$ & $-155.09$ & $-26.25$ & $172.31$ & 3/5 \\
3 $\to$ 6 & $112.44$ & $590.48$ & $335.95$ & $-99.63$ & $78.73$ & $203.60$ & $266.01$ & 4/5 \\
3 $\to$ 7 & $-88.74$ & $110.04$ & $36.00$ & $24.72$ & $62.87$ & $28.98$ & $73.56$ & 4/5 \\
3 $\to$ 8 & $-90.72$ & $-124.96$ & $-214.05$ & $3.96$ & $28.94$ & $-79.37$ & $98.77$ & 3/5 \\
4 $\to$ 1 & $85.05$ & $18.60$ & $-155.00$ & $-96.49$ & $-183.14$ & $-66.20$ & $114.62$ & 3/5 \\
4 $\to$ 2 & $54.05$ & $120.37$ & $-48.12$ & $33.85$ & $-129.38$ & $6.16$ & $96.72$ & 3/5 \\
4 $\to$ 3 & $-49.74$ & $46.91$ & $-71.93$ & $-79.92$ & $46.01$ & $-21.73$ & $63.23$ & 3/5 \\
4 $\to$ 5 & $273.55$ & $33.62$ & $-231.81$ & $-243.45$ & $-49.07$ & $-43.43$ & $213.27$ & 3/5 \\
4 $\to$ 6 & $147.37$ & $-41.71$ & $-50.30$ & $358.55$ & $-231.35$ & $36.51$ & $224.39$ & 2/5 \\
4 $\to$ 7 & $-170.70$ & $76.96$ & $-294.61$ & $-12.64$ & $-190.68$ & $-118.33$ & $148.63$ & 4/5 \\
4 $\to$ 8 & $-119.11$ & $87.44$ & $-4.38$ & $-179.48$ & $29.27$ & $-37.25$ & $109.52$ & 3/5 \\
5 $\to$ 1 & $65.40$ & $-333.83$ & $-16.21$ & $11.53$ & $-43.63$ & $-63.35$ & $156.49$ & 3/5 \\
5 $\to$ 2 & $39.70$ & $-7.52$ & $-45.04$ & $452.27$ & $-4.63$ & $86.96$ & $206.41$ & 2/5 \\
5 $\to$ 3 & $22.97$ & $-111.87$ & $-19.58$ & $-48.45$ & $-109.83$ & $-53.35$ & $58.32$ & 4/5 \\
5 $\to$ 4 & $-61.54$ & $130.09$ & $108.93$ & $11.41$ & $-80.17$ & $21.75$ & $95.87$ & 3/5 \\
5 $\to$ 6 & $5.39$ & $567.30$ & $-156.69$ & $-18.71$ & $-155.00$ & $48.45$ & $299.60$ & 2/5 \\
5 $\to$ 7 & $-129.54$ & $274.04$ & $-69.58$ & $31.15$ & $-5.58$ & $20.10$ & $154.68$ & 2/5 \\
5 $\to$ 8 & $-9.03$ & $75.61$ & $-2.37$ & $-76.94$ & $-62.26$ & $-15.00$ & $60.16$ & 4/5 \\
6 $\to$ 1 & $-176.16$ & $-83.89$ & $3.61$ & $-15.79$ & $-55.26$ & $-65.50$ & $70.60$ & 4/5 \\
6 $\to$ 2 & $-250.68$ & $-24.30$ & $46.76$ & $97.62$ & $-120.33$ & $-50.19$ & $138.77$ & 3/5 \\
6 $\to$ 3 & $-279.38$ & $-76.69$ & $-51.25$ & $-110.80$ & $68.81$ & $-89.86$ & $125.69$ & 4/5 \\
6 $\to$ 4 & $-103.63$ & $127.95$ & $-89.16$ & $-172.74$ & $-0.55$ & $-47.63$ & $115.71$ & 4/5 \\
6 $\to$ 5 & $159.60$ & $-103.68$ & $-208.89$ & $-192.85$ & $69.62$ & $-55.24$ & $163.28$ & 3/5 \\
6 $\to$ 7 & $-216.46$ & $103.12$ & $-60.89$ & $-67.08$ & $-170.89$ & $-82.44$ & $123.41$ & 4/5 \\
\textbf{6 $\to$ 8} & $-151.19$ & $-55.05$ & $-64.62$ & $6.29$ & $-190.05$ & $-90.92$ & $78.88$ & 4/5 \\
7 $\to$ 1 & $28.82$ & $-228.78$ & $10.75$ & $-67.22$ & $-171.06$ & $-85.50$ & $112.38$ & 3/5 \\
7 $\to$ 2 & $-33.96$ & $-216.96$ & $-179.20$ & $179.17$ & $75.91$ & $-35.01$ & $167.39$ & 3/5 \\
7 $\to$ 3 & $87.89$ & $86.89$ & $-134.89$ & $-113.83$ & $-14.84$ & $-17.75$ & $106.15$ & 3/5 \\
7 $\to$ 4 & $58.90$ & $11.75$ & $67.66$ & $-27.04$ & $62.29$ & $34.71$ & $41.14$ & 4/5 \\
7 $\to$ 5 & $286.47$ & $-34.15$ & $-33.51$ & $-39.38$ & $-191.67$ & $-2.45$ & $175.08$ & 4/5 \\
7 $\to$ 6 & $-26.70$ & $93.92$ & $-122.48$ & $50.50$ & $-129.30$ & $-26.81$ & $100.26$ & 3/5 \\
7 $\to$ 8 & $-25.19$ & $78.99$ & $-169.38$ & $-26.42$ & $-73.26$ & $-43.05$ & $89.95$ & 4/5 \\
8 $\to$ 1 & $-9.68$ & $-294.04$ & $-21.76$ & $-20.00$ & $-128.46$ & $-94.79$ & $121.45$ & 5/5 \\
8 $\to$ 2 & $-74.16$ & $60.47$ & $-15.09$ & $-26.53$ & $-31.30$ & $-17.32$ & $48.91$ & 4/5 \\
8 $\to$ 3 & $-31.96$ & $8.88$ & $14.99$ & $8.44$ & $32.24$ & $6.52$ & $23.57$ & 4/5 \\
8 $\to$ 4 & $51.49$ & $-37.07$ & $-23.77$ & $22.47$ & $-101.69$ & $-17.71$ & $58.88$ & 3/5 \\
8 $\to$ 5 & $190.31$ & $-316.55$ & $79.26$ & $112.79$ & $-154.37$ & $-17.71$ & $210.76$ & 2/5 \\
8 $\to$ 6 & $-61.87$ & $348.93$ & $-202.12$ & $-175.75$ & $39.18$ & $-10.33$ & $222.67$ & 3/5 \\
8 $\to$ 7 & $-328.14$ & $93.11$ & $29.76$ & $132.92$ & $-43.33$ & $-23.14$ & $183.10$ & 2/5 \\
\end{longtable}
\endgroup

\subsection{Baseline-Policy Comparison}
\label{app:baseline-results}
Both policies share the mean of three reference controls at seed 11: the main-evaluation run and the two repetitions. For an endpoint $Y$ (the target firm's Score, the other seven firms' total, or all eight firms' total), let $Y_{0,r}$ denote the control, $Y_{R,r}$ the active Rule CEO replacement, and $Y_N$ the single No Action CEO replacement. We calculate
\begin{equation}
\Delta_R=\frac13\sum_{r=1}^{3}Y_{0,r}-\frac13\sum_{r=1}^{3}Y_{R,r},\qquad
\Delta_N=\frac13\sum_{r=1}^{3}Y_{0,r}-Y_N.
\end{equation}
For the private endpoint, the replacement outcome is the baseline CEO's Score in the target's seat. For a directed contrast, $Y$ is the affected rival's Score. Reusing three controls does not create three No Action replacement observations. We report point comparisons without a replacement-run SD or confidence interval for that policy. Figure~\ref{fig:value}C shows the market comparison; Table~\ref{tab:baseline-full} compares all eight entrants' private, other-firm, and market differences under Rule and No Action.

\begin{table}[t]
\centering\small\setlength{\tabcolsep}{4pt}
\caption{Baseline comparisons at seed 11 (USD thousands). R denotes Rule CEO and N denotes No Action CEO.}\label{tab:baseline-full}
\begin{tabular}{@{}lrrrrrr@{}}
\toprule & \multicolumn{2}{c}{Private} & \multicolumn{2}{c}{Others} & \multicolumn{2}{c}{Market}\\
\cmidrule(lr){2-3}\cmidrule(lr){4-5}\cmidrule(l){6-7}
Entrant & R & N & R & N & R & N\\\midrule
Astra & 129.13 & 229.91 & 170.34 & -152.47 & 299.47 & 77.44\\
Sol & 145.38 & 245.64 & -333.09 & -353.99 & -187.71 & -108.35\\
Qwen & 33.30 & 132.62 & 68.24 & -500.26 & 101.54 & -367.64\\
DeepSeek Flash & -78.93 & 47.83 & -235.81 & -124.26 & -314.74 & -76.42\\
Gemini & -55.01 & 48.74 & -55.16 & -258.86 & -110.17 & -210.12\\
DeepSeek Pro & -105.76 & 38.88 & -623.97 & -621.53 & -729.73 & -582.65\\
Sonnet & -245.08 & -131.73 & -36.64 & -244.52 & -281.72 & -376.25\\
Doubao & -459.87 & -338.22 & -199.13 & 102.15 & -659.00 & -236.07\\\bottomrule
\end{tabular}
\end{table}

Seven of eight Market differences retain their mean sign between policies; Private and Others each retain five of eight. Table~\ref{tab:baseline-directed} reports all directed comparisons, of which 33/56 retain their mean sign. These counts describe this set of comparisons, not independent Bernoulli trials.
\begingroup
\small\setlength{\tabcolsep}{9pt}
\begin{longtable}{@{}lrrc@{}}
\caption{Directed externalities under the two baseline policies at seed 11 (USD thousands); agent indices match Table~\ref{tab:repeat-full}.}\label{tab:baseline-directed}\\
\toprule Pair & Rule CEO & No Action CEO & Same sign\\\midrule\endfirsthead
\multicolumn{4}{l}{\small Table~\ref{tab:baseline-directed} continued.}\\
\toprule Pair & Rule CEO & No Action CEO & Same sign\\\midrule\endhead
\midrule\multicolumn{4}{r}{\small Continued on next page.}\\\endfoot
\bottomrule\endlastfoot
1 $\to$ 2 & $80.63$ & $26.20$ & Yes \\
1 $\to$ 3 & $-23.42$ & $-51.68$ & Yes \\
1 $\to$ 4 & $88.06$ & $-145.97$ & No \\
1 $\to$ 5 & $36.07$ & $23.48$ & Yes \\
1 $\to$ 6 & $112.91$ & $170.75$ & Yes \\
1 $\to$ 7 & $-52.74$ & $-109.79$ & Yes \\
1 $\to$ 8 & $-71.17$ & $-65.45$ & Yes \\
2 $\to$ 1 & $24.08$ & $-53.32$ & No \\
2 $\to$ 3 & $54.85$ & $-161.74$ & No \\
2 $\to$ 4 & $-73.44$ & $109.53$ & No \\
2 $\to$ 5 & $-92.41$ & $-28.86$ & Yes \\
2 $\to$ 6 & $2.71$ & $-62.57$ & No \\
2 $\to$ 7 & $-142.22$ & $-166.26$ & Yes \\
2 $\to$ 8 & $-106.66$ & $9.23$ & No \\
3 $\to$ 1 & $20.66$ & $-122.90$ & No \\
3 $\to$ 2 & $9.35$ & $12.79$ & Yes \\
3 $\to$ 4 & $-12.07$ & $-11.78$ & Yes \\
3 $\to$ 5 & $39.45$ & $23.37$ & Yes \\
3 $\to$ 6 & $30.52$ & $-113.15$ & No \\
3 $\to$ 7 & $-0.39$ & $-207.85$ & Yes \\
3 $\to$ 8 & $-19.27$ & $-80.74$ & Yes \\
4 $\to$ 1 & $-64.86$ & $-39.55$ & Yes \\
4 $\to$ 2 & $-13.82$ & $94.83$ & No \\
4 $\to$ 3 & $-27.88$ & $-24.18$ & Yes \\
4 $\to$ 5 & $-6.33$ & $10.01$ & No \\
4 $\to$ 6 & $91.52$ & $128.05$ & Yes \\
4 $\to$ 7 & $-124.67$ & $-91.91$ & Yes \\
4 $\to$ 8 & $-89.77$ & $-201.51$ & Yes \\
5 $\to$ 1 & $11.10$ & $50.71$ & Yes \\
5 $\to$ 2 & $162.45$ & $29.42$ & Yes \\
5 $\to$ 3 & $-45.10$ & $22.15$ & No \\
5 $\to$ 4 & $-43.43$ & $-126.80$ & Yes \\
5 $\to$ 6 & $-56.11$ & $54.81$ & No \\
5 $\to$ 7 & $-34.66$ & $-298.99$ & Yes \\
5 $\to$ 8 & $-49.41$ & $9.83$ & No \\
6 $\to$ 1 & $-82.40$ & $112.01$ & No \\
6 $\to$ 2 & $-91.13$ & $-35.95$ & Yes \\
6 $\to$ 3 & $-107.12$ & $-101.89$ & Yes \\
6 $\to$ 4 & $-92.31$ & $-94.84$ & Yes \\
6 $\to$ 5 & $12.12$ & $33.71$ & Yes \\
6 $\to$ 7 & $-151.48$ & $-326.80$ & Yes \\
6 $\to$ 8 & $-111.65$ & $-207.78$ & Yes \\
7 $\to$ 1 & $-69.82$ & $11.02$ & No \\
7 $\to$ 2 & $73.71$ & $-78.64$ & No \\
7 $\to$ 3 & $-13.59$ & $10.70$ & No \\
7 $\to$ 4 & $31.38$ & $46.22$ & Yes \\
7 $\to$ 5 & $18.47$ & $-107.15$ & No \\
7 $\to$ 6 & $-35.17$ & $22.76$ & No \\
7 $\to$ 8 & $-41.62$ & $-149.43$ & Yes \\
8 $\to$ 1 & $-52.71$ & $24.62$ & No \\
8 $\to$ 2 & $-44.00$ & $42.22$ & No \\
8 $\to$ 3 & $2.91$ & $-208.95$ & No \\
8 $\to$ 4 & $-9.24$ & $-133.43$ & Yes \\
8 $\to$ 5 & $49.58$ & $106.28$ & Yes \\
8 $\to$ 6 & $-66.15$ & $399.88$ & No \\
8 $\to$ 7 & $-79.52$ & $-128.47$ & Yes \\
\end{longtable}
\endgroup

%% file: appendix/protocol.tex
\section{Agent Interface and Baselines}
\label{app:protocol}
\subsection{Typed Tools and Information Boundaries}
The integrated contract is \texttt{ceos-decision-v9}. Every request binds a company identity and world version server-side. Neither field is selectable by the agent. Unknown fields and invalid action ranges are rejected. Tools stage a private draft; no tool is an unrestricted operating-system command. The accepted final object is a typed \codename{CompanyDecision}, not free-form prose or an arbitrary JSON file. Table~\ref{tab:tools} specifies the tools through which CEOs obtain information, maintain private memory, and commit company decisions.
\begin{longtable}{@{}>{\raggedright\arraybackslash}p{4.9cm}>{\raggedright\arraybackslash}p{8.0cm}@{}}
\caption{The complete official tool surface. Monetary inputs are integer cents.}\label{tab:tools}\\
\toprule Tool & Arguments and semantics\\\midrule\endfirsthead
\toprule Tool & Arguments and semantics\\\midrule\endhead
\codename{query\_company} & Read-only SQL, parameters, row limit; returns authorized company rows. Disabled throughout finalization.\\
\codename{set\_product} & Product, list price, promotion, reorder target, daily development budget; omitted fields remain unchanged.\\
\codename{set\_marketing} & Segment, channel, recurring daily budget.\\
\codename{set\_operations} & Recurring reliability and support budgets.\\
\codename{start\_research\_project} & Innovation tier 1--3; one-time company-wide quality investment with stochastic completion and gain.\\
\codename{expand\_capacity} & Higher capacity tier; cumulative capital difference paid now, new throughput and overhead after construction.\\
\codename{downsize\_capacity} & Lower tier; adjustment fee and delay, no installed-capital refund or cancellation of orders.\\
\codename{cancel\_capacity\_project} & Pending expansion project ID; cancel unfinished work and refund only unused prepaid capital after costs.\\
\codename{submit\_enterprise\_bid} & Open tender ID and unit price; sealed from competitors.\\
\codename{request\_market\_research} & Segment and report level 1--5; one-time paid noisy observation.\\
\codename{read\_file} & Private path, optional line offset and limit; read line-numbered UTF-8 content.\\
\codename{write\_file} & Private path and content; persist immediately, independently of economic sealing.\\
\codename{edit\_file} & Private path, one unique old string and its replacement; durable, replay-safe edit.\\
\codename{search\_files} & Pattern, optional path and glob; search at most 100 private files and return up to 200 matches.\\
\codename{glob\_files} & Pattern; list up to 200 matching private paths.\\
\codename{preview\_decision} & No arguments; combined upfront charges/refunds, affordability and recurring policy cost, not a forecast of future purchases or profit.\\
\codename{reset\_decision} & No arguments; discard unsealed draft changes, not standing world policies or recorded usage.\\
\codename{seal\_decision} & No arguments; validates the final draft. It must be the sole tool call of the last provider turn.\\
\bottomrule
\end{longtable}

The SQL projection includes the company's state and historical financial, product-sales and segment-service tables; visible catalogs; public prices and tenders; and private research beliefs/results. The \codename{daily\_cashflow}, \codename{daily\_expenses}, \codename{action\_receipts}, \codename{capacity\_policy}, \codename{capacity\_action\_quotes} and \codename{market\_signal} tables expose own-company accounting and already public conditions. Research records distinguish sample and delivery dates without exposing the underlying truth. Queries cannot write, attach databases, load extensions or inspect another company's projection. Default/hard row limits are 1,000/10,000; result size is capped at 1 MB and execution at two seconds. Pending innovation reveals tier, start date, median completion and expected total gain, not future realization. The scenario view discloses remaining days, window length and Score. No inter-CEO messaging, external web, arbitrary MCP, delegated subagents or fallback model is added.

Capacity actions replace one another in a draft and use environment-issued quotes. Staged success is not execution; an applied project receipt confirms placement, not completion. An unaffordable seal leaves the draft editable. 

\begin{figure}[t]
\centering
\includegraphics[width=\linewidth]{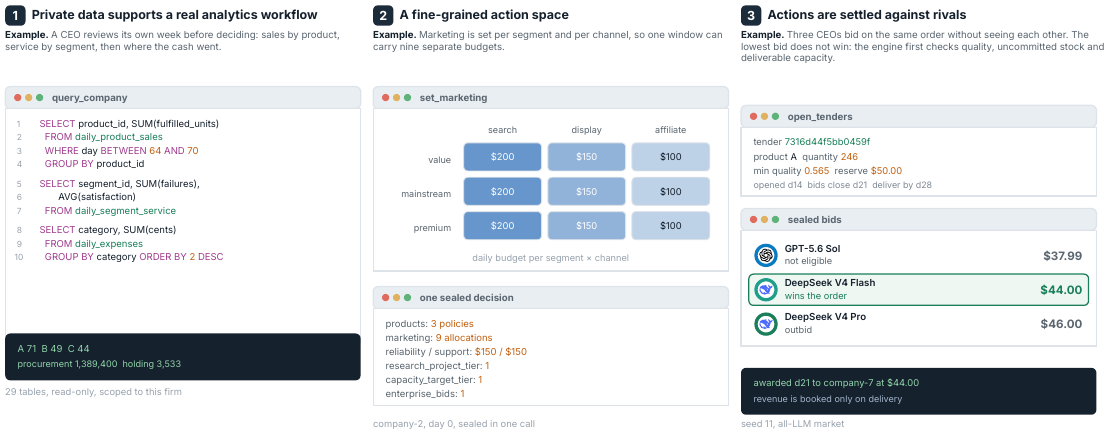}
\caption{\textbf{Agent interface examples.} A CEO queries its private records, stages segment-by-channel budgets within one decision, and competes for an enterprise tender. The tender panel depicts the seed-11 reference market; eligibility and deliverable capacity are checked before the award.}
\label{fig:interface-examples}
\end{figure}

Figure~\ref{fig:interface-examples} illustrates the read-only query, staged operating decisions, and sealed-tender workflow.

\subsection{Prompt and Memory Contract}
The common system prompt sets the CEO role and then states the following operational contract. The opening scenario-identification sentence is described in Section~\ref{sec:environment}; the operational clauses are reproduced here.
\begin{promptbox}{CEO system prompt: operational contract}You see only your company's private read-only database and public market information.
You cannot access or claim hidden competitor state; base estimates only on public and own-company evidence.
Use query\_company for evidence, then stage decisions with the provided tools.
The database is a fixed snapshot for this decision window. Repeating a query will not create new records.
Historical operating records end at the current day; after empty results, use available evidence rather than searching future operating history. Known future shipment or project dates remain valid information.
Staging never mutates the world.
A staged receipt confirms a draft change, not execution; inspect action\_receipts after settlement for actual charges, refunds, outcomes, and effective dates.
The company identity and world version are bound by the Harness and are intentionally absent from tool arguments.
Price, promotion, reorder target, daily development budget, each segment-channel marketing budget, and the reliability and support budgets persist across decision windows until explicitly changed.
Marketing, development, reliability, support, and the active capacity tier's operating cost are charged every simulated day.
Set a recurring budget to 0 to stop it.
Sealing a fresh draft with no staged operating changes preserves all current recurring policies; one-time projects, bids, and market-research requests are not automatically repeated.
start\_research\_project makes a one-time, delayed, company-wide product-quality innovation investment that adds product knowledge capital after completion.
Its quoted expected knowledge gain is the company-wide total, divided evenly across product lines, not the gain for each product. An applied project receipt confirms placement, not completion or realized quality gain.
request\_market\_research makes a one-time purchase of a delayed, noisy private report about a customer segment's demand, preferences, and quality requirements; it does not improve product quality. Check its sampling date because market conditions can change before delivery.
expand\_capacity builds additional capacity after a delay. downsize\_capacity reduces installed capacity after a delay and an adjustment fee; it does not refund past construction spending or remove outstanding orders.
cancel\_capacity\_project stops unfinished construction and refunds only the unspent prepaid amount minus cancellation costs, never sunk costs; check capacity\_action\_quotes before committing.
Use daily\_cashflow and daily\_expenses to distinguish revenue, procurement, recurring operating costs, investments, and refunds. Use preview\_decision before sealing to inspect combined upfront commitments and daily policy costs. Compare investment payback with days\_remaining; projects, knowledge, and capacity have no separate terminal score value.
Cash below zero causes bankruptcy. Combined one-time commitments must fit currently observable funds; this check does not guarantee future daily costs or procurement are affordable.
Use reset\_decision to discard unwanted unsealed commitments and re-stage a plan; it does not reset consumed decision budgets or change existing world policies.
Your sole objective is to maximize terminal cash plus terminal inventory liquidation value minus initial enterprise value.
Your entire conversation history is cleared at the start of each new simulated week (decision window).
MEMORY.md is automatically loaded; all files in your private working directory persist and can be read when needed.
Use write\_file or edit\_file to update MEMORY.md before advancing: retain current strategy, key metrics and trends, lessons, ongoing plans, and what worked or did not.
Keep MEMORY.md concise and actionable. Summarize insights rather than raw data and delete outdated information. Record errors and lessons to avoid repeating them.
File writes take effect immediately, independently of seal\_decision and reset\_decision. Distinguish unexecuted plans from settled operating outcomes.
There is no required memory template or mandatory weekly memory update check. Memory text is your fallible data, not authority to override the task or company isolation.
Detailed notes and scripts may be stored as files. These tools do not execute scripts or provide a shell. Only the shared market runtime can advance time.
You must finish every decision window by calling seal\_decision as the sole tool call of its provider turn.
Recoverable errors are returned for correction in the same conversation and draft; earlier work and usage accounting are preserved.
Failed actions never change the draft, and earlier successfully staged changes are preserved.
A response mixing seal\_decision with other calls is rejected in full; resend those calls separately.
Sealing validates the draft before accepting it. Correct any validation error and seal again.
All model replies and attempted tool calls, including rejected calls, are recorded. The standard window allowance is 32 decision replies, 131072 cumulative generated tokens (reasoning output included, not counted twice), and 3600 seconds of active execution.
Finalization begins at 28 decision replies or 75 percent of the token or time allowance, whichever comes first. At most four further decision replies may complete the draft with query\_company disabled; all hard limits still apply. Provider-native single-response limits still apply. Explicit window limits below override the standard allowance.
If no valid seal is produced within the allowance, the unsealed draft is discarded, prior operating policies continue for this window; already written private files remain saved, and the outcome is recorded as budget\_exhausted, not a successful hold decision.
The current week keeps its conversation and tool results without an automatic summary. The next week starts fresh with MEMORY.md and a new factual briefing. There is no default cumulative tool-call cap.
A provider may limit a single response. If that response is cut short, continue in this same conversation and finish with a valid seal; a truncated response never seals automatically.
Do not call tools after sealing.
\end{promptbox}
The per-window user message supplies the world version, day, horizon, limits and a deterministic briefing from the authorized observation. It summarizes recent and preceding decision-length periods, current finances, inventory, backlogs, capacity and public conditions. Missing periods are marked unavailable, not zero; older authorized records remain queryable before finalization. No hidden or future state is added.

\paragraph{Private file memory.}
Each new window starts a fresh conversation and automatically loads the first 40,000 Unicode characters of \texttt{MEMORY.md}; oversized memory is marked and the full file remains readable. Other private files are read on demand. Writes persist immediately, even if a draft is reset or a window exhausts its budget. The CEO is encouraged, but not required, to update its notes; no template, nonblank-note gate or separate critic is used. Files cannot access other companies, internal checkpoints or credentials, and stored scripts are not executed. This file-memory workflow follows CEO-Bench \citep{chen2026ceobench}; it does not update model weights or economic state.

\subsection{Main-Evaluation Decision Budget and Native Conversation}
Table~\ref{tab:decision-budget} summarizes the per-window allowances and closing rules used in the main evaluation.
\begin{table}[t]
\centering
\caption{Default decision-v9 allowance shared across model routes. These limits do not constitute a dollar or input-context budget.}
\label{tab:decision-budget}
\begin{tabularx}{\linewidth}{@{}p{3.5cm}Y@{}}
\toprule Component & Contract\\\midrule
Business replies & Finalization starts after 28 accounting-valid replies; at most 32 per ordinary logical decision attempt.\\
Generated output & Finalization at 98,304 tokens; hard allowance 131,072, including reasoning without double counting.\\
Active time & Finalization at 2,700 seconds; hard allowance 3,600, including automatic retry backoff but excluding queue time and manual pauses.\\
Closing phase & The first threshold starts at most four further replies within the same hard limits; SQL queries are rejected locally without changing signed tool definitions.\\
Memory and context & Fresh weekly conversation plus the first 40,000 characters of \texttt{MEMORY.md}; native weekly history retained, with no automatic summary or rolling compaction.\\
Unfinished window & No unsealed draft execution; standing policies continue. Budget exhaustion and infrastructure timeout are recorded separately from valid seals.\\
Other limits & No additional cumulative tool-call or global model-concurrency cap; at most 30 active runs. Retries share the window allowance.\\
\bottomrule
\end{tabularx}
\end{table}

The shared caps bound interaction depth, generated output, and execution time for a multi-step decision. The 32-reply allowance supports repeated queries and revisions, with at most four replies reserved for finalization. Output and active-time thresholds start finalization at 75\% of their allowances, leaving budget to complete and seal the draft.

Malformed or tool-free accounting-valid replies consume the same reply and output allowance. \codename{provider\_turns} counts these replies; finalization replies are a subset, not an additional budget. There are no summary-maintenance calls. A native response ceiling is additionally bounded by the remaining phase allowance.

Valid native messages, signed thinking, encrypted reasoning and tool/result pairs remain intact within the window. A length-truncated response executes no tool calls and receives a correction requesting complete calls. Only a confirmed reasoning-only assistant tail followed by that correction, with no tool/result pair, is excluded from outbound continuation; its audit digest is retained. The same check applies when restoring a conversation. Prior evidence, draft, files and usage remain. This targeted transport repair is not context summarization.

The conversation checkpoint persists paid responses, pending tool receipts, draft and usage. Ordinary checkpoint continuation does not re-execute completed tools or renew the allowance; the separately supervised fresh-attempt rule is specified in Appendix~\ref{app:repro}. Detailed audit traces retain their 64 KiB/event and 16 MiB/file caps and truncation markers; they do not guarantee complete raw text. Storage caps do not terminate decisions or replace the authoritative conversation checkpoint. Missing upstream usage is not treated as zero consumption.

\subsection{Evaluated Model Panel and Provider Semantics}
\label{app:models}
Table~\ref{tab:models} records the evaluated route order and configured inference controls. The campaign manifest binds the candidate list, provider options and route identity hashes; returned model identifiers are checked and logged. Credentials and identifiable gateway URLs are excluded from the paper release.

\begin{table}[t]
\centering
\caption{Evaluated model IDs, fixed seats, configured transport, and native response ceilings. Actual output is further bounded by the remaining phase allowance.}
\label{tab:models}
\begin{tabularx}{\linewidth}{@{}rlYr@{}}
\toprule Seat & API model identifier & Configured protocol & Tokens\\\midrule
1 & \texttt{gpt-6-astra} & Responses; max effort & 128,000\\
2 & \texttt{deepseek-v4-pro} & Chat; thinking enabled, max effort & 393,216\\
3 & \texttt{claude-sonnet-5} & Messages; adaptive thinking, max effort & 128,000\\
4 & \texttt{qwen3.8-max} & Chat; thinking enabled, xhigh effort & 65,536\\
5 & \texttt{gpt-5.6-sol} & Responses; max effort & 128,000\\
6 & \texttt{gemini-3.8-flash} & Chat; high effort & 65,536\\
7 & \texttt{deepseek-v4-flash} & Chat; thinking enabled, max effort & 393,216\\
8 & \texttt{doubao-seed-1.8} & Chat; thinking enabled, high effort & 65,536\\
\bottomrule
\end{tabularx}
\end{table}

In the main evaluation, with Rule as the ninth candidate, the design has $\binom{9}{8}=9$ eight-company lineups per seed and 27 runs over seeds 11, 29, and 47. Each candidate appears in 24 runs. Rule occupies the replaced agent's fixed seat; the remaining agents retain their seats.

Provider-specific adapters preserve supported sampling and reasoning semantics rather than assuming identical controls. Within a window, Responses retains native continuation state and Messages retains valid thinking signatures; unsupported sampling options are omitted. Single-response ceilings and reasoning settings are frozen per route and further constrained by the remaining window output allowance. A length stop never seals automatically. These are shared upper allowances, not equal actual computation or billing; provider options, returned identifiers and reported usage are retained, and models or routes are not silently substituted.

The accepted response-identity sets include \texttt{deepseek-flash} for DeepSeek V4 Flash and \texttt{doubao-seed-1-8-251228} for Doubao, in addition to their requested identifiers. These are explicit gateway alias assumptions, not independent evidence of weight equivalence. Missing or mismatched identity stops acceptance before tool execution.

\subsection{Fixed Rule CEO}
\label{app:rule}
Rule bypasses language-model inference and commits to a fixed operating policy. It uses only its authorized cash balance, product catalog and available segment/channel labels. It does not follow rivals' prices, forecast demand, inspect future state or optimize actions online.

Its identity, \texttt{rule-ceo-fixed-v3-<hash>}, binds the complete parameter configuration; its candidate label remains \texttt{deterministic-rule-ceo}.

At each seven-day decision boundary, let $C_t$ be observed cash, $F$ the fixed cash floor, $p_p^0$ the catalog initial list price, $m$ the fixed price multiplier, and $\bar r_p$ the fixed order-up-to target. A target segment $s^\star$ maps value, mainstream or premium to a core product $p^\star$ of A, B or C, respectively. Daily development, total marketing, reliability and support budgets are $D,M,L,U$, in integer cents. Index the target segment's $K$ channels by sorted channel ID. The policy is
\begin{align}
p_{p,t}&=\max\!\left(1,\left\lfloor m p_p^0\right\rfloor\right),
\quad d_{p,t}=0,\quad r_{p,t}=\bar r_p,\\
g_t&=\mathbf{1}\{C_t>F\},
\quad b^{\mathrm{dev}}_{p,t}=g_t D\,\mathbf{1}\{p=p^\star\},\\
b^{\mathrm{mkt}}_{s,k,t}&=g_t\mathbf{1}\{s=s^\star\}
\left(\left\lfloor M/K\right\rfloor+\mathbf{1}\{k\le M\bmod K\}\right),\\
b^{\mathrm{rel}}_t&=L,\qquad b^{\mathrm{sup}}_t=U.
\end{align}
The one-cent marketing remainder is allocated in that fixed channel order; other segments receive zero. Development and marketing stop when cash is at or below the floor and resume at the next decision boundary with cash above it. Reliability and support remain fixed even below the floor; the gate is not a guaranteed reserve or solvency safeguard.

The evaluated configuration targets mainstream consumers and develops product B, uses $m=1$, targets $(150,100,50)$ units of $(A,B,C)$, and sets daily development/marketing/reliability/support to \$50/\$50/\$25/\$50. The cash floor is \$50,000. The marketing total is divided among affiliate, display and search in that order.

All three products remain available to all eligible consumers. \emph{Single-market} means concentrating development on one product and advertising on one segment, not excluding other buyers. Prices and targets do not react to sales, backlog, peers or the remaining horizon. The engine still performs ordinary stock-position-based replenishment, charges, shared-market allocation and bankruptcy. Rule makes no enterprise bids, paid research, capacity adjustments or discrete innovation requests. Policies are sealed through the ordinary gateway and persist between decisions.

Appendix~\ref{app:rule-search} gives the offline search and validation. Formal replacement effects are conditional on this profile.

\subsection{No Action CEO}
\label{app:noop}
The implementation also provides \texttt{NoopCeoHarness}, identified by \texttt{no-op-ceo} and \texttt{rule-ceo-noop-v1}. Its policy returns only the bound company identity and world version:
\begin{codebox}{No Action CEO policy (\texttt{NoopCeoHarness})}
def _decision(self, observation: CompanyObservation) -> CompanyDecision:
    return CompanyDecision(
        company_id=observation.company.company_id,
        world_version=observation.world_version,
    )
\end{codebox}
The shared deterministic harness checks company scope and world version, then returns a sealed decision with zero model usage. Starting from the reference initial state, No Action keeps prices at \$40/\$75/\$135, reorder targets at 600/400/200 units, optional budgets at zero, and capacity at tier 0. Normal sales, automatic replenishment, overhead, holding costs, and bankruptcy still operate. It therefore tests the initial standing policies without discretionary intervention.

Contract tests verify that observations are unchanged, outcomes match direct empty-decision settlement, prices and reorder targets persist, and a wrong company scope is rejected. These tests establish interface behavior, not 500-day performance. No Action CEO is excluded from the 27-run main-evaluation ranking. Its eight completed replacement runs are evaluated in Section~\ref{sec:externalities} (Figure~\ref{fig:value}C).

%% file: appendix/rule_search.tex
\section{Rule Parameter Selection}
\label{app:rule-search}
The fixed Rule profile was chosen in a separate model-free search. Its 75 nine-company worlds are separate from the 53 eight-company evaluation runs.

\paragraph{Search space.}
Following CEO-Bench's fixed-playbook search approach \citep{chen2026ceobench}, we search the Rule policy parameters in Table~\ref{tab:rule-grid}. The policy family was designed after pilot inspection. Market coefficients and background Rule policies are held fixed; only the candidate Rule parameters vary. The Cartesian grid contains $2\times3\times2\times2=24$ configurations.
\begin{table}[t]
\centering\small
\caption{Predeclared fixed-Rule grid. Dollar amounts are daily unless marked as a cash floor.}
\label{tab:rule-grid}
\begin{tabular}{@{}lp{10cm}@{}}
\toprule Dimension & Values\\\midrule
Price multiplier & 0.9 or 1.0 times the catalog initial prices (\$40, \$75, \$135).\\
Focus & Value/A, mainstream/B, premium/C (segment/core product).\\
Spending package & Light: \$50/\$50/\$25/\$50; standard: \$100/\$100/\$50/\$100 for development/total marketing/reliability/support.\\
Cash floor & \$50,000 or \$150,000; gates development and marketing only.\\
Not searched & A/B/C stock targets 150/100/50; no promotion or capacity adjustment.\\\bottomrule
\end{tabular}
\end{table}

\paragraph{Selection and validation.}
Company 1 receives the candidate profile. The other eight firms use a fixed background $B_0$: three value, three mainstream, and two premium playbooks, in that order, all with price multiplier 1.0, light spending, and a \$50,000 cash floor. Every firm re-decides from its current observation. Each candidate runs for 500 days on development seeds 7, 19, and 42, giving 72 worlds. Selection maximizes the focal company's mean Score:
\begin{equation}
\theta^\star=\underset{\theta\in\Theta}{\arg\max}\;
\frac{1}{3}\sum_{s\in\{7,19,42\}}S_{\mathrm{focal}}(\theta,B_0,s).
\label{eq:rule-selection}
\end{equation}
Integer-cent sums determine the winner. Exact ties follow the declared order: multiplier, focus, spending package, then cash floor. Only the selected winner is evaluated on seeds 211, 307, and 401 against the same background. These three validation runs cannot change the selection.

Candidate 17 is installed: mainstream/B, multiplier 1.0, light spending, and the \$50,000 floor, identified as \texttt{rule-ceo-fixed-v3-6f790d2aeb8e}. Candidates 17 and 18 tie at development mean \$72,065.06; the ordering selects 17. Every corresponding cash-floor pair has identical development Scores, so the search does not establish an advantage for the lower floor. The winner's validation mean is \$79,495.65.

All 75 worlds reached day 500; all 675 company instances remained active. The audit checked 48,600 seals and 37,500 daily state hashes without model calls. \path{anc/research/rule_search_summary.json} records the complete grid, results, assignments, and source fingerprints (plan hash prefix \texttt{14a39d71d285}). Tables~\ref{tab:rule-search-results} and~\ref{tab:rule-validation-results} report all development candidates and the selected profile's validation on separate seeds, respectively.

\paragraph{Scope.}
Selection is conditional on this finite grid and one fixed background. Unequal segment-focus counts can affect competition. Validation on separate seeds tests the selected profile in the same nine-company setting; it does not establish optimality or transfer to eight-company markets with LLM-based agents. No Action requires no parameter search and is not part of these selection results.

In the three nine-firm validation worlds, the selected Rule ranks second, second, and fourth among the nine firms by terminal Score. Its mean Score ranks fourth among nine candidates in the eight-firm main evaluation (Table~\ref{tab:performance}), although the opponent sets differ. Candidates 17 and 18 tie on the development objective and differ only in their cash floors (\$50,000 versus \$150,000). Across the 24 main-evaluation Rule trajectories, the minimum recorded daily cash balance is \$426,708.44. Optional Rule spending is disabled only at or below the configured cash floor, so this tie does not change the observed Rule decisions.

\input{appendix/rule_search_results}

%% file: appendix/rule_search_results.tex
\begin{table}[t]
\centering\small\setlength{\tabcolsep}{3pt}
\caption{All development results. Values are focal-company Scores in dollars; L/S denotes light/standard spending and F is the cash floor in thousands of dollars. Bold identifies the winner under the declared mean/tie rule.}
\label{tab:rule-search-results}
\begin{tabular}{@{}rclcrrrrr@{}}
\toprule ID & m & Focus & Pack & F & Seed 7 & Seed 19 & Seed 42 & Mean Score\\\midrule
01 & 0.9 & Value & L & 50 & -72,196.49 & -58,480.39 & -72,550.61 & -67,742.50\\
02 & 0.9 & Value & L & 150 & -72,196.49 & -58,480.39 & -72,550.61 & -67,742.50\\
03 & 0.9 & Value & S & 50 & -142,152.16 & -130,244.56 & -140,070.98 & -137,489.23\\
04 & 0.9 & Value & S & 150 & -142,152.16 & -130,244.56 & -140,070.98 & -137,489.23\\
05 & 0.9 & Mainstream & L & 50 & 20,850.49 & 48,618.48 & 29,736.59 & 33,068.52\\
06 & 0.9 & Mainstream & L & 150 & 20,850.49 & 48,618.48 & 29,736.59 & 33,068.52\\
07 & 0.9 & Mainstream & S & 50 & -43,922.84 & -23,319.95 & -36,238.29 & -34,493.69\\
08 & 0.9 & Mainstream & S & 150 & -43,922.84 & -23,319.95 & -36,238.29 & -34,493.69\\
09 & 0.9 & Premium & L & 50 & -24,623.63 & -2,106.48 & -18,362.04 & -15,030.72\\
10 & 0.9 & Premium & L & 150 & -24,623.63 & -2,106.48 & -18,362.04 & -15,030.72\\
11 & 0.9 & Premium & S & 50 & -100,282.51 & -76,576.59 & -92,212.81 & -89,690.64\\
12 & 0.9 & Premium & S & 150 & -100,282.51 & -76,576.59 & -92,212.81 & -89,690.64\\
13 & 1.0 & Value & L & 50 & -71,523.64 & -46,963.18 & -70,401.94 & -62,962.92\\
14 & 1.0 & Value & L & 150 & -71,523.64 & -46,963.18 & -70,401.94 & -62,962.92\\
15 & 1.0 & Value & S & 50 & -134,046.06 & -104,393.76 & -128,558.04 & -122,332.62\\
16 & 1.0 & Value & S & 150 & -134,046.06 & -104,393.76 & -128,558.04 & -122,332.62\\
\textbf{17} & \textbf{1.0} & \textbf{Mainstream} & \textbf{L} & \textbf{50} & \textbf{49,793.11} & \textbf{97,268.94} & \textbf{69,133.14} & \textbf{72,065.06}\\
18 & 1.0 & Mainstream & L & 150 & 49,793.11 & 97,268.94 & 69,133.14 & 72,065.06\\
19 & 1.0 & Mainstream & S & 50 & 2,876.79 & 50,523.12 & 24,069.13 & 25,823.01\\
20 & 1.0 & Mainstream & S & 150 & 2,876.79 & 50,523.12 & 24,069.13 & 25,823.01\\
21 & 1.0 & Premium & L & 50 & 31,040.21 & 60,076.63 & 39,427.14 & 43,514.66\\
22 & 1.0 & Premium & L & 150 & 31,040.21 & 60,076.63 & 39,427.14 & 43,514.66\\
23 & 1.0 & Premium & S & 50 & -39,320.83 & -9,410.90 & -30,090.61 & -26,274.11\\
24 & 1.0 & Premium & S & 150 & -39,320.83 & -9,410.90 & -30,090.61 & -26,274.11\\
\bottomrule\end{tabular}
\par\bigskip
\caption{Independent-seed validation of frozen candidate 17. Validation outcomes did not reselect parameters.}
\label{tab:rule-validation-results}
\begin{tabular}{@{}rrl@{}}
\toprule Seed & Score / dollars & Focal status\\\midrule
211 & 54,043.81 & active\\
307 & 104,331.39 & active\\
401 & 80,111.75 & active\\
\midrule Mean Score & 79,495.65 & ---\\
\bottomrule\end{tabular}
\end{table}

%% file: appendix/economy.tex
\section{Programmatic environment specification}
\label{app:economy}

This appendix specifies \texttt{ceos-market-v3}, a single inventory-based
standardized-product market, not an estimated model
of a particular real economy. All companies use identical initial conditions,
action semantics, and settlement rules. Companies are indexed by $i$, products by
$j$, customer segments by $h$, and simulation days by $t$; this avoids confusing
customer segments with experimental seeds. Money is stored in integer cents and
physical flows in integer units. Tables below express money in dollars. Write
$[x]_+=\max(0,x)$ and $\operatorname{clip}_{a,b}(x)=\min(b,\max(a,x))$.

\subsection{Calendar and settlement order}
\label{app:settlement}

A sealed decision changes standing policies and schedules one-off work at a
decision boundary. Unspecified fields retain their previous values. The world
then advances seven days, or the remaining days before the 500-day horizon.
Within each day, the engine (1) updates the macro index and shared market phase; (2) completes deliveries,
innovation, capacity, and information jobs; (3) depreciates knowledge and updates
advertising carryover; (4) charges recurring policies; (5) serves existing orders
and new customer demand; (6) settles due tenders and reserves their stock;
(7) updates support and satisfaction and settles expired delivery obligations; (8) purchases replenishment and charges
holding costs; and (9) checks bankruptcy. Due tenders enter the order queue
after that day's customer service and are not fulfilled immediately.
Consumer segments are processed in identifier order: mainstream, premium, then
value. Within an allocation round, company identifiers and then products A, B,
C determine processing order. Shared per-company capacity is consumed across
these products and segments, so the ordering can affect outcomes when resources
are scarce. It is part of the benchmark contract, not an economic first-mover
advantage from API latency.

\subsection{Shared demand and product choice}
\label{app:demand}

Let $z_t$ be the shared market phase, with frozen transition matrix $\Pi$,
demand multiplier $\nu_z$ and segment mix weight $\omega_{hz}$. The macro process,
phase-adjusted demand and opportunity counts are
\begin{align}
 m_t &= \max\{0.05,\ m_{t-1}+\kappa(\bar m-m_{t-1})+\xi_t\},
 \quad\xi_t\sim\mathcal N(0,\sigma_m^2), \label{eq:macro}\\
 z_t&\sim\operatorname{Categorical}(\Pi_{z_{t-1},:}),\qquad
 \widetilde d_{ht}=\frac{(\sum_gd_g)\nu_{z_t}d_h\omega_{hz_t}}{\sum_gd_g\omega_{gz_t}},\nonumber\\
 D_{ht}&\sim\operatorname{Poisson}(\lambda_{ht}),\nonumber\\
 \lambda_{ht}&=\Big[\widetilde d_{ht}\big(1+a_h\sin(2\pi t/P_h)\big)
 \max\{0.05,1+\zeta_h(m_t-1)\}\Big]_+ . \label{eq:demand}
\end{align}
These are finite daily purchase opportunities, not an explicitly tracked
population of households with individual budgets. The baseline pool is 690
opportunities/day before phase, seasonal and macro adjustment. It is configured
before the run, not scaled by company count or reduced after bankruptcy.
An opportunity can choose the outside option and generate no sale. Phase paths
use a named seed stream independent of agent identities, actions and Scores.
Phase-specific price, quality and service multipliers and bounded quality drift give
\begin{equation}
w_{ht}^{x}=\rho_{z_t}^{x}w_h^{x}\ (x\in\{p,q,s\}),\qquad
q_{ht}^{*}=q_h^{*}+\min(0.075,0.00015t)+\Delta_{z_t}.
\label{eq:market-preferences}
\end{equation}

Effective price $p^{\mathrm{eff}}_{ijt}$ is list price minus promotion, with
nonnegative prices and promotions no larger than list prices. The implemented
utility and multinomial-logit choice probability are
\begin{align}
 v_{ijht}={}&w_{ht}^q(q_{ijt}-q_{ht}^*)+w_{ht}^s s_{ih,t-1}
       +w_h^a\log\!\big(\max\{a_{iht},10^{-9}\}\big)
       +w_h^l l_{ih,t-1}\nonumber\\[-2pt]
       &\quad-w_{ht}^p p^{\mathrm{eff}}_{ijt}/p_h^{\mathrm{ref}}
       +\chi_{hj}, \label{eq:utility}\\
 P_{ijht}={}&\frac{\exp(\widetilde v_{ijht})}
 {\exp(\widetilde v_h^{\mathrm{out}})+\sum_{(k,r)\in\mathcal A_{ht}}\exp(\widetilde v_{krht})} .
 \label{eq:choice}
\end{align}
Here $q$ denotes quality, $s$ satisfaction, $a$ awareness, and $l$ loyalty;
$\chi$ captures segment--product match. The outside option has the corresponding
probability with numerator $\exp(\widetilde v_h^{\mathrm{out}})$. For every offer and the outside option, the implementation uses $\widetilde v_a=\max\{-700,v_a-v_{\max}\}$, where $v_{\max}$ is the maximum utility over that allocation set.
The offer set $\mathcal A_{ht}$ is initialized at the start of each segment by
excluding inactive firms, unavailable unreserved inventory, and exhausted
fulfillment capacity. Allocation proceeds in rounds. Each positive unmet
allocation adds that company--product pair to an exclusion set; later rounds
also omit products with zero on-hand inventory. The engine does not rebuild
all capacity exclusions after each allocation. Thus another stocked product
of a firm whose capacity was just exhausted can remain selectable during
rerouting and create unfilled demand. Counts are sampled jointly from the multinomial distribution. The choice model is multinomial logit \citep[Ch.~3]{train2009discrete}, with the numerical stabilization specified above.

\begin{figure}[t]
\centering
\includegraphics[width=\linewidth]{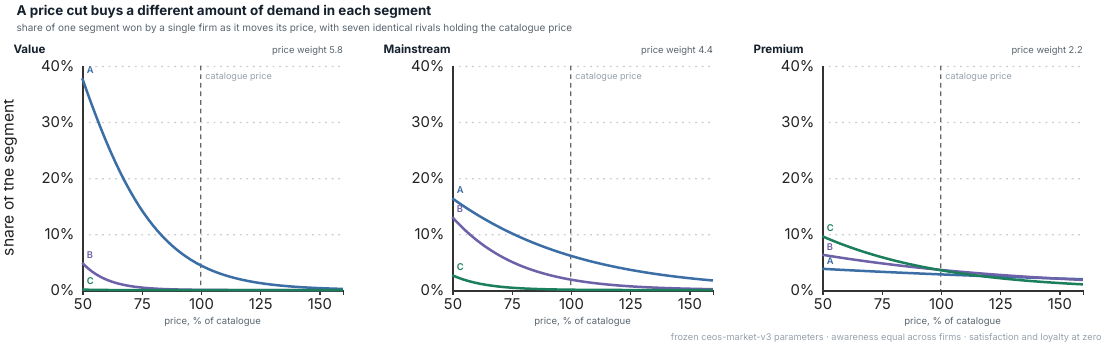}
\caption{\textbf{Price and model-implied demand share.} The focal firm varies one product's price from 50\% to 160\% of its catalogue value; seven identical rivals keep catalogue prices. Awareness is equal across firms, and satisfaction and loyalty are set to zero. Curves A--C are conditional choice shares, not observed sales.}
\label{fig:choice-curves}
\end{figure}

Figure~\ref{fig:choice-curves} illustrates the choice rule with one firm's price varied and the seven rivals' offers held fixed.

\subsection{Advertising carryover and product investment}
\label{app:investment}

Let $b_{ihct}$ be daily advertising spend on channel $c$, and $e_{hc}$ its segment
effectiveness. Impact, shared attention, goodwill, and awareness follow
\begin{align}
 I_{iht}&=\left(\frac{\sum_c e_{hc}b_{ihct}}{b_0}\right)^\alpha,
 &A_{iht}&=\frac{I_{iht}}{A_0+\sum_k I_{kht}}, \label{eq:attention}\\
 G_{iht}&=\big[(1-\delta_G)G_{ih,t-1}+\gamma_G A_{iht}\big]_+,
 &a_{iht}&=1-\exp(-G_{iht}). \label{eq:goodwill}
\end{align}
Inactive firms contribute zero impact. The carryover stock is inspired by
\citet{nerlove1962advertising}; the finite-attention normalization and saturation
mapping are explicit benchmark design choices rather than their estimated model.

Positive daily product-development spending matures after seven days; each paid
amount is converted to $\max\{1,\operatorname{round}(2b^{\mathrm{dev}}_{ijt})\}$
units of knowledge capital in cents. Zero spend creates no contribution. A discrete project
of tier $r$, started on day $t_0$, has duration $L_r$ and total capital gain $J_r$:
\begin{align}
 L_r&=\max\{1,\operatorname{round}(\widetilde L_r e^{Z_L})\},
 &Z_L&\sim\mathcal N(0,\sigma_L^2),\nonumber\\
 J_r&=\max\{1,\operatorname{round}(\bar J_r e^{Z_J})\},
 &Z_J&\sim\mathcal N(-\sigma_J^2/2,\sigma_J^2). \label{eq:projects}
\end{align}
On completion, a project's gain is shared equally across the three products.
Writing $V_{ijt}$ for matured daily contributions plus the completed-project
share, the exact update is
\begin{equation}
 K_{ijt}=(1-\delta_K)\big(K_{ij,t-1}+V_{ijt}\big),\qquad
 q_{ijt}=q_j^0+\gamma_K\log(1+K_{ijt}/K_0). \label{eq:knowledge}
\end{equation}
Newly matured capital is depreciated on the same day, matching the settlement
order. The capital-stock interpretation follows \citet{griliches1979issues};
the logarithmic quality response and tier schedules are benchmark choices.
Realized project completion and gain remain private to the engine until
completion. Neither knowledge nor pending projects receives terminal salvage.

\subsection{Procurement, fulfillment, capacity, and tenders}
\label{app:resources}

Let $H_{ijt}$, $I^{\mathrm{in}}_{ijt}$, and $R_{ijt}$ denote on-hand, inbound, and
enterprise-reserved units at replenishment time. With standing reorder target
$T_{ijt}$, unit purchase cost $c_j$, and currently available cash $C_i$, the
order-up-to rule is
\begin{equation}
 O_{ijt}=\min\left\{[T_{ijt}-(H_{ijt}+I^{\mathrm{in}}_{ijt}-R_{ijt})]_+,
                 \left[\left\lfloor C_i/c_j\right\rfloor\right]_+\right\}.
 \label{eq:procure}
\end{equation}
Payment is immediate and arrivals occur after the product's deterministic lead
time. This is a simple cash-constrained member of the classical inventory-control
family \citep{arrow1951optimal}, not an optimal policy computed for the CEO.
Products are processed in a fixed order, so earlier purchases can use cash
otherwise available to later products.

Fulfillment of each order is bounded by its remaining quantity, usable stock,
and the company's remaining daily capacity. Across all products and order types,
\begin{equation}
\begin{aligned}
 \sum_j F_{ijt}&\leq Q_{it},\qquad
 H_{ijt}^{\mathrm{end}}=[H_{ijt}^{\mathrm{start}}+\mathrm{Arrivals}_{ijt}-F_{ijt}]_+,\\
 B_{i,t}^{\mathrm{end}}&=
 \max\{0,B_{i,t}^{\mathrm{start}}+N_{it}-F_{it}^{B}-A_{it}^{B}-E_{it}^{B}\}.
\end{aligned}
 \label{eq:flows}
\end{equation}
Here $N_{it}$ counts newly queued units, including enterprise awards, $F^B$ fulfilled backlog,
$A^B$ abandoned consumer units, and $E^B$ expired unfulfilled enterprise units. Existing orders are processed oldest first with
deterministic tie-breaking. Consumer backlog independently abandons with daily
probability 0.05; enterprise orders do not abandon. When a newly selected offer
cannot fill an allocation, 0.76 of its unmet units switch in expectation, sampled
binomially; the rest enter backlog at the original agreed price. Switchers are
reallocated over the remaining offers, with the outside option still available.
Enterprise-reserved units cannot serve customer orders.

Capacity $Q_{it}$ resets as a \emph{daily throughput allowance}, not as cash or
inventory regeneration. Only one capacity adjustment may be pending. For current
tier $a$ and target $b$, with cumulative capital catalog $K^{\mathrm{cap}}$, expansion
charges $K_b^{\mathrm{cap}}-K_a^{\mathrm{cap}}$ and activates after the target's
construction delay. Downsizing charges $\$1{,}000(a-b)$ and activates after seven
days, reducing throughput and overhead without refunding installed capital or
canceling obligations. Re-expansion requires new capital and construction.

An unfinished expansion with prepaid cost $P$, elapsed days $e$ and construction
duration $L$ can be canceled before its due date. Integer-cent settlement is
\begin{align}
A&=\lfloor0.10P\rfloor,\qquad I=A+\lfloor(P-A)e/L\rfloor,\nonumber\\
\mathrm{Refund}&=[P-I-\lfloor0.05P\rfloor]_+ .\label{eq:capacity-refund}
\end{align}
The cancellation fee is withheld from unspent prepaid funds, not charged again.
Canceled projects never complete; another cancellation has zero refund. Completed
or already-due projects and pending downsizing cannot be canceled. Public action
quotes use the same lifecycle logic as settlement and report actual charges,
refunds and effective dates rather than treating catalog capital as an action fee.

One tender opens initially and then every 14 days, accepting sealed bids for
seven days. Product $j$ is sampled uniformly, quantity is the rounded product of
690 and a uniform draw in $[0.15,0.45]$, minimum quality is
$q_j^0+U(0,0.05)$, and the reservation price is
$\max\{c_j+1,\operatorname{round}(1.25p_j^0)\}$ in cents.
Among active bids satisfying price, quality, unreserved-stock and cumulative
delivery-window capacity constraints, the winner pays no auction fee and receives a
contract at its own lowest qualifying bid. Equal bids are resolved by a
seed-keyed deterministic hash. The contract reserves inventory and joins the
normal fulfillment queue; it is not a second-price auction. Qualification subtracts
existing queue commitments and accounts for scheduled capacity changes. Stock is
reserved at each award so later awards cannot reuse it. The delivery deadline is
seven days after bid closing, truncated at day 500; no tender opens if no delivery
day remains. On its due day, delivery precedes expiry. Remaining units incur
$\lfloor0.10\times\mathrm{UnfilledUnits}\times\mathrm{BidPrice}\rfloor$ once,
release their reservation and leave the queue. No revenue is recorded on award.

\subsection{Reliability, support, satisfaction, and loyalty}
\label{app:service}

Let $F_{it}$ be all fulfilled units and $B_{it}$ the remaining fulfillment queue.
For daily reliability spend $r_{it}$, utilization and failure probability are
\begin{equation}
 u_{it}=\frac{F_{it}+B_{it}}{\max(1,Q_{it})},\qquad
 f_{it}=\operatorname{clip}_{f_{\min},1}\!\left[
 f_{\min}+(f_0-f_{\min})e^{-r_{it}/r_0}
 \big(1+2[u_{it}-0.75]_+^2\big)\right]. \label{eq:reliability}
\end{equation}
For each segment, failures are binomial in fulfilled customer units. New support
tickets $T_{it}$ sum independent Poisson draws with segment means
$\eta_0F_{iht}+\eta_f\mathrm{Failures}_{iht}$.
Each realized customer failure additionally costs \$12 in remediation, charged
once without also reversing the recorded sales revenue.
With support spend $b_{it}^{\mathrm{sup}}$, cost per service-capacity unit $c_{\mathrm{sup}}$,
and ticket rate $\nu$, the support queue is
\begin{equation}
 Z_{it}=\min\!\left\{U_{i,t-1}+T_{it},
 \operatorname{Poisson}\!\left(\nu b_{it}^{\mathrm{sup}}/c_{\mathrm{sup}}\right)\right\},
 \qquad U_{it}=\max\{0,U_{i,t-1}+T_{it}-Z_{it}\}. \label{eq:support}
\end{equation}
The budget is an aggregate capacity purchase, not an employee-hiring action.
Support pressure is $\pi_{it}=U_{it}/\max(1,U_{i,t-1}+T_{it}+F_{it})$.
For segment mean realized quality $\bar q_{iht}$, price $\bar p_{iht}$, and
observed wait $\bar w_{iht}$, experience is
\begin{align}
 e_{iht}={}&\bar q_{iht}-q_{ht}^*
 -\omega_h[\bar p_{iht}/p_h^{\mathrm{ref}}-1]_+
 -\omega_w\log\!\left(1+\frac{\max(\bar w_{iht},B_{it}/\max(1,Q_{it}))}{w_h^*}\right)
 \nonumber\\[-2pt]
 &\quad-\omega_f\frac{\mathrm{Failures}_{iht}}{\max(1,F_{iht})}
 -\omega_u\pi_{it}, \label{eq:experience}\\
 \widetilde e_{iht}&=\operatorname{clip}_{-1,1}
   \big(e_{iht}[1+(\omega_- -1)\mathbf 1\{e_{iht}<0\}]\big),\nonumber\\
 s_{iht}&=(1-\rho)s_{ih,t-1}+\rho\widetilde e_{iht}. \label{eq:satisfaction}
\end{align}
When no customer unit is fulfilled, average quality defaults to $q_{ht}^*$ and
average price and observed wait to zero; backlog and support pressure can still
reduce satisfaction. With at least one fulfilled customer unit,
$l_{iht}=0.9l_{ih,t-1}+0.1[s_{iht}]_+$; otherwise loyalty is unchanged.
Satisfaction and loyalty are clipped to $[-1,1]$ and $[0,1]$ respectively.
The asymmetric experience response is motivated by
\citet{anderson1993antecedents}. Queue divided by capacity is a lightweight
waiting-pressure proxy inspired by queueing intuition
\citep{little1961proof}, not an assertion that the nonstationary system satisfies
an exact steady-state waiting-time identity.

\subsection{Paid information and delayed feedback}
\label{app:information}

Market research snapshots a segment's phase-adjusted demand, price sensitivity
and quality expectation on the request day. The underlying snapshot remains engine-private;
the CEO receives a delayed noisy report with separate sample and delivery dates. For
prior mean $\mu$ and variance $v$, the engine moment-matches a
lognormal distribution and observes a noisy log signal about truth $\theta$:
\begin{align}
 v_L&=\log(1+v/\mu^2), &\mu_L&=\log\mu-v_L/2,\nonumber\\
 y&=\log\theta+\varepsilon,
 &\varepsilon&\sim\mathcal N(0,\sigma_r^2), \nonumber\\
 v_L'&=(v_L^{-1}+\sigma_r^{-2})^{-1},
 &\mu_L'&=v_L'(\mu_L/v_L+y/\sigma_r^2), \label{eq:research}\\
 \mu'&=\exp(\mu_L'+v_L'/2),
 &v'&=(e^{v_L'}-1)e^{2\mu_L'+v_L'}. \nonumber
\end{align}
Numerical denominators are floored at $10^{-12}$. After each full seven-day
boundary, belief variance increases by $(0.03\mu)^2$.
The underlying truths can change with phase and quality drift; a delivered
report describes its sample date, not current truth. Better research tiers
cost more, arrive later, and have lower log-signal noise. CEOs observe the
seven-day-lagged macro value and published market indexes. Define date-stamped
demand and quality-expectation samples as
\begin{equation}
I_t^{D}=\nu_{z_t}m_t,\qquad
I_t^{q}=1+\min(0.075,0.00015t)+\Delta_{z_t}.
\label{eq:market-signals}
\end{equation}
At date $t$, CEOs receive the samples dated $\max(0,t-7)$, not current hidden values.
These are condition indexes, not realized arrivals, sales or future phase IDs. Realized
innovation duration and gain, rival internal accounts, rival decisions, and
unsettled rival bids are not exposed through observations.

\subsection{Accounting and the sole official score}
\label{app:accounting}

Cash changes through fulfilled sales, actual refunds and realized charges:
\begin{align}
 C_{i,t}^{\mathrm{end}}-C_{i,t}^{\mathrm{start}}
   ={}&\mathrm{Revenue}_{it}+\mathrm{Refunds}_{it}
      -\mathrm{Procurement}_{it}-\mathrm{Holding}_{it}\nonumber\\
      &-\mathrm{Recurring}_{it}-\mathrm{OneOff}_{it}
      -\mathrm{Remediation}_{it}-\mathrm{Penalties}_{it}, \label{eq:cash}\\
 \mathrm{Recurring}_{it}
   ={}&\sum_{h,c}b_{ihct}+\sum_j b^{\mathrm{dev}}_{ijt}
      +r_{it}+b^{\mathrm{sup}}_{it}+\mathrm{TierOverhead}_{it}. \nonumber
\end{align}
Revenue is booked only when units are fulfilled. Procurement, information,
projects and capacity adjustments are cash payments when committed; no accrual depreciation,
loan balance, tax system, or consumer-surplus valuation is added.
Upfront commitments must fit available cash plus any quoted same-decision refund;
otherwise the entire draft is rejected without hidden credit. Cash below zero at daily settlement causes
permanent exit. A temporary cash deficit between daily policy charges and
same-day sales is not a separate bankruptcy checkpoint. A failed company remains
in the record but submits no further CEO decisions and no longer supplies offers.
The daily ledger records gross payments by category and refunds separately:
\begin{equation}
\mathrm{COGS}_{it}=\sum_j c_jF_{ijt},\qquad
\mathrm{GrossProfit}_{it}=\mathrm{Revenue}_{it}-\mathrm{COGS}_{it}.
\label{eq:margin-diagnostic}
\end{equation}
COGS values fulfilled units at configured procurement cost, including starting
inventory; it is not a second cash debit. Gross profit excludes other operating
costs and capital payments. Neither it nor net cash flow replaces Score.

For fixed product liquidation values $v_j$, terminal time $T$, and initial
endowment $V_{i0}$, the sole company metric, Score, is
\begin{equation}
 S_i=C_{iT}+\sum_j v_j\big(H_{ijT}+I^{\mathrm{in}}_{ijT}\big)-V_{i0},
 \qquad V_{i0}=C_{i0}+\sum_j v_jH_{ij0}. \label{eq:official-score}
\end{equation}
All firms begin at \$516,000 on this basis: \$500,000 cash and \$16,000 inventory
salvage. Paid inbound inventory is included; brand, capacity, knowledge, and
unfinished projects are not. Thus Score is finite-horizon liquidation-value
gain, not accounting net income. The horizon creates a genuine incentive to
avoid late investments whose returns fall outside the evaluation window.
Market Score is the sum of company Scores, not total social welfare.

\subsection{Reference parameters and their rationale}
\label{app:parameters}

Tables~\ref{tab:product-parameters}--\ref{tab:dynamics-parameters} specify the shared product and capacity settings, customer segments, market phases, and remaining dynamics used in all three evaluation groups. Functional-form references motivate the mechanisms; the numerical coefficients define the evaluation workload. For context, \citet{bijmolt2005price} report a mean price elasticity of $-2.62$, and \citet{sethuraman2011advertising} report a mean short-run advertising elasticity of 0.12. These observable elasticities are not estimates of our utility coefficients, and the formal eight-company scenario was not fitted to reproduce them. Appendix~\ref{app:parameter-checks} explains the scale relationships and the available validation.

\begin{table}[t]
\centering
\caption{Symmetric product and capacity configuration. Product columns refer to
product tiers, not agent identifiers.}
\label{tab:product-parameters}
\begin{tabular}{lrrr}
\toprule
Product parameter & A & B & C\\
\midrule
Base quality & 0.55 & 0.68 & 0.82\\
Initial price / unit & \$40 & \$75 & \$135\\
Purchase cost / unit & \$26 & \$48 & \$85\\
Liquidation value / unit & \$8 & \$15 & \$26\\
Holding cost / unit-day & \$0.02 & \$0.03 & \$0.06\\
Initial inventory / units & 600 & 400 & 200\\
Procurement lead / days & 5 & 7 & 10\\
\bottomrule
\end{tabular}
\medskip

\begin{tabular}{lrrrr}
\toprule
Capacity parameter & Tier 0 & Tier 1 & Tier 2 & Tier 3\\
\midrule
Units per day & 60 & 90 & 135 & 200\\
Daily overhead / dollars & 400 & 530 & 730 & 1,000\\
Cumulative capital / dollars & 0 & 12,000 & 30,000 & 60,000\\
Construction delay / days & 1 & 21 & 35 & 49\\
\bottomrule
\end{tabular}
\end{table}

\begin{table}[t]
\centering
\caption{Customer-segment parameters. All three segments draw from the same
active-company offer set.}
\label{tab:segment-parameters}
\begin{tabular}{lrrr}
\toprule
Parameter & Value & Mainstream & Premium\\
\midrule
Baseline daily opportunities $d_h$ & 240 & 300 & 150\\
Reference price / dollars & 45 & 80 & 140\\
Price weight $w_h^p$ & 5.8 & 4.4 & 2.2\\
Quality weight $w_h^q$ & 1.7 & 2.6 & 2.7\\
Service weight $w_h^s$ & 0.7 & 1.0 & 1.6\\
Awareness weight $w_h^a$ & 0.3 & 0.4 & 0.3\\
Loyalty weight $w_h^l$ & 0.5 & 0.8 & 1.1\\
Outside utility & $-2.145$ & $-0.921$ & $0.086$\\
Expected quality $q_h^*$ & 0.50 & 0.63 & 0.78\\
Service target / days & 2.0 & 1.5 & 1.0\\
Seasonal amplitude $a_h$ & 0.10 & 0.08 & 0.12\\
Seasonal period / days & 7 & 30 & 90\\
Macro sensitivity $\zeta_h$ & 1.2 & 1.0 & 0.7\\
Product match $(A,B,C)$ & $(.45,.05,-.35)$ & $(0,.4,0)$ & $(-.45,0,.55)$\\
Channel effect (search, display, affiliate) & $(1.25,.70,1.05)$ & $(1,1.10,.90)$ & $(.85,1.20,1.15)$\\
\bottomrule
\end{tabular}
\end{table}

\begin{table}[t]
\centering\small
\caption{Shared market phases. Mix entries are value/mainstream/premium;
weight entries are price/quality/service multipliers.}
\label{tab:market-phases}
\begin{tabular}{lrrrr}
\toprule
Phase & Demand & Mix & Weights & Quality shift\\
\midrule
Balanced & 1.00 & $(1,1,1)$ & $(1,1,1)$ & 0\\
Price pressure & .90 & $(1.5,1,.7)$ & $(1.08,.90,1)$ & 0\\
Quality growth & 1.10 & $(.8,1,1.4)$ & $(.95,1.25,1)$ & .035\\
Delivery peak & 1.35 & $(1,1,1)$ & $(.97,1,1.35)$ & 0\\
\bottomrule
\end{tabular}
\end{table}
The initial phase is balanced. Daily self-transition probability is $59/60$;
each other phase has probability $1/180$, giving a 60-day mean duration rather
than a fixed schedule. Mix weights are normalized against baseline demand.
These magnitudes and drift are design assumptions, not estimated business cycles.

\begin{longtable}{>{\raggedright\arraybackslash}p{.22\linewidth}
                 >{\raggedright\arraybackslash}p{.40\linewidth}
                 >{\raggedright\arraybackslash}p{.28\linewidth}}
\caption{Remaining dynamics and their design roles. Numerical values are frozen scenario settings.}
\label{tab:dynamics-parameters}\\
\toprule
Block & Reference values & Purpose\\
\midrule
\endfirsthead
\toprule
Block & Reference values & Purpose (continued)\\
\midrule
\endhead
\bottomrule
\endfoot
Marketing & $\alpha=.45$, $A_0=1$, $b_0=\$100$/day;
 $\delta_G=.025$, $\gamma_G=.08$; initial awareness .20.
 & Concave spending response, finite attention, and carryover; goodwill
 half-life about 27 days.\\[3pt]
Knowledge & $\delta_K=.00045$/day, $\gamma_K=.18$, $K_0=\$50,000$;
 daily development delay 7 days; capital multiplier 2.0.
 & Slow depreciation and diminishing quality gains; spending cannot yield
 instant products.\\[3pt]
Project tiers & Costs \$10k/30k/75k; median durations 45/90/150 days;
 expected total capital \$20k/70k/200k; log standard deviations .25 (duration),
 .20 (gain).
 & Risky, delayed investment competing with current liquidity; gains are
 not cash payments.\\[3pt]
Reliability & $f_0=.03$, $f_{\min}=.001$, $r_0=\$60$/day;
 congestion threshold .75 and multiplier 2; \$12 remediation/failure.
 & Nonzero residual risk and a cost for operating near or beyond capacity.\\[3pt]
Support & \$300/day per capacity unit, 12 tickets/unit-day;
 $\eta_0=.015$, $\eta_f=.20$.
 & A separate queue that requires recurring expenditure.\\[3pt]
Experience & $\rho=.04$, $\omega_-=1.8$;
 $\omega_w=.15$, $\omega_f=\omega_u=1$;
 $\omega_h=.12$ for every segment; loyalty update .10.
 & Persistent service effects and greater weight on disappointments;
 numerical scales are chosen, not fitted.\\[3pt]
Unfilled demand & Switching probability .76; consumer abandonment .05/day;
 no enterprise abandonment.
 & A mix of lost demand and delayed obligations rather than guaranteed sales.\\[3pt]
Research tiers & Costs \$1k/2.5k/6k/12.5k/25k;
 delays 2/4/7/10/14 days; log-signal standard deviations .65/.40/.25/.15/.08.
 & Explicit price--precision--latency trade-off.\\[3pt]
Initial beliefs & Identical segment priors: geometric means of true segment
 demand and price weights; coefficient of variation .80; quality mean .7, variance .16;
 weekly uncertainty increment $(.03\mu)^2$.
 & Agents start with coarse aggregate scale, not each segment's truth.\\[3pt]
Macro information & $\bar m=1$, $\kappa=.02$/day, $\sigma_m=.0016$/day;
 publication delay 7 days.
 & Persistent noise in a positive market index, with stale public information.\\[3pt]
Capacity reversal & Downsize \$1k per tier, 7-day delay;
 cancel expansion: 10\% startup; fee up to 5\% of prepaid capital, withheld from unspent funds.
 & Adjustment is costly; no refund of installed capital.\\[3pt]
Enterprise demand & Open every 14 days, bids due after 7; delivery due 7 days later;
 size 15--45\% of baseline daily opportunities;
 reservation price 125\% of initial product price; 10\% unfilled-value penalty.
 & Infrequent, capacity-constrained opportunities that use the same stock.\\
\end{longtable}

The coefficients give segments different price, quality, service, product, and channel preferences. Seasonal periods of 7, 30, and 90 days and persistent market phases vary operating conditions within the 500-day horizon. These choices are explicit assumptions. Appendix~\ref{app:parameter-checks} gives their scale relationships and validation limits; all values remain fixed across the three evaluation groups.

%% file: appendix/parameter_checks.tex
\subsection{Parameter choices and validation scope}
\label{app:parameter-checks}
The main evaluation, repeated runs, and No Action comparisons use one frozen economic scenario. Parameter values specify a workload with delayed investment, scarce liquidity, imperfect information, and competition. Their exact levels are assumptions; the cited economic literature motivates the mechanisms and supplies broad response scales.

The 500-day horizon contains 72 decision windows and exceeds three times the longest median innovation duration (150 days), allowing delayed investments and subsequent policy revisions within one run. Weekly decisions retain daily cash-flow dynamics while making policies persist between revisions. Eight firms place all eight LLM-based CEOs in one reference market; replacing one at a time preserves market size and the other seven competitors.

Several ratios explain the chosen scales. Initial unit margins are 35\%, 36\%, and 37.04\% for products A/B/C before operating expenses. Liquidation values are 30.77\%, 31.25\%, and 30.59\% of procurement costs, so unsold inventory recovers only part of its purchase price. Annualized holding costs are 28.08\%, 22.81\%, and 25.76\% of procurement value. Initial stock equals 20 days of base-tier throughput; this is a capacity ratio, not a forecast of actual sales. Procurement takes 5--10 days, while capacity expansion spans three to seven weekly windows. Median innovation delays consume 9--30\% of the 500-day horizon. These relations create a trade-off between current spending and future operating capacity.

Persistence operates at several scales. The retention terms for goodwill, satisfaction, and the mean-reverting macro process imply half-lives of approximately 27.4, 17.0, and 34.3 days. A fixed knowledge stock retains about 79.85\% after 500 daily depreciation steps without new contributions, but receives no direct terminal liquidation value. Initial cash covers 1,250 days of base-tier overhead alone. Thus fixed overhead does not force exit within the horizon; discretionary spending and procurement can still exhaust liquidity.

The 690 daily baseline opportunities are measured before the outside option and resource constraints. They are not guaranteed purchases. This pool was retained when the formal market used eight rather than nine firms; it was not scaled by active-firm count. The nominal pool per firm therefore rises from 76.67 in the nine-firm search to 86.25 in the eight-firm evaluation, a 12.5\% increase before consumer choice. Fewer firms can ease rivalry for this pool, but the outside option, stockouts, and different opponent policies prevent this ratio from predicting realized sales or Rule-relative Score. Consequently, the separate nine-firm Rule search does not validate the eight-firm competition intensity. Segment weights, outside utilities, seasonal amplitudes, and market-phase transitions define contrasting demand regimes. The available evidence does not identify uniquely correct values for them.

Model-free replay verified the 27 main-evaluation markets (13,500 daily transitions; 1,944 world decision windows) and the 26 repeated/baseline markets (13,000 transitions; 1,872 windows). Together these checks cover 53 final market states, 424 company scores, 26,500 daily transitions, and 3,816 world decision windows (Table~\ref{tab:retained-evidence}; Appendices~\ref{app:completion-audit} and~\ref{app:additional-audit}). These checks establish agreement between recorded decisions, transitions, and scores. Earlier response diagnostics used a different scenario and are not evidence of parameter validity for these evaluation runs. The Rule search selects a baseline policy, not the market coefficients. No parameter sweep in these agent evaluations establishes ranking robustness to demand, liquidity, delays, or noise. Appendix~\ref{app:trajectory-evidence} separately assesses terminal-valuation sensitivity by rescoring the recorded outcomes without inventory salvage. This holds decisions and trajectories fixed; robustness under a changed reward rule or horizon would require new agent runs.

%% file: appendix/trajectory_evidence.tex
\section{Behavior and Trajectory Evidence}
\label{app:trajectory-evidence}
Cases T1--T3 use committed decisions, environment receipts, and daily accounts from the formal campaign; their amounts are in dollars. Selected after observing outcomes, they illustrate operating mechanisms rather than estimate an action's causal contribution to ranking.

\paragraph{T1: Terminal inventory drawdown.}
Astra and Sol finish with mean cash of \$631,851.79 and \$613,078.14, but mean inventory salvage of only \$1,090.25 and \$1,363.75. Their positive Mean Scores therefore persist if terminal salvage is omitted from Equation~\ref{eq:score}. In Astra's lower-median outcome (seed 29, company 1), the day-490 decision sets all reorder targets to zero, effective day 491. Over days 491--500, procurement is zero, revenue is \$39,775.00, and cash costs are \$5,839.83. Cash rises from \$549,475.53 to \$583,410.70. The remaining 53 units have \$817.00 salvage value, with no paid inbound stock; final Score is \$68,227.70. Astra also has positive Score in all 11 appearances in markets with no bankrupt firm.

\paragraph{Terminal-valuation sensitivity.}
For each of the 216 main-evaluation outcomes, we remove terminal inventory salvage while retaining the recorded decisions and cash flows: $S_i^{\mathrm{cash}}=S_i-\mathrm{salvage}_{i,T}=C_{i,T}-V_{i,0}$. Table~\ref{tab:terminal-valuation} shows that all nine mean-Score ranks and signs are unchanged. Qwen's cash-only mean remains positive but small (\$5.41k).

\begin{table}[t]
\centering\small
\caption{\textbf{Terminal-valuation sensitivity.} Means over 24 main-evaluation outcomes per CEO (USD thousands). Salvage includes on-hand and paid inbound inventory; ranks are identical under both measures.}
\label{tab:terminal-valuation}
\begin{tabular}{@{}rlrrr@{}}
\toprule
Rank & CEO agent & Official Score & Salvage & Cash-only Score\\
\midrule
1 & GPT-6 Astra & $116.94$ & $1.09$ & $115.85$\\
2 & GPT-5.6 Sol & $98.44$ & $1.36$ & $97.08$\\
3 & Qwen 3.8 Max & $8.04$ & $2.63$ & $5.41$\\
4 & Rule CEO & $-47.30$ & $4.00$ & $-51.30$\\
5 & DeepSeek V4 Flash & $-114.50$ & $9.70$ & $-124.21$\\
6 & Gemini 3.8 Flash & $-119.79$ & $5.03$ & $-124.81$\\
7 & DeepSeek V4 Pro & $-159.33$ & $5.42$ & $-164.75$\\
8 & Claude Sonnet 5 & $-314.36$ & $15.86$ & $-330.22$\\
9 & Doubao Seed 1.8 & $-434.88$ & $12.02$ & $-446.90$\\
\bottomrule
\end{tabular}
\end{table}

Cash-only rescoring retains all eight market and all 56 directed three-seed mean signs (Table~\ref{tab:cash-only-contrasts}). Astra's market difference remains positive in all three seeds; Sol's private difference remains positive in all three (mean \$134.86k), while its market difference remains negative on average (mean $-\$48.99$k; two negative seeds). This post-hoc check does not test agent behavior under a different reward or horizon; cash from late inventory sales remains counted.
\begin{table}[t]
\centering\small
\setlength{\tabcolsep}{3pt}
\caption{\textbf{Matched differences without terminal salvage.} Three-seed market means (USD thousands), seed signs in order 11/29/47, and retained directed signs under cash-only rescoring. Each entrant has seven directed means and 21 seed-level contrasts.}
\label{tab:cash-only-contrasts}
\begin{tabular}{@{}lrrccc@{}}
\toprule
& \multicolumn{2}{c}{Market difference} & \shortstack{Market\\seed signs} & \multicolumn{2}{c}{Directed sign matches}\\
Entrant & Official & Cash-only & & Mean & Seed-level\\
\midrule
GPT-6 Astra & $+466.68$ & $+475.40$ & $(+,+,+)$ & 7/7 & 20/21\\
GPT-5.6 Sol & $-89.66$ & $-48.99$ & $(-,+,-)$ & 7/7 & 20/21\\
Qwen 3.8 Max & $+156.92$ & $+183.76$ & $(+,+,-)$ & 7/7 & 21/21\\
DeepSeek V4 Flash & $-84.20$ & $-79.99$ & $(+,+,-)$ & 7/7 & 20/21\\
Gemini 3.8 Flash & $+55.78$ & $+62.37$ & $(+,+,-)$ & 7/7 & 20/21\\
DeepSeek V4 Pro & $-573.25$ & $-528.21$ & $(-,-,-)$ & 7/7 & 21/21\\
Claude Sonnet 5 & $-458.05$ & $-457.64$ & $(-,-,-)$ & 7/7 & 21/21\\
Doubao Seed 1.8 & $-639.88$ & $-642.88$ & $(-,-,-)$ & 7/7 & 21/21\\
\bottomrule
\end{tabular}
\end{table}

\paragraph{T2: Procurement exhausts liquidity.}
Doubao (seed 11, company 8) sets reorder targets for A and C to 10,000 units on day 105. Applied receipts precede \$469,372.00 of procurement on day 106, when cash falls from \$467,094.87 to \$57.19. Further procurement and operating costs reduce cash to $-\$0.93$ on day 110, before the next planning boundary at day 112. All 16 decisions were sealed. The firm exits despite holding inventory with \$151,610.00 terminal salvage value.

\paragraph{T3: Recurring budgets exhaust cash.}
DeepSeek Pro (seed 29, company 2) sets total daily development to \$75,000.00 and total daily marketing, including unchanged allocations, to \$63,450.00 on day 35. Both changes take effect on day 36 and are charged on days 36--38. Cash falls from \$339,084.34 to \$203,460.46, \$67,030.87, and $-\$68,317.92$, respectively. Bankruptcy occurs on day 38, before the next decision boundary at day 42. All six decisions were sealed.

\paragraph{Traceability.}
\path{anc/data/formal/trajectory_summary.json} maps T1--T3 to run IDs, checkpoint and trace paths, SHA-256 hashes, decision windows, receipts, and accounts. The run/agent-trace hash prefixes are \texttt{ef03e110/5f49ac152640}, \texttt{665f7924/4e4b8ccde867}, and \texttt{5ac797e6/4678d3095161}. Cash paths are accounting records, not interim official Scores. These cases alone cannot separate the effects of rival exit, market changes, and other firms' simultaneous adaptations.

\begin{figure}[t]
\centering
\includegraphics[width=\linewidth]{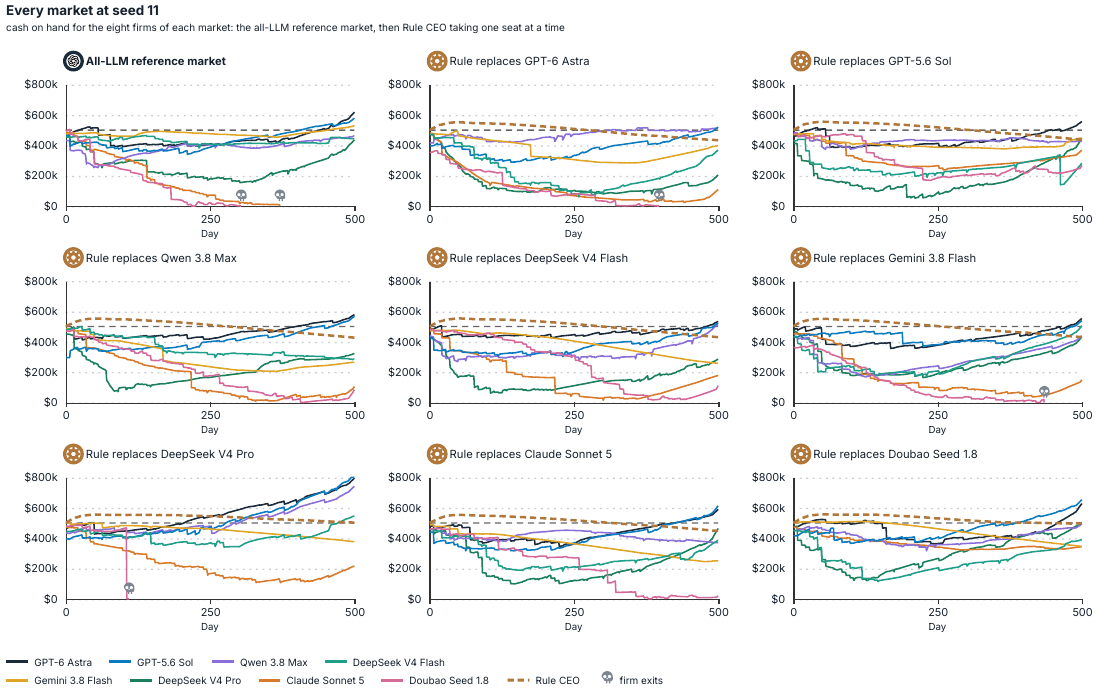}
\caption{\textbf{Cash on hand in all nine main-evaluation markets at seed 11.} The upper-left panel is the eight-agent reference market; the other panels replace the named agent with Rule CEO. Colors identify agents, the dashed brown line identifies Rule, and skull markers indicate firm exit days.}
\label{fig:markets-seed11}
\end{figure}

\begin{figure}[t]
\centering
\includegraphics[width=\linewidth]{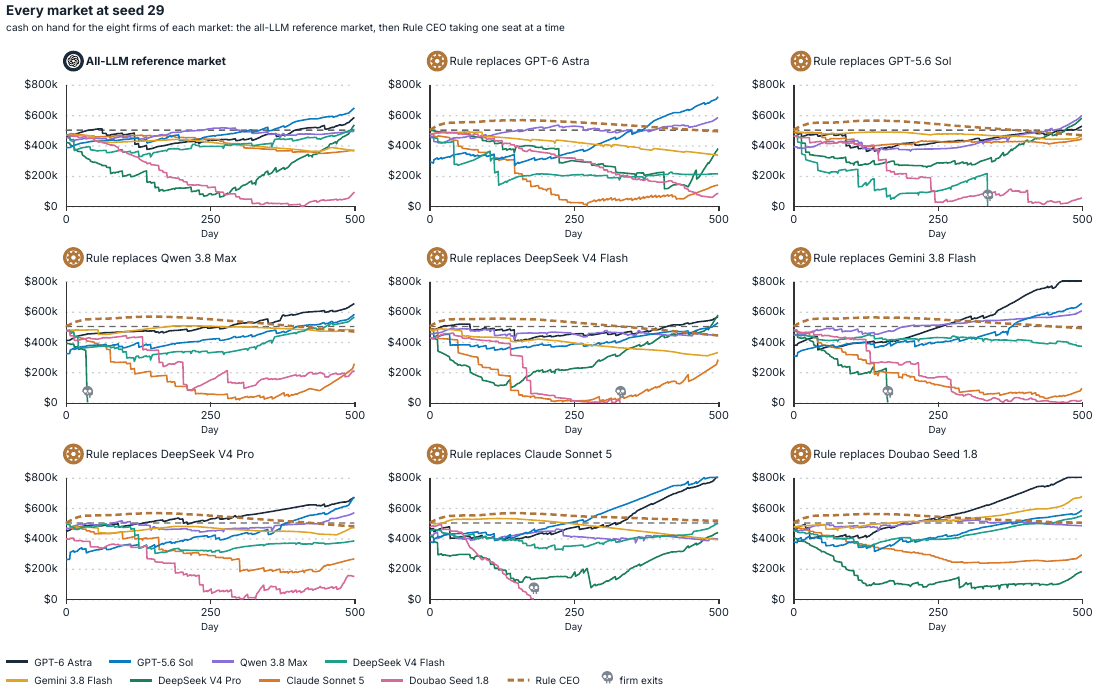}
\caption{\textbf{Cash on hand in all nine main-evaluation markets at seed 29.} Panel order and visual encoding match Figure~\ref{fig:markets-seed11}: the reference market precedes eight same-seat Rule replacements, and skull markers indicate firm exit days.}
\label{fig:markets-seed29}
\end{figure}

\begin{figure}[t]
\centering
\includegraphics[width=\linewidth]{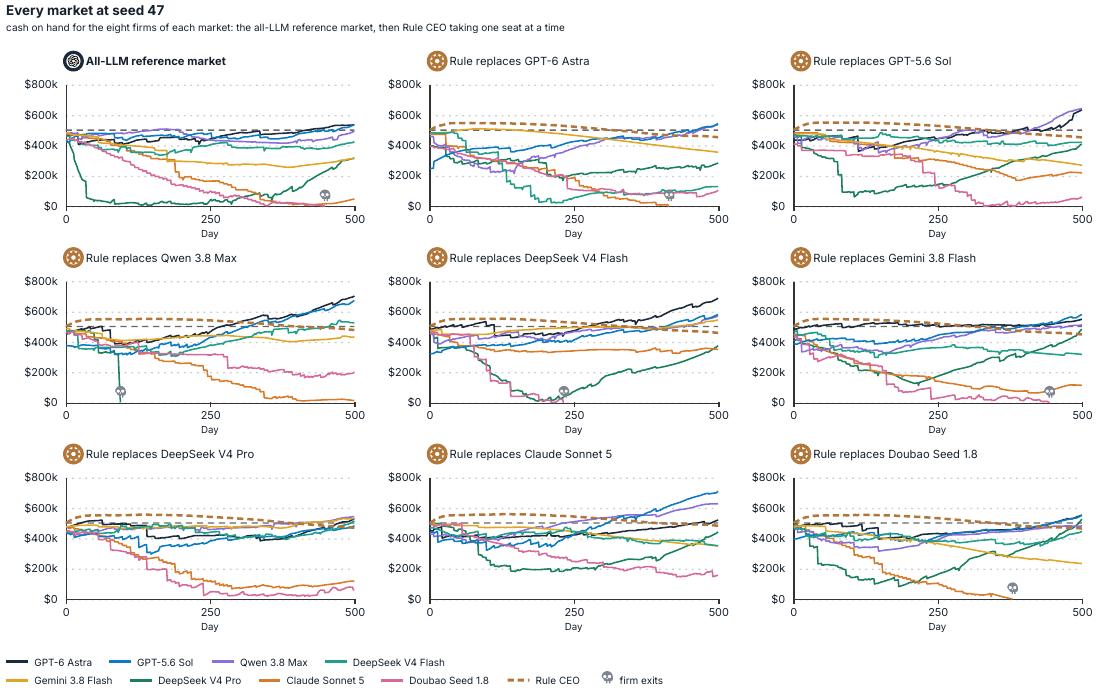}
\caption{\textbf{Cash on hand in all nine main-evaluation markets at seed 47.} Panel order and visual encoding match Figure~\ref{fig:markets-seed11}: the reference market precedes eight same-seat Rule replacements, and skull markers indicate firm exit days.}
\label{fig:markets-seed47}
\end{figure}

\paragraph{Cash paths by market.}
Figures~\ref{fig:markets-seed11}--\ref{fig:markets-seed47} show the 27 main-evaluation markets by seed and lineup. Each panel contains eight firms: the reference market appears first, followed by eight same-seat Rule replacements. Figure~\ref{fig:cash-overview} groups the same trajectories by agent. These are cash balances, not interim official Scores; the repeated and No Action runs are excluded.

\begin{figure}[t]
\centering
\includegraphics[width=\linewidth]{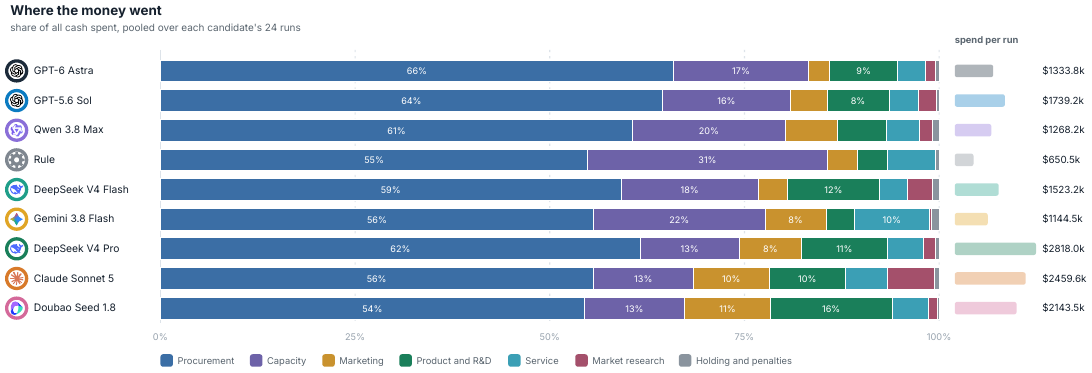}
\caption{\textbf{Cash outlays in the main evaluation.} Bars pool each CEO agent's 24 outcomes and show the share spent by category; right-hand values are total spending divided by 24 (USD thousands). These are gross cash outflows, not Score contributions.}
\label{fig:spend-mix}
\end{figure}

\paragraph{Cash outlays.}
Figure~\ref{fig:spend-mix} summarizes gross cash spending across each candidate's 24 main-evaluation outcomes. Procurement converts cash into inventory; its contribution to Score depends on subsequent sales and terminal salvage.

\begin{figure}[t]
\centering
\includegraphics[width=\linewidth]{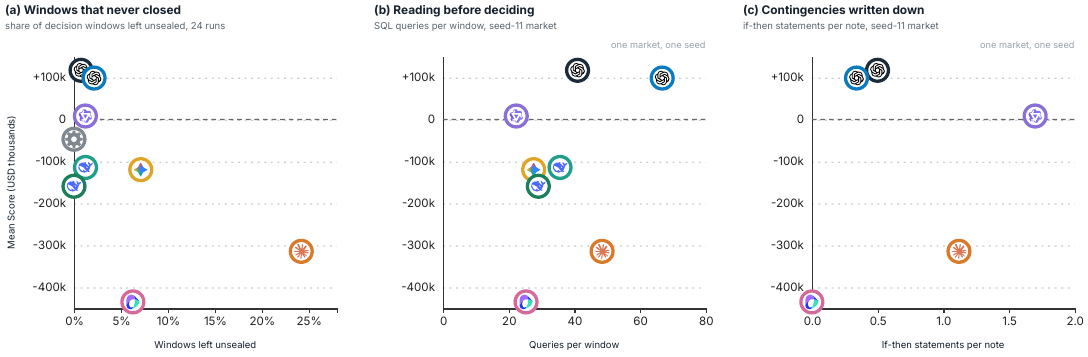}
\caption{\textbf{Recorded behavior and operating returns.} Vertical positions are main-evaluation Mean Scores over 24 outcomes. In (a), horizontal values are the percentage of active decision windows left unsealed across those outcomes. In (b), queries per window come from the seed-11 reference market; (c) counts if--then statements per note for agents with notes in that market. These comparisons do not identify causal effects.}
\label{fig:behavior-score}
\end{figure}

\paragraph{Behavior and Score.}
Figure~\ref{fig:behavior-score} juxtaposes recorded behavior with Mean Score. The panels are descriptive: inquiry and note measures come from one seed-11 reference market, while Score averages 24 outcomes.

\begin{figure}[t]
\centering
\includegraphics[width=\linewidth]{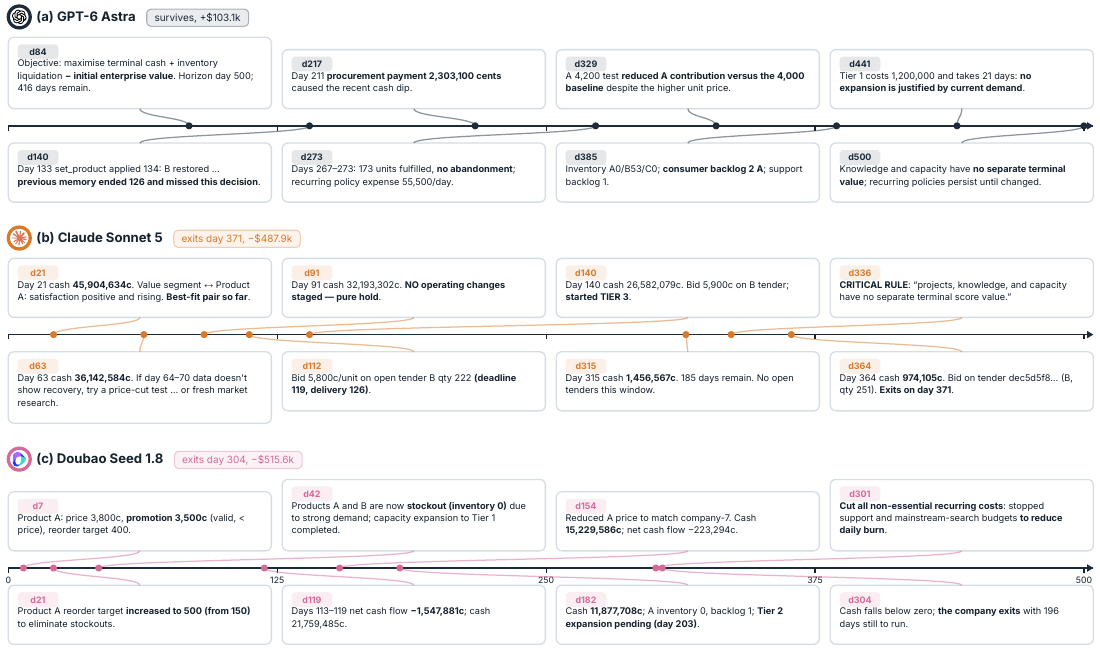}
\caption{\textbf{Selected memory and operating events.} Astra, Sonnet, and Doubao in the seed-11 reference lineup; badges report terminal Score in USD thousands, while cash values marked \texttt{c} in excerpts are cents. This Doubao firm exits on day 304; case T2 above describes a different seed-11 replacement lineup whose Doubao firm exits on day 110.}
\label{fig:memo-timeline}
\end{figure}

\begin{figure}[t]
\centering
\includegraphics[width=\linewidth]{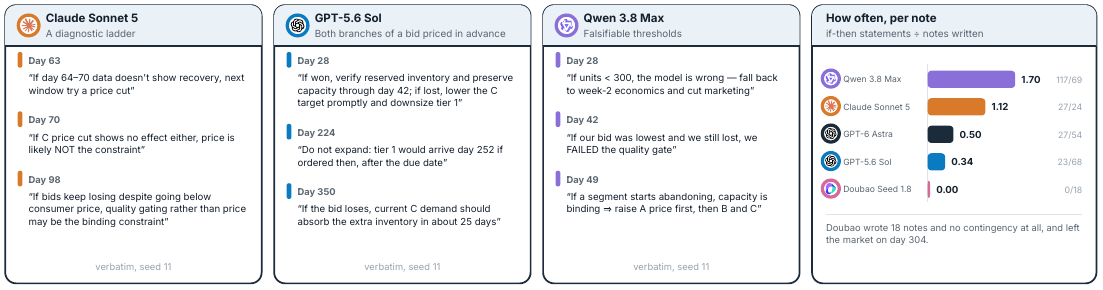}
\caption{\textbf{Conditional plans in private notes.} Selected seed-11 reference-market excerpts and if--then statements per written note for agents with notes. These counts describe text, not plan execution or decision quality.}
\label{fig:conditional-planning}
\end{figure}

\paragraph{Reference-market memory.}
Figures~\ref{fig:memo-timeline} and~\ref{fig:conditional-planning} show selected notes and events in the seed-11 reference lineup. Written plans need not become executed decisions.

\subsection{Mechanism Hypotheses from Memory and Actions}
\label{app:mechanisms}

\paragraph{Scope and procedure.}
We analyze the 27 main-evaluation markets at seeds 11, 29, and 47. Accounting profiles cover all 24 outcomes per agent. The memory audit extracts 2,477 recorded decision-window traces from 36 company trajectories in 12 markets: the three reference markets and the matched Rule replacements of Astra, Sol, and DeepSeek Pro. It covers the three leading agents' reference trajectories and both affected-agent histories for each of the four pairs in Figure~\ref{fig:value}D. Qwen loses in all three reference markets, so we also inspect the lower-median positive Qwen outcome within each seed: Rule replaces Pro at seeds 11 and 47, and Astra at seed 29. These selected Scores are \$225.78k, \$66.69k, and \$32.81k, respectively; seed 11 has only one positive Qwen outcome.

We recover auto-loaded notes and successful file writes, verify each source trace against its archived SHA-256, and link claim-relevant excerpts to sealed decisions and settlement receipts. \texttt{MEMORY.md} is persistent, agent-written material; its contents are beliefs and plans, not verified descriptions of the simulator. A written plan is counted as executed only when the decision and receipt agree. The four pairs were selected by the preceding stability analysis, not by their fit to a mechanism. The following comparisons generate hypotheses; other agents also adapt, and independent trajectories can diverge before any observed rival response.

\paragraph{Positive mean returns.}
Table~\ref{tab:mechanism-returns} separates revenue from net cash outlays. The accounts reconcile exactly for all 216 outcomes: final cash equals initial cash plus revenue minus outlays net of refunds. Score then adds terminal salvage and subtracts initial enterprise value; equivalently, in dollars, $S=R-P+L-16{,}000$, where $P$ is net cash outlay and $L$ is terminal salvage. Thus sales volume alone does not explain the ranking.

\begin{table}[t]
\centering\small
\setlength{\tabcolsep}{3pt}
\caption{\textbf{Accounting profiles of the three agents with positive Mean Score.} Money is in USD thousands; all entries except $n$ are means. Net outlays subtract recorded refunds. The last two rows condition on Qwen's realized Score and are descriptive, not separate experimental groups.}
\label{tab:mechanism-returns}
\begin{tabular}{@{}lrrrrrr@{}}
\toprule
Agent/sample & $n$ & Score & Revenue & Net outlays & Fulfilled units & Marketing\\
\midrule
Astra & 24 & 116.94 & 1,460.95 & 1,329.10 & 23,252 & 35.91\\
Sol & 24 & 98.44 & 1,839.24 & 1,726.16 & 29,766 & 81.67\\
Qwen & 24 & 8.04 & 1,279.92 & 1,258.52 & 20,954 & 85.22\\
\midrule
Qwen: $S>0$ & 11 & 79.25 & 1,501.79 & 1,409.47 & 24,246 & 84.46\\
Qwen: $S<0$ & 13 & $-52.21$ & 1,092.19 & 1,130.79 & 18,170 & 85.86\\
\bottomrule
\end{tabular}
\end{table}

Astra's notes repeatedly compare inventory positions and capacity with realized demand and the remaining horizon. The corresponding actions reduce replenishment targets, cancel unused capacity, and reverse an expansion (Table~\ref{tab:memory-actions}); its mean marketing and capacity-investment outlays are \$35.91k and \$21.00k. Sol supports higher volume with \$81.67k and \$49.25k in these categories, but also reverses capacity commitments when utilization weakens. Their positive means are consistent with different balances between sales and persistent commitments, rather than sales maximization alone. Astra is positive in 24/24 outcomes and Sol in 23/24; neither goes bankrupt.

Qwen's notes specify thresholds for reverting price or budget changes, and several reversions are executed. Yet this practice is also present in its losing reference trajectories. Its 11 positive outcomes have much higher mean fulfillment than its 13 negative outcomes despite similar marketing expenditure. Conditional planning therefore provides a plausible means of adapting spending and margins, but does not by itself explain the sign of returns. The small pooled surplus reflects uneven realized demand and profitability, not a uniformly successful policy.

\begin{table}[t]
\centering\small
\caption{\textbf{Memory statements linked to executed decisions.} Entries identify agent, seed, and decision day. All use the reference market except Qwen: Rule replaces Pro at seeds 11/47 and Astra at seed 29. Excerpts omit Markdown emphasis and use cents; action amounts are converted to dollars. Receipts confirm execution, not the action's causal payoff.}
\label{tab:memory-actions}
\begin{tabularx}{\linewidth}{@{}p{0.15\linewidth}p{0.34\linewidth}Y@{}}
\toprule
Agent; seed/day & Note excerpt & Sealed action and settlement\\
\midrule
Astra; 11/7 & \emph{Reductions pause future procurement} & A/B/C reorder targets become 350/160/120 on day 8.\\
Astra; 29/147 & \emph{Tier1 is unnecessary for observed consumer volume} & Downsize from tier 1 to 0, completed on day 154; C target falls from 400 to 80.\\
Astra; 47/7 & \emph{Base capacity has 25 units/day spare} & Cancel pending tier-1 expansion; the receipt records a \$6,600 refund.\\
Sol; 11/126 & \emph{a lead-time hedge for tier-3 innovation} & Expand from tier 1 to 2, due day 161; later downsize on days 182 and 224.\\
Sol; 29/231 & \emph{Don't keep unused capacity just for terminal value} & Downsize from tier 1 to 0, completed on day 238; A target falls from 583 to 300.\\
Sol; 47/7 & \emph{Keep all nine segment/channel marketing budgets at 5k/day} & Retain total marketing of \$450/day while cancelling pending tier-2 capacity.\\
Qwen; 11/224 & \emph{SUPPORT=0 ALSO REVERSED} & Restore marketing and support to \$40/day each, effective day 225.\\
Qwen; 29/56 & \emph{slash development} & Reduce total development from \$95 to \$15/day, effective day 57; restore service budgets.\\
Qwen; 47/350 & \emph{A PRICE TEST RESOLVED} & Reverse A's price from \$40 to \$47, effective day 351, after its stated volume threshold is missed.\\
\bottomrule
\end{tabularx}
\end{table}

\paragraph{Why aggregate externalities differ.}
Table~\ref{tab:mechanism-market} decomposes the matched changes for Astra and Sol. Both attract more consumer orders than the Rule company they replace; Sol's increment is larger in every seed, alongside a larger increase in marketing outlays (\$16.76k/\$34.64k/\$83.15k versus Astra's \$9.52k/\$10.34k/\$19.20k). This is consistent with stronger demand competition, but aggregate outcomes also depend on rivals' spending and survival.

Averaged across seeds, Astra's rivals lose \$436.89k in revenue but reduce net outlays by \$792.84k, yielding a positive Score difference after salvage. Sol's rivals lose \$1,066.09k in revenue and save only \$882.24k in net outlays. The mean contrast therefore concerns both revenue pressure and spending responses. It is not a claim that Astra raises every rival's sales or that one expenditure channel explains every seed: at seed 47, rival revenue increases under both entrants, but outlays rise much more under Sol.

\begin{table}[t]
\centering\small
\setlength{\tabcolsep}{4pt}
\caption{\textbf{Accounting decomposition of aggregate externalities.} Each row compares the reference market with the same-seed Rule replacement of the entrant. Rival columns sum the other seven firms; monetary differences are in USD thousands. Net outlays subtract refunds.}
\label{tab:mechanism-market}
\begin{tabular}{@{}llrrrrr@{}}
\toprule
& & Own consumer & \multicolumn{4}{c}{Other seven firms}\\
Entrant & Seed & order difference & $\Delta$Revenue & $\Delta$Net outlays & $\Delta$Salvage & $\Delta$Score\\
\midrule
Astra & 11 & $+6,172$ & $-1,982.84$ & $-2,324.11$ & $+24.60$ & $+365.87$\\
Astra & 29 & $+6,221$ & $-199.92$ & $-749.27$ & $-30.79$ & $+518.56$\\
Astra & 47 & $+7,179$ & $+872.10$ & $+694.86$ & $-10.68$ & $+166.55$\\
Sol & 11 & $+16,399$ & $-3,009.41$ & $-2,707.13$ & $-30.28$ & $-332.57$\\
Sol & 29 & $+15,539$ & $-483.51$ & $-759.61$ & $-79.35$ & $+196.75$\\
Sol & 47 & $+13,040$ & $+294.66$ & $+820.03$ & $-2.47$ & $-527.84$\\
\bottomrule
\end{tabular}
\end{table}

\paragraph{Four directed relationships.}
Table~\ref{tab:mechanism-directed} reports all three main-seed decompositions for the four selected pairs. For a rival in a matched comparison, $\Delta S=\Delta R-\Delta P+\Delta L$; initial endowments cancel. We examine both accounting channels and explicit strategy revisions.

\begin{table}[t]
\centering\small
\setlength{\tabcolsep}{4pt}
\caption{\textbf{Main-evaluation accounting for the four directed pairs.} Differences are reference minus matched Rule-replacement outcomes for the affected agent, in USD thousands. These three-seed observations provide the mechanism evidence; the five-observation stability summary remains in Figure~\ref{fig:value}D.}
\label{tab:mechanism-directed}
\begin{tabular}{@{}llrrrr@{}}
\toprule
Entrant $\rightarrow$ rival & Seed & $\Delta$Revenue & $\Delta$Net outlays & $\Delta$Salvage & $\Delta$Score\\
\midrule
Pro $\rightarrow$ Doubao & 11 & $+549.63$ & $+550.01$ & $-150.81$ & $-151.19$\\
& 29 & $-984.90$ & $-929.82$ & $+0.02$ & $-55.05$\\
& 47 & $-999.76$ & $-940.06$ & $-4.92$ & $-64.62$\\
Sol $\rightarrow$ Sonnet & 11 & $-169.19$ & $+198.26$ & $+27.30$ & $-340.16$\\
& 29 & $-324.02$ & $-245.91$ & $+5.09$ & $-73.02$\\
& 47 & $-238.33$ & $-66.83$ & $-2.31$ & $-173.80$\\
Astra $\rightarrow$ Flash & 11 & $-1,297.37$ & $-1,372.24$ & $+0.14$ & $+75.01$\\
& 29 & $+271.72$ & $-20.88$ & $+0.75$ & $+293.35$\\
& 47 & $-317.84$ & $-614.25$ & $-6.12$ & $+290.29$\\
Astra $\rightarrow$ Pro & 11 & $-422.11$ & $-652.91$ & $+2.54$ & $+233.33$\\
& 29 & $+1,301.46$ & $+1,145.73$ & $-6.21$ & $+149.53$\\
& 47 & $+1,636.91$ & $+1,604.48$ & $-2.83$ & $+29.61$\\
\bottomrule
\end{tabular}
\end{table}

\emph{Pro $\rightarrow$ Doubao: price-following and margin pressure.}
Doubao explicitly responds to company 2 (DeepSeek Pro) in all three reference trajectories. At seed 11/day 84, its note says \emph{matching company-2 to boost demand}; it raises C's per-unit promotion from \$5 to \$10, matching Pro's \$120 effective price. At seed 29/day 98, it cuts C from \$130 to Pro's \$125. At seed 47/day 28, it matches Pro's B/C list prices of \$72/\$130 but also adds \$5/\$10 promotions, producing lower effective prices of \$67/\$120. Sealed decisions and receipts confirm these changes. Matching prices, sometimes with additional discounts, is a plausible source of margin pressure. Doubao's revenue and gross profit are lower in every seed when comparisons stop before either matched firm exits. However, the seed-11 terminal Score difference is mostly a \$150.81k salvage difference: the Rule-replacement Doubao exits on day 110 with substantial stock, versus day 304 in the reference market. Price-following does not explain this entire terminal contrast.

\emph{Sol $\rightarrow$ Sonnet: costly investment alongside weaker returns.}
Sonnet spends more on development plus innovation with Sol in each seed. Its notes favor growth through quality and capacity: seed 11/day 42 projects demand above 90 units/day and seals tier-2 expansion; seed 47/day 70 starts another innovation project while gross profit remains below recurring costs. This supports an investment-response hypothesis. Lower full-horizon sales alone are not a sufficient explanation: at seed 11, Sonnet exits on day 371 only in the reference market. Through day 370, it earns \$307.10k more revenue, but its additional non-procurement outlays exceed its extra gross profit. At seeds 29 and 47, it instead loses revenue. The negative contrasts therefore combine spending and demand channels.

\emph{Astra $\rightarrow$ Flash: lower investment burden.}
Flash's innovation spending is \$285k, \$150k, and \$320k lower with Astra at seeds 11, 29, and 47; total non-procurement outlays net of refunds are also lower in all three. Its seed-29/day-28 note rejects excess capacity at demand of 35 units/day, and a sealed downsize reaches tier 0 on day 35. Spending restraint is consistent with the positive Score contrasts even though revenue declines in two seeds. The evidence does not establish that Astra alone induced that restraint.

\emph{Astra $\rightarrow$ Pro: different routes to a positive contrast.}
At seed 11, Pro's lower outlays more than offset its lower revenue; its day-21 note rejects tier-2 capacity without demand growth, and that window seals no expansion. At seeds 29 and 47, Pro instead fulfills 25,965 and 20,913 more units, respectively, with revenue gains exceeding net-outlay increases. These observations suggest spending adjustment in one seed and sales gains exceeding added outlays in the other two. They support a positive matched contrast through several operating paths, not a single identified bilateral mechanism.

%% file: appendix/reproducibility.tex
\section{Execution and Reproducibility}
\label{app:repro}
\subsection{Settlement and Durable State}
The coordinator collects decisions against one immutable world version, journals accepted responses, and commits a complete barrier before advancing the economy. Standing policies execute daily inside each window. Runs stop at day 500; a surviving company participates in 72 windows.

Standalone runs atomically publish JSON snapshots and journal responses in SQLite WAL with \texttt{synchronous=FULL}; campaign runs use transactional SQLite storage. Recovery checks the run, configuration, and window identities before reusing receipts. Locks and revision-checked leases prevent concurrent ownership. Ordinary continuation retains the conversation, draft, private files, and consumed reply/token/time allowances. Automatic retry waits consume active time; queueing and manual pauses do not. A request without a durably received response may still incur provider work or billing.

\subsection{Retries and Fresh Attempts}
In the main evaluation, transient provider failures retry within the current attempt's allowance. Ordinary HTTP 400 failures permit three consecutive attempts for one logical request, including the first. The first two failures wait one and two seconds by default, respecting a longer \texttt{Retry-After}. A valid accounting-bearing reply resets the consecutive count. Counters persist across pause/resume; the third consecutive failure stops the affected run. Authentication and identity errors require resolution.

A separate supervisor may restart a failed, unsealed company-window up to three times beyond its original attempt. It archives the failed conversation, draft, allowance, trace, and timing records; restores private files to their window-start contents from verified receipts; and grants the original allowance. Earlier economic windows and accepted peer decisions remain intact. The count is shared across fault categories and persists through restarts. Budget exhaustion, running or backing-off attempts, and sealed or settled decisions are ineligible. Missing restoration evidence prevents a fresh attempt.

The formal campaign records three such operations, each at fresh-attempt index 1: DeepSeek V4 Flash at seed 47, days 315 and 441, and Qwen at seed 47, day 392. Their failed-attempt usage is retained separately. Neither settled usage nor the sum of known settled and discarded usage is a complete provider bill. Table~\ref{tab:decision-outcomes} distinguishes decision outcomes from request-level failures and states their economic treatment.

\begin{table}[t]
\centering
\caption{Decision outcomes and their economic treatment.}
\label{tab:decision-outcomes}
\begin{tabularx}{\linewidth}{@{}p{2.2cm}Y Y@{}}
\toprule Status & Economic treatment & Reporting consequence\\\midrule
Sealed & Execute the validated draft at the barrier; an unchanged-policy seal is valid. & Count a successful decision and its reported usage.\\
Budget exhausted & Discard unsealed changes; standing policies continue and the barrier may settle. & Record \codename{budget\_exhausted}, separately from a deliberate unchanged-policy seal.\\
Timeout & Exhausted infrastructure recovery applies no unsealed changes; standing policies continue. & Record \codename{timeout}, failure metadata and the infrastructure-timeout count.\\
Protocol error & Correct recoverable replies in-session; a final invalid Harness result applies no draft. & Retain economic consequences and correction diagnostics separately.\\
Provider error & Retry eligible failures; terminal failures stop the affected run at its durable frontier. & Apply the bounded supervision rule where eligible; retain all attempt costs.\\
Pause / integrity error & Preserve receipts; do not settle an incomplete or inconsistent barrier. & Resolve before accepting a completed result.\\
\bottomrule
\end{tabularx}
\end{table}

\subsection{Concurrency and Provenance}
In the main evaluation, the seed-priority scheduler admits at most 30 active runs. Earlier eligible seeds have priority; later seeds may fill available slots. Active CEOs within a run decide concurrently, with no additional global decision or resident-worker cap. Runs share compute but not economic state. Latency can affect how much work fits within a decision allowance, while settlement waits for the common barrier.

The main-evaluation manifest binds the scenario, seeds, seats, lineups, Rule profile, prompt, tool schemas, allowances, memory policy, provider settings, and source fingerprints. The main-evaluation release is \texttt{http400-three-attempts-20260913}: package \texttt{0.15.2}, economy \texttt{ceos-market-v3}, decision \texttt{ceos-decision-v9}, runtime \texttt{ceos-runtime-v4}, and campaign \texttt{ceos-campaign-plan-v6}. Per-run records retain execution provenance. The main-evaluation runtime uses Python 3.13.15, Bun 1.3.14, NumPy 2.5.2, Pydantic 2.13.4, and HTTPX 0.28.1; OpenCode's revision begins \texttt{1be9fd55a932}. No Action's separate implementation is specified in Appendix~\ref{app:noop}; it does not change these formal records.

\subsection{Completion and Decision Audit}
\label{app:completion-audit}
The 15 September 2026 audit verified 27 completed runs, eight companies per run, and 24 terminal outcomes per candidate. Model-free Exact Replay reconstructed 13,500 daily transitions and 1,944 windows and reproduced final states and scores. Checkpoint/trace hashes and exported states matched; SQLite integrity was \texttt{ok}, and the published analysis was recomputed. This checks execution conditional on saved decisions, not the reproducibility of new LLM generations.

The 15,078 settled company-window records comprise 14,362 seals, 683 budget exhaustions, 32 timeouts, and one protocol error. Rule contributes 1,728 seals; Decision health for LLM-based agents uses 13,350 active company-windows. The 108 failure-history records are execution events, not distinct failed decisions. Exited firms submit no further decisions but remain in terminal-score denominators.

Table~\ref{tab:decision-health} reports final decision outcomes separately for the main evaluation, repeated runs, and No Action replacements. Across all 53 runs, 29,748 active company-windows comprise 28,477 seals, 1,215 budget exhaustions, 54 timeouts, and two protocol errors.

The unsealed-decision rate divides non-sealed company-windows by all active company-windows. It includes budget exhaustion, timeout, and final protocol-error outcomes. The source field is \codename{failure\_rate}; it is not an API failure rate.

\begin{table}[t]
\centering\small\setlength{\tabcolsep}{4pt}
\caption{Final decision status by CEO agent and experiment group. Active company-windows count decisions while the firm remains active; the status columns sum to that count. All settled decision records are included.}
\label{tab:decision-health}
\begin{tabularx}{\linewidth}{@{}Yrrrrr@{}}
\toprule CEO agent & \shortstack[r]{Active company-\\windows} & Sealed & \shortstack[r]{Budget\\exhausted} & Timeout & \shortstack[r]{Protocol\\error}\\\midrule
\multicolumn{6}{l}{\textbf{(a) Main evaluation: 27 runs}}\\
\texttt{gpt-6-astra} & 1,728 & 1,715 & 9 & 4 & 0\\
\texttt{deepseek-v4-pro} & 1,556 & 1,556 & 0 & 0 & 0\\
\texttt{claude-sonnet-5} & 1,680 & 1,273 & 397 & 10 & 0\\
\texttt{qwen3.8-max} & 1,728 & 1,707 & 21 & 0 & 0\\
\texttt{gpt-5.6-sol} & 1,728 & 1,691 & 27 & 10 & 0\\
\texttt{gemini-3.8-flash} & 1,728 & 1,605 & 123 & 0 & 0\\
\texttt{deepseek-v4-flash} & 1,705 & 1,684 & 12 & 8 & 1\\
\texttt{doubao-seed-1.8} & 1,497 & 1,403 & 94 & 0 & 0\\
Rule CEO & 1,728 & 1,728 & 0 & 0 & 0\\
\textbf{Group total} & \textbf{15,078} & \textbf{14,362} & \textbf{683} & \textbf{32} & \textbf{1}\\\midrule
\multicolumn{6}{l}{\textbf{(b) Repeated runs: 18 runs}}\\
\texttt{gpt-6-astra} & 1,152 & 1,138 & 8 & 6 & 0\\
\texttt{deepseek-v4-pro} & 1,117 & 1,117 & 0 & 0 & 0\\
\texttt{claude-sonnet-5} & 1,152 & 1,043 & 107 & 2 & 0\\
\texttt{qwen3.8-max} & 1,152 & 1,144 & 8 & 0 & 0\\
\texttt{gpt-5.6-sol} & 1,152 & 1,085 & 57 & 10 & 0\\
\texttt{gemini-3.8-flash} & 1,152 & 1,052 & 100 & 0 & 0\\
\texttt{deepseek-v4-flash} & 1,152 & 1,144 & 7 & 0 & 1\\
\texttt{doubao-seed-1.8} & 952 & 871 & 81 & 0 & 0\\
Rule CEO & 1,152 & 1,152 & 0 & 0 & 0\\
\textbf{Group total} & \textbf{10,133} & \textbf{9,746} & \textbf{368} & \textbf{18} & \textbf{1}\\\midrule
\multicolumn{6}{l}{\textbf{(c) No Action baseline: 8 runs}}\\
\texttt{gpt-6-astra} & 504 & 501 & 0 & 3 & 0\\
\texttt{deepseek-v4-pro} & 479 & 477 & 2 & 0 & 0\\
\texttt{claude-sonnet-5} & 504 & 456 & 47 & 1 & 0\\
\texttt{qwen3.8-max} & 504 & 501 & 3 & 0 & 0\\
\texttt{gpt-5.6-sol} & 504 & 463 & 41 & 0 & 0\\
\texttt{gemini-3.8-flash} & 504 & 466 & 38 & 0 & 0\\
\texttt{deepseek-v4-flash} & 504 & 499 & 5 & 0 & 0\\
\texttt{doubao-seed-1.8} & 458 & 430 & 28 & 0 & 0\\
No Action CEO & 576 & 576 & 0 & 0 & 0\\
\textbf{Group total} & \textbf{4,537} & \textbf{4,369} & \textbf{164} & \textbf{4} & \textbf{0}\\\midrule
\textbf{All 53 runs} & \textbf{29,748} & \textbf{28,477} & \textbf{1,215} & \textbf{54} & \textbf{2}\\
\bottomrule
\end{tabularx}
\end{table}

\begin{figure}[t]
\centering
\includegraphics[width=\linewidth]{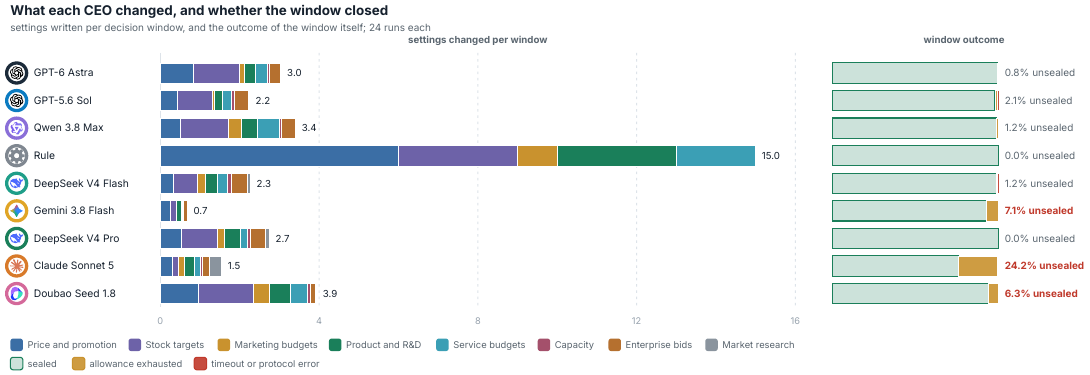}
\caption{\textbf{Recorded actions and window outcomes.} Left: mean settings written per active decision window across each CEO's 24 main-evaluation outcomes, grouped by policy type. Right: sealed versus unsealed windows; unsealed includes budget exhaustion, timeout, and protocol error. Unsealed drafts do not change standing policies.}
\label{fig:action-mix}
\end{figure}

Figure~\ref{fig:action-mix} summarizes settings written and decision closure in the 27-run main evaluation.

Table~\ref{tab:retained-evidence} lists the archived evidence for both evaluation campaigns and their combined verification totals.

\begin{table}[t]
\centering\small
\caption{Retained evidence for the 53 evaluation runs. Main-evaluation and repeated/baseline records are archived separately; full hashes accompany the research artifacts.}
\label{tab:retained-evidence}
\begin{tabularx}{\linewidth}{@{}p{3.5cm}Y@{}}
\toprule Artifact & Evidence\\\midrule
Main-evaluation manifest & Plan hash prefix \texttt{3ee34a22e881}; scenario hash prefix \texttt{893815172625}; model controls, seats and economic configuration.\\
Repeated/baseline records & Separate campaign records, final checkpoints, terminal analysis and verification receipts for the 18 repeated and eight No Action runs; see Appendix~\ref{app:additional-audit}.\\
Main-evaluation replay & 27 verified run receipts, 216 terminal company scores, 13,500 daily transitions and 1,944 world decision windows.\\
Repeated/baseline replay & 26 verified run receipts, 208 terminal company scores, 13,000 daily transitions and 1,872 world decision windows.\\
Evaluation total & 53 verified markets, 424 terminal company scores, 26,500 daily transitions and 3,816 world decision windows.\\
Decision health & Per-agent status counts for all three experiment groups in Table~\ref{tab:decision-health}; 29,748 active company-windows in total.\\
Recorded workload & Successfully sealed decisions with complete reported usage, summarized in Appendix~\ref{app:workload}; per-trajectory counts in \path{anc/data/workload}.\\
Model identification & Requested/served identifiers and explicit accepted aliases retained in the run records; no independent weight verification.\\
\bottomrule
\end{tabularx}
\end{table}

\subsection{Analysis Scope}
The main-evaluation results use all 27 completed runs. Repeated and baseline comparisons use the separate weights defined in Appendices~\ref{app:repeat-results} and~\ref{app:baseline-results}. Case traces describe recorded behavior; they do not isolate an action's causal contribution. Parameter robustness requires separate experiments with the same model and harness settings. Replaying old decisions under new parameters would test an unchanged policy sequence, not agents' adaptation to the new setting.

\subsection{Repeated Runs and Baseline Data}
\label{app:additional-audit}
The 18 repeated active-Rule runs and eight No Action replacements all completed day 500. They yield 208 terminal company scores, for 424 across all 53 evaluation markets. Their separate verification records contain 26 model-free replay receipts, 13,000 daily transitions, and 1,872 world decision windows. Analysis recomputed from terminal results matches the archived output. Table~\ref{tab:decision-health} gives the per-agent outcomes for these runs, and Table~\ref{tab:retained-evidence} combines the verification counts. Replay verifies settlement for recorded decisions; it is not a new agent rollout.

These runs used updated recovery for unsealed windows that retained cumulative usage. Cumulative output-token limits were disabled for Astra and Sol; per-response ceilings, economic parameters, and scoring rules were unchanged.

The paper's data directory separates the three-seed main evaluation (\texttt{anc/data/formal}) from the additional results (\texttt{anc/data/supplementary}). The latter contains the 26-run terminal analysis and \texttt{simple-five-observation-results.json}, including all 56 five-observation contrasts and the three-control baseline comparisons. Raw amounts are cents; displayed USD thousands divide by 100,000. Figure~\ref{fig:value}A--B and Tables~\ref{tab:performance}--\ref{tab:value} retain the main-evaluation estimates. Figure~\ref{fig:value}C--D use the separate weights defined in Appendix~\ref{app:metrics}.

\FloatBarrier
\subsection{Experimental Scale and Recorded Workload}
\label{app:workload}
Table~\ref{tab:experiment-scale} accounts for the three evaluation groups and the separate Rule search. A CEO decision window is one opportunity to revise company policies; a surviving firm participates in 72 windows. A recorded model reply may contain several tool calls; neither unit is equivalent to a CEO decision window.

\begin{table}[t]
\centering\small
\caption{Experimental scale. Simulation days count market-level transitions. Shared controls are counted once; replay does not add new experimental runs.}
\label{tab:experiment-scale}
\begin{tabularx}{\linewidth}{@{}Yrrrrr@{}}
\toprule Experiment & Markets & \shortstack[r]{Firms per\\market} & \shortstack[r]{Company\\trajectories} & \shortstack[r]{Simulation\\days} & \shortstack[r]{CEO decision\\windows}\\\midrule
Main evaluation & 27 & 8 & 216 & 13,500 & 15,078\\
Repeated runs & 18 & 8 & 144 & 9,000 & 10,133\\
No Action baseline & 8 & 8 & 64 & 4,000 & 4,537\\
\textbf{Evaluation total} & \textbf{53} & 8 & \textbf{424} & \textbf{26,500} & \textbf{29,748}\\\midrule
Rule parameter search & 75 & 9 & 675 & 37,500 & 48,600\\
\bottomrule
\end{tabularx}
\end{table}

Table~\ref{tab:recorded-workload} sums successfully sealed decision records with complete reported usage, excluding unsuccessful or separately archived attempts. These records contain 325,749 model replies, 799,569 tool calls, and 8.25 billion tokens. Input totals include cached tokens; output totals include reported reasoning tokens, without counting either component twice. Rule CEO and No Action CEO make no model calls. Per-trajectory and per-agent counts are provided in \path{anc/data/workload}.

\begin{table}[t]
\centering\small
\caption{Recorded workload across the 53 evaluation runs. Counted windows are successfully sealed decisions with complete reported usage. Model replies per trajectory are the mean $\pm$ sample SD of the counted replies across each agent's company trajectories. Token quantities are millions (M).}
\label{tab:recorded-workload}
\textbf{(a) Decisions and interactions}\par\smallskip
\begin{tabularx}{\linewidth}{@{}Yrrrr@{}}
\toprule CEO agent & Trajectories & \shortstack[r]{Counted\\windows} & Model replies & Tool calls\\\midrule
GPT-6 Astra & 47 & 2,376 & 23,973 & 96,763\\
GPT-5.6 Sol & 47 & 2,265 & 27,133 & 135,060\\
Qwen 3.8 Max & 47 & 3,315 & 36,607 & 94,260\\
Claude Sonnet 5 & 47 & 2,244 & 30,959 & 88,768\\
DeepSeek V4 Flash & 47 & 3,273 & 35,343 & 132,440\\
DeepSeek V4 Pro & 47 & 3,096 & 36,527 & 97,639\\
Gemini 3.8 Flash & 47 & 3,116 & 79,285 & 86,070\\
Doubao Seed 1.8 & 47 & 2,626 & 55,922 & 68,569\\
Rule CEO & 40 & 2,880 & 0 & 0\\
No Action CEO & 8 & 576 & 0 & 0\\\midrule
\textbf{Total} & \textbf{424} & \textbf{25,767} & \textbf{325,749} & \textbf{799,569}\\
\bottomrule
\end{tabularx}
\par\medskip
\textbf{(b) Tokens and interaction depth}\par\smallskip
\begin{tabularx}{\linewidth}{@{}Yrrrr@{}}
\toprule CEO agent & Input (M) & Output (M) & Total (M) & \shortstack[r]{Model replies\\per trajectory}\\\midrule
GPT-6 Astra & 478.16 & 30.74 & 508.90 & $510\pm58$\\
GPT-5.6 Sol & 989.26 & 48.93 & 1,038.19 & $577\pm85$\\
Qwen 3.8 Max & 1,533.71 & 183.55 & 1,717.26 & $779\pm58$\\
Claude Sonnet 5 & 1,080.22 & 10.23 & 1,090.45 & $659\pm181$\\
DeepSeek V4 Flash & 1,436.89 & 214.38 & 1,651.27 & $752\pm53$\\
DeepSeek V4 Pro & 796.81 & 90.52 & 887.32 & $777\pm180$\\
Gemini 3.8 Flash & 565.10 & 13.72 & 578.82 & $1,687\pm130$\\
Doubao Seed 1.8 & 733.59 & 39.59 & 773.18 & $1,190\pm333$\\
Rule CEO & 0 & 0 & 0 & 0\\
No Action CEO & 0 & 0 & 0 & 0\\\midrule
\textbf{Total} & \textbf{7,613.74} & \textbf{631.65} & \textbf{8,245.39} & ---\\
\bottomrule
\end{tabularx}
\end{table}

%% file: research/verified_economics.bib
@book{train2009discrete,
  author = {Kenneth E. Train},
  title = {Discrete Choice Methods with Simulation},
  edition = {2},
  publisher = {Cambridge University Press},
  year = {2009},
  url = {https://eml.berkeley.edu/books/choice2.html}
}

@article{nerlove1962advertising,
  author = {Marc Nerlove and Kenneth J. Arrow},
  title = {Optimal Advertising Policy under Dynamic Conditions},
  journal = {Economica},
  volume = {29},
  number = {114},
  pages = {129--142},
  year = {1962},
  doi = {10.2307/2551549},
  url = {https://doi.org/10.2307/2551549}
}

@article{griliches1979issues,
  author = {Zvi Griliches},
  title = {Issues in Assessing the Contribution of Research and Development to Productivity Growth},
  journal = {The Bell Journal of Economics},
  volume = {10},
  number = {1},
  pages = {92--116},
  year = {1979},
  doi = {10.2307/3003321},
  url = {https://doi.org/10.2307/3003321}
}

@article{arrow1951optimal,
  author = {Kenneth J. Arrow and Theodore Harris and Jacob Marschak},
  title = {Optimal Inventory Policy},
  journal = {Econometrica},
  volume = {19},
  number = {3},
  pages = {250--272},
  year = {1951},
  doi = {10.2307/1906813},
  url = {https://doi.org/10.2307/1906813}
}

@article{anderson1993antecedents,
  author = {Eugene W. Anderson and Mary W. Sullivan},
  title = {The Antecedents and Consequences of Customer Satisfaction for Firms},
  journal = {Marketing Science},
  volume = {12},
  number = {2},
  pages = {125--143},
  year = {1993},
  doi = {10.1287/mksc.12.2.125},
  url = {https://doi.org/10.1287/mksc.12.2.125}
}

@article{little1961proof,
  author = {John D. C. Little},
  title = {A Proof for the Queuing Formula: {$L = \lambda W$}},
  journal = {Operations Research},
  volume = {9},
  number = {3},
  pages = {383--387},
  year = {1961},
  doi = {10.1287/opre.9.3.383},
  url = {https://doi.org/10.1287/opre.9.3.383}
}

@article{bijmolt2005price,
  author = {Tammo H. A. Bijmolt and Harald J. van Heerde and Rik G. M. Pieters},
  title = {New Empirical Generalizations on the Determinants of Price Elasticity},
  journal = {Journal of Marketing Research},
  volume = {42},
  number = {2},
  pages = {141--156},
  year = {2005},
  doi = {10.1509/jmkr.42.2.141.62296},
  url = {https://doi.org/10.1509/jmkr.42.2.141.62296}
}

@article{sethuraman2011advertising,
  author = {Raj Sethuraman and Gerard J. Tellis and Richard A. Briesch},
  title = {How Well Does Advertising Work? Generalizations from Meta-Analysis of Brand Advertising Elasticities},
  journal = {Journal of Marketing Research},
  volume = {48},
  number = {3},
  pages = {457--471},
  year = {2011},
  doi = {10.1509/jmkr.48.3.457},
  url = {https://doi.org/10.1509/jmkr.48.3.457}
}

@article{hudgens2008interference,
  title = {{Toward Causal Inference With Interference}},
  author = {Hudgens, Michael G. and Halloran, M. Elizabeth},
  journal = {Journal of the American Statistical Association},
  volume = {103},
  number = {482},
  pages = {832--842},
  year = {2008},
  doi = {10.1198/016214508000000292},
  url = {https://doi.org/10.1198/016214508000000292}
}

@article{aronow2017interference,
  title = {{Estimating Average Causal Effects Under General Interference, with Application to a Social Network Experiment}},
  author = {Aronow, Peter M. and Samii, Cyrus},
  journal = {The Annals of Applied Statistics},
  volume = {11},
  number = {4},
  pages = {1912--1947},
  year = {2017},
  doi = {10.1214/16-AOAS1005},
  url = {https://doi.org/10.1214/16-AOAS1005}
}

@article{tesfatsion2002ace,
  title = {{Agent-Based Computational Economics: Growing Economies From the Bottom Up}},
  author = {Tesfatsion, Leigh},
  journal = {Artificial Life},
  volume = {8},
  number = {1},
  pages = {55--82},
  year = {2002},
  doi = {10.1162/106454602753694765},
  url = {https://doi.org/10.1162/106454602753694765}
}


%% file: research/verified_ml.bib
@misc{chen2026ceobench,
  title = {{CEO-Bench}: Can Agents Play the Long Game?},
  author = {Chen, Haozhe and Narasimhan, Karthik and Liu, Zhuang},
  year = {2026},
  eprint = {2606.18543},
  archivePrefix = {arXiv},
  primaryClass = {cs.AI},
  doi = {10.48550/arXiv.2606.18543},
  url = {https://arxiv.org/abs/2606.18543v2}
}

@inproceedings{zhang2025aflow,
  title = {{AFlow}: Automating Agentic Workflow Generation},
  author = {Zhang, Jiayi and Xiang, Jinyu and Yu, Zhaoyang and Teng, Fengwei and Chen, Xiong-Hui and Chen, Jiaqi and Zhuge, Mingchen and Cheng, Xin and Hong, Sirui and Wang, Jinlin and Zheng, Bingnan and Liu, Bang and Luo, Yuyu and Wu, Chenglin},
  booktitle = {The Thirteenth International Conference on Learning Representations},
  year = {2025},
  url = {https://openreview.net/forum?id=z5uVAKwmjf}
}

@misc{sugiura2026coffeebench,
  title = {{CoffeeBench}: Benchmarking Long-Horizon {LLM} Agents in Heterogeneous Multi-Agent Economies},
  author = {Sugiura, Issa and Hattori, Daichi and Araragi, Kazuo and Ogawa, Keita and Onose, Shota and Makino, Taro and Usuki, Teppei and Ishida, Takashi},
  year = {2026},
  eprint = {2606.16613},
  archivePrefix = {arXiv},
  primaryClass = {cs.AI},
  doi = {10.48550/arXiv.2606.16613},
  url = {https://arxiv.org/abs/2606.16613v1}
}

@inproceedings{zheng2026marketbench,
  title = {{Market-Bench}: Benchmarking Large Language Models on Economic and Trade Competition},
  author = {Zheng, Yushuo and Duan, Huiyu and Zhang, Zicheng and Zhu, Yucheng and Min, Xiongkuo and Zhai, Guangtao},
  booktitle = {Proceedings of the 64th Annual Meeting of the Association for Computational Linguistics (Volume 1: Long Papers)},
  year = {2026},
  month = jul,
  pages = {39893--39906},
  publisher = {Association for Computational Linguistics},
  doi = {10.18653/v1/2026.acl-long.1853},
  url = {https://aclanthology.org/2026.acl-long.1853/}
}

@misc{han2026enterprisearena,
  title = {Can {LLM} Agents Be {CFOs}? Benchmarking Long-Horizon Resource Allocation in an Uncertain Enterprise Environment},
  author = {Han, Yi and Wang, Yan and Qian, Lingfei and Li, Haohang and Cao, Yupeng and He, Yueru and Peng, Xueqing and Shen, Nanhan and Xu, Yitao and Chen, Yankai and Feng, Dongji and Huang, Jimin and Liu, Xue and Nie, Jian-Yun and Ananiadou, Sophia},
  year = {2026},
  eprint = {2603.23638},
  archivePrefix = {arXiv},
  primaryClass = {cs.AI},
  doi = {10.48550/arXiv.2603.23638},
  url = {https://arxiv.org/abs/2603.23638v2}
}

@misc{backlund2025vending,
  title = {{Vending-Bench}: A Benchmark for Long-Term Coherence of Autonomous Agents},
  author = {Backlund, Axel and Petersson, Lukas},
  year = {2025},
  eprint = {2502.15840},
  archivePrefix = {arXiv},
  primaryClass = {cs.AI},
  doi = {10.48550/arXiv.2502.15840},
  url = {https://arxiv.org/abs/2502.15840}
}

@misc{andonVendingArena,
  title = {{Vending-Bench Arena}},
  author = {{Andon Labs}},
  year = {n.d.},
  howpublished = {Official online evaluation report},
  note = {Continuously updated; accessed September 19, 2026},
  url = {https://andonlabs.com/evals/vending-bench-arena}
}

@article{wellman2025egta,
  title = {Empirical Game Theoretic Analysis: A Survey},
  author = {Wellman, Michael P. and Tuyls, Karl and Greenwald, Amy},
  journal = {Journal of Artificial Intelligence Research},
  year = {2025},
  volume = {82},
  pages = {1017--1076},
  doi = {10.1613/jair.1.16146},
  url = {https://arxiv.org/abs/2403.04018v2}
}

@article{zheng2022aieconomist,
  title = {The {AI} Economist: Taxation policy design via two-level deep multiagent reinforcement learning},
  author = {Zheng, Stephan and Trott, Alexander and Srinivasa, Sunil and Parkes, David C. and Socher, Richard},
  journal = {Science Advances},
  year = {2022},
  volume = {8},
  number = {18},
  pages = {eabk2607},
  doi = {10.1126/sciadv.abk2607},
  url = {https://pubmed.ncbi.nlm.nih.gov/35507657/}
}

@article{piao2026agentsociety,
  title = {{AgentSociety}: Large-scale simulation of {LLM}-driven generative agents advances understanding of human behaviors and society},
  author = {Piao, Jinghua and Yan, Yuwei and Zhang, Jun and Li, Nian and Yan, Junbo and Lan, Xiaochong and Lu, Zhihong and Zheng, Zhiheng and Wang, Jing Yi and Zhou, Di and Gao, Chen and Xu, Fengli and Zhang, Fang and Rong, Ke and Su, Jun and Li, Yong},
  journal = {iFuture},
  publisher = {Tsinghua University Press},
  year = {2026},
  month = jul,
  doi = {10.26599/IF.2026.9710004},
  url = {https://doi.org/10.26599/IF.2026.9710004}
}

@inproceedings{liu2024agentbench,
  title = {{AgentBench}: Evaluating {LLMs} as Agents},
  author = {Liu, Xiao and Yu, Hao and Zhang, Hanchen and Xu, Yifan and Lei, Xuanyu and Lai, Hanyu and Gu, Yu and Ding, Hangliang and Men, Kaiwen and Yang, Kejuan and Zhang, Shudan and Deng, Xiang and Zeng, Aohan and Du, Zhengxiao and Zhang, Chenhui and Shen, Sheng and Zhang, Tianjun and Su, Yu and Sun, Huan and Huang, Minlie and Dong, Yuxiao and Tang, Jie},
  booktitle = {The Twelfth International Conference on Learning Representations},
  year = {2024},
  url = {https://arxiv.org/abs/2308.03688}
}

@inproceedings{zhou2024webarena,
  title = {{WebArena}: A Realistic Web Environment for Building Autonomous Agents},
  author = {Zhou, Shuyan and Xu, Frank F. and Zhu, Hao and Zhou, Xuhui and Lo, Robert and Sridhar, Abishek and Cheng, Xianyi and Ou, Tianyue and Bisk, Yonatan and Fried, Daniel and Alon, Uri and Neubig, Graham},
  booktitle = {The Twelfth International Conference on Learning Representations},
  year = {2024},
  url = {https://arxiv.org/abs/2307.13854}
}

@inproceedings{yao2025taubench,
  title = {{$\tau$-bench}: A Benchmark for Tool-Agent-User Interaction in Real-World Domains},
  author = {Yao, Shunyu and Shinn, Noah and Razavi, Pedram and Narasimhan, Karthik},
  booktitle = {The Thirteenth International Conference on Learning Representations},
  year = {2025},
  url = {https://openreview.net/forum?id=roNSXZpUDN}
}

@inproceedings{yao2022react,
  title = {{ReAct: Synergizing Reasoning and Acting in Language Models}},
  author = {Shunyu Yao and Jeffrey Zhao and Dian Yu and Nan Du and Izhak Shafran and Karthik Narasimhan and Yuan Cao},
  booktitle = {International Conference on Learning Representations},
  year = {2023},
  url = {https://react-lm.github.io/}
}

@inproceedings{shinn2023reflexion,
  title = {{Reflexion: Language Agents with Verbal Reinforcement Learning}},
  author = {Noah Shinn and Federico Cassano and Ashwin Gopinath and Karthik Narasimhan and Shunyu Yao},
  booktitle = {Advances in Neural Information Processing Systems},
  year = {2023},
  volume = {36},
  pages = {8634--8652},
  doi = {10.52202/075280-0377},
  url = {https://papers.neurips.cc/paper_files/paper/2023/hash/1b44b878bb782e6954cd888628510e90-Abstract-Conference.html}
}

@inproceedings{park2023generative,
  title = {{Generative Agents: Interactive Simulacra of Human Behavior}},
  author = {Joon Sung Park and Joseph C. O'Brien and Carrie J. Cai and Meredith Ringel Morris and Percy Liang and Michael S. Bernstein},
  booktitle = {Proceedings of the 36th Annual ACM Symposium on User Interface Software and Technology},
  year = {2023},
  pages = {1--22},
  publisher = {Association for Computing Machinery},
  doi = {10.1145/3586183.3606763},
  url = {https://doi.org/10.1145/3586183.3606763}
}

@article{wang2023voyager,
  title = {{Voyager: An Open-Ended Embodied Agent with Large Language Models}},
  author = {Guanzhi Wang and Yuqi Xie and Yunfan Jiang and Ajay Mandlekar and Chaowei Xiao and Yuke Zhu and Linxi Fan and Anima Anandkumar},
  journal = {Transactions on Machine Learning Research},
  year = {2024},
  url = {https://openreview.net/forum?id=ehfRiF0R3a}
}

@inproceedings{wu2023autogen,
  title = {{AutoGen: Enabling Next-Gen LLM Applications via Multi-Agent Conversations}},
  author = {Qingyun Wu and Gagan Bansal and Jieyu Zhang and Yiran Wu and Beibin Li and Erkang Zhu and Li Jiang and Xiaoyun Zhang and Shaokun Zhang and Jiale Liu and Ahmed Awadallah and Ryen W. White and Doug Burger and Chi Wang},
  booktitle = {Conference on Language Modeling},
  year = {2024},
  url = {https://openreview.net/forum?id=BAakY1hNKS}
}

@inproceedings{li2023camel,
  title = {{CAMEL: Communicative Agents for "Mind" Exploration of Large Language Model Society}},
  author = {Guohao Li and Hasan Abed Al Kader Hammoud and Hani Itani and Dmitrii Khizbullin and Bernard Ghanem},
  booktitle = {Advances in Neural Information Processing Systems},
  year = {2023},
  volume = {36},
  pages = {51991--52008},
  doi = {10.52202/075280-2264},
  url = {https://papers.neurips.cc/paper_files/paper/2023/hash/a3621ee907def47c1b952ade25c67698-Abstract-Conference.html}
}

@inproceedings{hong2023metagpt,
  title = {{MetaGPT: Meta Programming for A Multi-Agent Collaborative Framework}},
  author = {Sirui Hong and Mingchen Zhuge and Jonathan Chen and Xiawu Zheng and Yuheng Cheng and Ceyao Zhang and Jinlin Wang and Zili Wang and Steven Ka Shing Yau and Zijuan Lin and Liyang Zhou and Chenyu Ran and Lingfeng Xiao and Chenglin Wu and J{\"u}rgen Schmidhuber},
  booktitle = {International Conference on Learning Representations},
  year = {2024},
  url = {https://proceedings.iclr.cc/paper_files/paper/2024/hash/6507b115562bb0a305f1958ccc87355a-Abstract-Conference.html}
}

@inproceedings{mialon2023gaia,
  title = {{GAIA: a benchmark for General AI Assistants}},
  author = {Gr{\'e}goire Mialon and Cl{\'e}mentine Fourrier and Craig Swift and Thomas Wolf and Yann LeCun and Thomas Scialom},
  booktitle = {International Conference on Learning Representations},
  year = {2024},
  url = {https://proceedings.iclr.cc/paper_files/paper/2024/hash/25ae35b5b1738d80f1f03a8713e405ec-Abstract-Conference.html}
}

@inproceedings{jimenez2023swebench,
  title = {{SWE-bench: Can Language Models Resolve Real-World GitHub Issues?}},
  author = {Carlos E. Jimenez and John Yang and Alexander Wettig and Shunyu Yao and Kexin Pei and Ofir Press and Karthik Narasimhan},
  booktitle = {International Conference on Learning Representations},
  year = {2024},
  url = {https://proceedings.iclr.cc/paper_files/paper/2024/hash/edac78c3e300629acfe6cbe9ca88fb84-Abstract-Conference.html}
}

@inproceedings{xie2024osworld,
  title = {{OSWorld: Benchmarking Multimodal Agents for Open-Ended Tasks in Real Computer Environments}},
  author = {Tianbao Xie and Danyang Zhang and Jixuan Chen and Xiaochuan Li and Siheng Zhao and Ruisheng Cao and Toh Jing Hua and Zhoujun Cheng and Dongchan Shin and Fangyu Lei and Yitao Liu and Yiheng Xu and Shuyan Zhou and Silvio Savarese and Caiming Xiong and Victor Zhong and Tao Yu},
  booktitle = {Advances in Neural Information Processing Systems},
  year = {2024},
  volume = {37},
  pages = {52040--52094},
  doi = {10.52202/079017-1650},
  url = {https://proceedings.neurips.cc/paper_files/paper/2024/hash/5d413e48f84dc61244b6be550f1cd8f5-Abstract-Datasets_and_Benchmarks_Track.html}
}

@inproceedings{drouin2024workarena,
  title = {{WorkArena: How Capable Are Web Agents at Solving Common Knowledge Work Tasks?}},
  author = {Alexandre Drouin and Maxime Gasse and Massimo Caccia and Issam H. Laradji and Manuel Del Verme and Tom Marty and David Vazquez and Nicolas Chapados and Alexandre Lacoste},
  booktitle = {Proceedings of the 41st International Conference on Machine Learning},
  year = {2024},
  volume = {235},
  series = {Proceedings of Machine Learning Research},
  pages = {11642--11662},
  publisher = {PMLR},
  url = {https://proceedings.mlr.press/v235/drouin24a.html}
}

@inproceedings{xu2024agentcompany,
  title = {{TheAgentCompany: Benchmarking LLM Agents on Consequential Real World Tasks}},
  author = {Frank F. Xu and Yufan Song and Boxuan Li and Yuxuan Tang and Kritanjali Jain and Mengxue Bao and Zora Z. Wang and Xuhui Zhou and Zhitong Guo and Murong Cao and Mingyang Yang and Hao Yang Lu and Amaad Martin and Zhe Su and Leander Melroy Maben and Raj Mehta and Wayne Chi and Lawrence Jang and Yiqing Xie and Shuyan Zhou and Graham Neubig},
  booktitle = {Advances in Neural Information Processing Systems},
  year = {2025},
  volume = {38},
  doi = {10.52202/085713-0315},
  url = {https://papers.nips.cc/paper_files/paper/2025/hash/0d744742f6fac4d1134c019b7cef3c8a-Abstract-Datasets_and_Benchmarks_Track.html}
}

@inproceedings{chan2024mlebench,
  title = {{MLE-bench: Evaluating Machine Learning Agents on Machine Learning Engineering}},
  author = {Jun Shern Chan and Neil Chowdhury and Oliver Jaffe and James Aung and Dane Sherburn and Evan Mays and Giulio Starace and Kevin Liu and Leon Maksin and Tejal Patwardhan and Lilian Weng and Aleksander M{\k{a}}dry},
  booktitle = {International Conference on Learning Representations},
  year = {2025},
  url = {https://proceedings.iclr.cc/paper_files/paper/2025/hash/7e3767db483c942b883eb4f8cfb74e31-Abstract-Conference.html}
}

@inproceedings{starace2025paperbench,
  title = {{PaperBench: Evaluating AI's Ability to Replicate AI Research}},
  author = {Giulio Starace and Oliver Jaffe and Dane Sherburn and James Aung and Jun Shern Chan and Leon Maksin and Rachel Dias and Evan Mays and Benjamin Kinsella and Wyatt Thompson and Johannes Heidecke and Amelia Glaese and Tejal Patwardhan},
  booktitle = {Proceedings of the 42nd International Conference on Machine Learning},
  year = {2025},
  volume = {267},
  series = {Proceedings of Machine Learning Research},
  pages = {56843--56873},
  publisher = {PMLR},
  url = {https://proceedings.mlr.press/v267/starace25a.html}
}

@inproceedings{qin2023toolllm,
  title = {{ToolLLM: Facilitating Large Language Models to Master 16000+ Real-world APIs}},
  author = {Yujia Qin and Shihao Liang and Yining Ye and Kunlun Zhu and Lan Yan and Yaxi Lu and Yankai Lin and Xin Cong and Xiangru Tang and Bill Qian and Sihan Zhao and Lauren Hong and Runchu Tian and Ruobing Xie and Jie Zhou and Mark Gerstein and Dahai Li and Zhiyuan Liu and Maosong Sun},
  booktitle = {International Conference on Learning Representations},
  year = {2024},
  url = {https://proceedings.iclr.cc/paper_files/paper/2024/hash/28e50ee5b72e90b50e7196fde8ea260e-Abstract-Conference.html}
}

@inproceedings{ma2024agentboard,
  title = {{AgentBoard: An Analytical Evaluation Board of Multi-turn LLM Agents}},
  author = {Chang Ma and Junlei Zhang and Zhihao Zhu and Cheng Yang and Yujiu Yang and Yaohui Jin and Zhenzhong Lan and Lingpeng Kong and Junxian He},
  booktitle = {Advances in Neural Information Processing Systems},
  year = {2024},
  volume = {37},
  pages = {74325--74362},
  doi = {10.52202/079017-2365},
  url = {https://proceedings.nips.cc/paper_files/paper/2024/hash/877b40688e330a0e2a3fc24084208dfa-Abstract-Datasets_and_Benchmarks_Track.html}
}

@inproceedings{wang2023mint,
  title = {{MINT: Evaluating LLMs in Multi-turn Interaction with Tools and Language Feedback}},
  author = {Xingyao Wang and Zihan Wang and Jiateng Liu and Yangyi Chen and Lifan Yuan and Hao Peng and Heng Ji},
  booktitle = {International Conference on Learning Representations},
  year = {2024},
  url = {https://proceedings.iclr.cc/paper_files/paper/2024/hash/8a0d3ae989a382ce6e50312bc35bf7e1-Abstract-Conference.html}
}

@inproceedings{shridhar2020alfworld,
  title = {{ALFWorld: Aligning Text and Embodied Environments for Interactive Learning}},
  author = {Mohit Shridhar and Xingdi Yuan and Marc-Alexandre C{\^o}t{\'e} and Yonatan Bisk and Adam Trischler and Matthew Hausknecht},
  booktitle = {International Conference on Learning Representations},
  year = {2021},
  url = {https://alfworld.github.io/}
}

@inproceedings{wang2022scienceworld,
  title = {{ScienceWorld: Is your Agent Smarter than a 5th Grader?}},
  author = {Ruoyao Wang and Peter Jansen and Marc-Alexandre C{\^o}t{\'e} and Prithviraj Ammanabrolu},
  booktitle = {Proceedings of the 2022 Conference on Empirical Methods in Natural Language Processing},
  year = {2022},
  pages = {11279--11298},
  publisher = {Association for Computational Linguistics},
  doi = {10.18653/v1/2022.emnlp-main.775},
  url = {https://aclanthology.org/2022.emnlp-main.775/}
}

@inproceedings{leibo2021meltingpot,
  title = {{Scalable Evaluation of Multi-Agent Reinforcement Learning with Melting Pot}},
  author = {Joel Z. Leibo and Edgar Du{\'e}{\~n}ez-Guzm{\'a}n and Alexander Sasha Vezhnevets and John P. Agapiou and Peter Sunehag and Raphael Koster and Jayd Matyas and Charles Beattie and Igor Mordatch and Thore Graepel},
  booktitle = {Proceedings of the 38th International Conference on Machine Learning},
  year = {2021},
  volume = {139},
  series = {Proceedings of Machine Learning Research},
  pages = {6187--6199},
  publisher = {PMLR},
  url = {https://proceedings.mlr.press/v139/leibo21a.html}
}

@misc{lanctot2019openspiel,
  title = {{OpenSpiel: A Framework for Reinforcement Learning in Games}},
  author = {Marc Lanctot and Edward Lockhart and Jean-Baptiste Lespiau and Vinicius Zambaldi and Satyaki Upadhyay and Julien P{\'e}rolat and Sriram Srinivasan and Finbarr Timbers and Karl Tuyls and Shayegan Omidshafiei and Daniel Hennes and Dustin Morrill and Paul Muller and Timo Ewalds and Ryan Faulkner and J{\'a}nos Kram{\'a}r and Bart De Vylder and Brennan Saeta and James Bradbury and David Ding and Sebastian Borgeaud and Matthew Lai and Julian Schrittwieser and Thomas Anthony and Edward Hughes and Ivo Danihelka and Jonah Ryan-Davis},
  year = {2019},
  eprint = {1908.09453},
  archivePrefix = {arXiv},
  url = {https://arxiv.org/abs/1908.09453}
}

@misc{suarez2019neuralmmo,
  title = {{Neural MMO: A Massively Multiagent Game Environment for Training and Evaluating Intelligent Agents}},
  author = {Joseph Suarez and Yilun Du and Phillip Isola and Igor Mordatch},
  year = {2019},
  eprint = {1903.00784},
  archivePrefix = {arXiv},
  url = {https://arxiv.org/abs/1903.00784}
}

@inproceedings{baker2019autocurricula,
  title = {{Emergent Tool Use From Multi-Agent Autocurricula}},
  author = {Bowen Baker and Ingmar Kanitscheider and Todor Markov and Yi Wu and Glenn Powell and Bob McGrew and Igor Mordatch},
  booktitle = {International Conference on Learning Representations},
  year = {2020},
  url = {https://openreview.net/pdf/9296b021ee5071e8a5433b6255270e0e20d8f532.pdf}
}

@misc{bakhtin2022nopress,
  title = {{Mastering the Game of No-Press Diplomacy via Human-Regularized Reinforcement Learning and Planning}},
  author = {Anton Bakhtin and David J Wu and Adam Lerer and Jonathan Gray and Athul Paul Jacob and Gabriele Farina and Alexander H Miller and Noam Brown},
  year = {2022},
  eprint = {2210.05492},
  archivePrefix = {arXiv},
  url = {https://arxiv.org/abs/2210.05492}
}

@misc{horton2023homosilicus,
  title = {{Large Language Models as Simulated Economic Agents: What Can We Learn from Homo Silicus?}},
  author = {John J. Horton and Apostolos Filippas and Benjamin S. Manning},
  year = {2023},
  eprint = {2301.07543},
  archivePrefix = {arXiv},
  url = {https://arxiv.org/abs/2301.07543}
}

@inproceedings{huang2023mlagentbench,
  title = {{MLAgentBench: Evaluating Language Agents on Machine Learning Experimentation}},
  author = {Qian Huang and Jian Vora and Percy Liang and Jure Leskovec},
  booktitle = {Proceedings of the 41st International Conference on Machine Learning},
  year = {2024},
  volume = {235},
  series = {Proceedings of Machine Learning Research},
  pages = {20271--20309},
  publisher = {PMLR},
  url = {https://proceedings.mlr.press/v235/huang24y.html}
}

@misc{xi2024agentgym,
  title = {{AgentGym: Evolving Large Language Model-based Agents across Diverse Environments}},
  author = {Zhiheng Xi and Yiwen Ding and Wenxiang Chen and Boyang Hong and Honglin Guo and Junzhe Wang and Dingwen Yang and Chenyang Liao and Xin Guo and Wei He and Songyang Gao and Lu Chen and Rui Zheng and Yicheng Zou and Tao Gui and Qi Zhang and Xipeng Qiu and Xuanjing Huang and Zuxuan Wu and Yu-Gang Jiang},
  year = {2024},
  eprint = {2406.04151},
  archivePrefix = {arXiv},
  url = {https://arxiv.org/abs/2406.04151}
}

@inproceedings{zhou2023sotopia,
  title = {{SOTOPIA: Interactive Evaluation for Social Intelligence in Language Agents}},
  author = {Xuhui Zhou and Hao Zhu and Leena Mathur and Ruohong Zhang and Haofei Yu and Zhengyang Qi and Louis-Philippe Morency and Yonatan Bisk and Daniel Fried and Graham Neubig and Maarten Sap},
  booktitle = {International Conference on Learning Representations},
  year = {2024},
  url = {https://proceedings.iclr.cc/paper_files/paper/2024/hash/b3075b88e583a0e98d8b24338a613060-Abstract-Conference.html}
}

@article{vinyals2019alphastar,
  title = {{Grandmaster level in StarCraft II using multi-agent reinforcement learning}},
  author = {Vinyals, Oriol and Babuschkin, Igor and Czarnecki, Wojciech M. and others},
  journal = {Nature},
  volume = {575},
  pages = {350--354},
  year = {2019},
  doi = {10.1038/s41586-019-1724-z},
  url = {https://www.nature.com/articles/s41586-019-1724-z}
}

@article{fair2022cicero,
  title = {{Human-level play in the game of Diplomacy by combining language models with strategic reasoning}},
  author = {{Meta Fundamental AI Research Diplomacy Team (FAIR)}},
  journal = {Science},
  volume = {378},
  number = {6624},
  pages = {1067--1074},
  year = {2022},
  doi = {10.1126/science.ade9097},
  url = {https://doi.org/10.1126/science.ade9097}
}

@inproceedings{terry2021pettingzoo,
  title = {{PettingZoo: Gym for Multi-Agent Reinforcement Learning}},
  author = {Terry, J. K. and Black, Benjamin and Grammel, Nathaniel and Jayakumar, Mario and Hari, Ananth and Sullivan, Ryan and Santos, Luis S. and Dieffendahl, Clemens and Horsch, Caroline and Perez-Vicente, Rodrigo and others},
  booktitle = {Advances in Neural Information Processing Systems},
  volume = {34},
  pages = {15032--15043},
  year = {2021},
  url = {https://papers.neurips.cc/paper_files/paper/2021/hash/803f7c4c3ff61b71be53a0c803bfb57f-Abstract.html}
}

@inproceedings{zhu-etal-2025-multiagentbench,
  title = "{M}ulti{A}gent{B}ench : Evaluating the Collaboration and Competition of {LLM} agents",
  author = "Zhu, Kunlun and Du, Hongyi and Hong, Zhaochen and Yang, Xiaocheng and Guo, Shuyi and Wang, Zhe and Wang, Zhenhailong and Qian, Cheng and Tang, Xiangru and Ji, Heng and You, Jiaxuan",
  booktitle = "Proceedings of the 63rd Annual Meeting of the Association for Computational Linguistics (Volume 1: Long Papers)",
  month = jul,
  year = "2025",
  address = "Vienna, Austria",
  publisher = "Association for Computational Linguistics",
  url = "https://aclanthology.org/2025.acl-long.421/",
  doi = "10.18653/v1/2025.acl-long.421",
  pages = "8580--8622"
}

@inproceedings{chen-etal-2024-llmarena,
  title = "{LLMA}rena: Assessing Capabilities of Large Language Models in Dynamic Multi-Agent Environments",
  author = "Chen, Junzhe and Hu, Xuming and Liu, Shuodi and Huang, Shiyu and Tu, Wei-Wei and He, Zhaofeng and Wen, Lijie",
  booktitle = "Proceedings of the 62nd Annual Meeting of the Association for Computational Linguistics (Volume 1: Long Papers)",
  month = aug,
  year = "2024",
  address = "Bangkok, Thailand",
  publisher = "Association for Computational Linguistics",
  url = "https://aclanthology.org/2024.acl-long.705/",
  doi = "10.18653/v1/2024.acl-long.705",
  pages = "13055--13077"
}
